\documentclass[submission,copyright,creativecommons]{eptcs}
\providecommand{\event}{TERMGRAPH 2014} % Name of the event you are submitting to

\usepackage{amsfonts,amsmath,amsthm,amscd,amssymb,float}
\usepackage[german,british,english]{babel}
\usepackage{stmaryrd}
\usepackage{comment}
\usepackage{xspace}
\usepackage{color,graphicx}
\usepackage{MnSymbol}
\usepackage{microtype}
\usepackage{tikz}
\usepackage{breakurl}      
\usepackage{tkz-berge}
\usepackage{enumerate}
\usepackage{fleqn}
\usepackage{wrapfig}
\usepackage{listings}
\usepackage{tikz}
\usepackage{pgf}
\usetikzlibrary{arrows,shapes,calc,%automata,backgrounds,calc,chains,fadings,fit,patterns,snakes,
                positioning,%shapes.geometric,shapes.symbols,through,trees,
                }
\theoremstyle{plain}
\newtheorem{theorem}{Theorem}

\newtheorem*{corollary*}{Corollary}
\newtheorem{lemma}[theorem]{Lemma}
\newtheorem{proposition}[theorem]{Proposition}

\theoremstyle{definition}
\newtheorem{definition}[theorem]{Definition}
\newtheorem{remark}[theorem]{Remark}
\newtheorem{example}[theorem]{Example}

\newcommand{\nb}{\nobreakdash}

\newcommand{\niks}{}

\definecolor{azure}{rgb}{0.94,1.00,1.00}
\definecolor{blue}{rgb}{0,0,0.5}
\definecolor{brown}{rgb}{.75,.25,.25}
\definecolor{cyan}{rgb}{0.25,0.88,0.82}
\definecolor{chocolate}{rgb}{0.82,0.41,0.12}
\definecolor{darkcyan}{rgb}{0.5,0,1}
\definecolor{darkgreen}{rgb}{0,0.39,0}
\definecolor{darkmagenta}{rgb}{0.5,0,0.5}
\definecolor{firebrick}{RGB}{175,25,25}
\definecolor{forestgreen}{rgb}{0.13,0.55,0.13}
\definecolor{lightcyan}{rgb}{0.88,1.00,1.00}
\definecolor{lightpink}{rgb}{1.00,0.71,0.76}
\definecolor{lightyellow}{rgb}{1.00,1.00,0.88}
\definecolor{lightgoldenrod}{rgb}{0.83,0.97,0.51}
\definecolor{lightgoldenrodyellow}{rgb}{0.98,0.98,0.82}
\definecolor{lightskyblue}{rgb}{0.53,0.81,0.98}
\definecolor{moccasin}{rgb}{1.00,0.89,0.71}
\definecolor{magenta}{rgb}{1,0,1}
\definecolor{navyblue}{rgb}{0,0,0.5}
\definecolor{orange}{rgb}{1.0,0.65,0.0}
\definecolor{orangered}{rgb}{1.0,0.27,0.0}
\definecolor{palegreen}{rgb}{0.60,0.98,0.60}
\definecolor{powderblue}{rgb}{0.69,0.88,0.90}
\definecolor{purple}{rgb}{1,0.5,1}
\definecolor{royalblue}{RGB}{65,105,225}
\definecolor{mediumblue}{RGB}{0,0,205}
\definecolor{cornflowerblue}{RGB}{100,149,237}
\definecolor{springgreen}{rgb}{0.0,1.0,0.5}
\definecolor{turquoise}{rgb}{0.25,0.88,0.82}
\definecolor{snow}{rgb}{1.00,0.98,0.98}
\definecolor{tan}{rgb}{0.82,0.71,0.55}
\definecolor{red}{rgb}{1,0,0}

\newcommand{\myparagraphbf}[1]{\vspace{1ex}\noindent{\textbf{{#1}.}}}

\newcommand{\funin}{\mathrel{:}}
\newcommand{\fap}[2]{#1({#2})}

\newcommand{\indap}[2]{#1 _{#2}}

\newcommand{\supap}[2]{#1 ^{#2}}

\newcommand{\pbap}[3]{#1 ^{#2}_{#3}}

\newcommand{\length}[1]{{\left|{#1}\right|}}

\newcommand{\sscompfuns}{\circ}
\newcommand{\scompfuns}[2]{{#1}\mathrel{\sscompfuns}{#2}}

\newcommand{\sidfun}{\text{\normalfont id}}

\newcommand{\sidfunon}[1]{\sidfun_{#1}}

\newcommand{\defd}[1]{{#1}{\downarrow}}

\newcommand{\sproji}{\indap{\pi}}

\newcommand{\tuple}[1]{\langle #1 \rangle}
\newcommand{\tuplespace}{\hspace*{0.5pt}}
\newcommand{\pair}[2]{\tuple{#1, \tuplespace #2}}

\newcommand{\quadruple}[4]{\tuple{#1, \tuplespace #2, \tuplespace #3, \tuplespace #4}}

\newcommand{\wordsover}[1]{{#1^*}}
\newcommand{\emptyword}{\epsilon}
\newcommand{\worddots}{\hspace*{0.2pt}\cdots\hspace{1.3pt}}
\newcommand{\wcns}{\cdot}

\newcommand{\descsetexpmid}{\mathrel{\vert}}

\newcommand{\descsetexp}[2]{\left\{{#1}\descsetexpmid{#2}\right\}}

\newcommand{\setexp}[1]{\left\{{#1}\right\}}

\newcommand{\spowersetof}{\powerset}
\newcommand{\powersetof}{\fap{\spowersetof}}
\renewcommand{\emptyset}{\varnothing}

\newcommand{\aset}{A}

\newcommand{\nat}{\mathbb{N}}

\newcommand{\slogand}{{\wedge}}
\newcommand{\logand}{\mathrel{\slogand}}
\newcommand{\slogor}{{\vee}}
\newcommand{\logor}{\mathrel{\slogor}}

\newcommand{\rootconnected}{root-con\-nect\-ed}
\newcommand{\lambdaterm}{$\sslabs$\nb-term}
\newcommand{\lambdaterms}{\lambdaterm{s}}
\newcommand{\lambdacalculus}{$\lambda$\nb-cal\-cu\-lus}

\newcommand{\lambdahigherordertg}{$\lambda$-higher-order-term-graph}
\newcommand{\lambdahigherordertgs}{\lambdahigherordertg s}
\newcommand{\lambdatg}{$\lambda$-term-graph}
\newcommand{\lambdatgs}{\lambdatg s}

\newcommand{\sgletrec}{\ensuremath{\mathsf{gletrec}}}
\newcommand{\gletrecexpression}{\sgletrec\nb-ex\-pres\-sion}
\newcommand{\gletrecexpressions}{\gletrecexpression s}
\newcommand{\gletrecnotation}{\sgletrec\nb-notation}

\newcommand{\rgsrepresentation}{\rgs\nb-re\-pre\-sen\-ta\-tion}

\newcommand{\rgs}{rgs{}}
\newcommand{\rgss}{\rgs's}
\newcommand{\ntg}{ntg}
\newcommand{\ntgs}{\ntg{s}}

\newcommand{\entg}{sntg{}}
\newcommand{\entgs}{\entg's}

\newcommand{\ntgsig}{\ntg\nb-sig\-na\-ture}
\newcommand{\ntgsigs}{\ntgsig{s}}

\newcommand{\ARS}{ARS}

\newcommand{\HRS}{HRS}

\newcommand{\HRSterms}{\HRS\nb-terms}

\newcommand{\txtletrec}{\ensuremath{\textsf{letrec}}}

\newcommand{\lambdalifting}{lambda-lif\-ting}

\newcommand{\sslabs}{\lambda}
\newcommand{\sslapp}{@}
\newcommand{\snlvar}{\mathsf{v}}
\newcommand{\snlvaracc}{\mathsf{v}^{\prime}}
\newcommand{\snlvarsucc}{\text{S}}
\newcommand{\srootsucc}{\text{S}_{\text{r}}}

\newcommand{\ssin}{\mathsf{i}}
\newcommand{\ssini}[1]{\ssin_{#1}}
\newcommand{\ssout}{\mathsf{o}}
\newcommand{\ssouti}[1]{\ssout_{#1}}
\newcommand{\ssinroot}{\ssin_{\mathsf{r}}}
\newcommand{\ssoutroot}{\ssout_{\mathsf{r}}}

\newcommand{\sinvert}{\mathsf{i}}
\newcommand{\soutvert}{\mathsf{o}}%{\scalebox{0.8}{$\mathsf{out}$}}
\newcommand{\sinroot}{\sinvert_{\mathsf{r}}}
\newcommand{\soutroot}{\soutvert_{\mathsf{r}}} %{\scalebox{0.8}{$\mathsf{ret}$}_{\mathsf{r}}}

\newcommand{\adbmal}{\ensuremath{\reflectbox{$\lambda$}}}

\newcommand{\avar}{x}

\newcommand{\avari}[1]{\avar_{#1}}

\newcommand{\slabsmindot}[1]{\sslabs{#1}}

\newcommand{\nats}{\mathbb{N}}

\newcommand{\asig}{\Sigma}

\newcommand{\asigacc}{\Sigma'}

\newcommand{\sarity}{{\mit ar}}
\newcommand{\arity}{\fap{\sarity}}

\newcommand{\asigi}[1]{\asig_{#1}}

\newcommand{\satomic}{\text{\normalfont at}}
\newcommand{\snested}{\text{\normalfont ne}}
\newcommand{\sconstant}{\text{\normalfont const}}
\newcommand{\sfun}{\text{\normalfont fun}}

\newcommand{\asigat}{\asig_{\satomic}}
\newcommand{\asigatconst}{\asig_{\satomic,\sconstant}}
\newcommand{\asigatconstprime}{\asig_{\satomic,\sconstant^{\prime}}}
\newcommand{\asigatfun}{\asig_{\satomic,\sfun}}

\newcommand{\asigne}{\asig_{\snested}}

\newcommand{\asigine}[1]{\asig_{{#1},\snested}}

\newcommand{\Ins}{\mathit{I}}
\newcommand{\Outs}{\mathit{O}}

\newcommand{\OutIns}{\mathit{OI}}

\newcommand{\afunsymvar}{f}
\newcommand{\bfunsymvar}{g}

\newcommand{\afunsymvari}{\indap{\afunsymvar}}

\newcommand{\afunsym}{\darkcyan{\mathsf{f}}}
\newcommand{\bfunsym}{\darkcyan{\mathsf{g}}}

\newcommand{\afunsymi}[1]{\indap{\afunsym}{\darkcyan{#1}}}
\newcommand{\bfunsymi}[1]{\indap{\bfunsym}{\darkcyan{#1}}}

\newcommand{\graphafunsym}{F}%{\mathsf{f}}
\newcommand{\graphbfunsym}{G}%{\mathsf{g}}
\newcommand{\graphafunsymi}{\indap{\graphafunsym}}
\newcommand{\graphbfunsymi}{\indap{\graphbfunsym}}

\newcommand{\avarsym}{\mathsf{x}}

\newcommand{\avarsymi}{\indap{\avarsym}}

\newcommand{\aconstsym}{\mathsf{c}}

\newcommand{\verts}{V}
\newcommand{\svlab}{\mathit{lab}}
\newcommand{\vlab}{\fap{\svlab}}
\newcommand{\svargs}{\mathit{args}}
\newcommand{\vargs}{\fap{\svargs}}
\newcommand{\vargsati}[1]{\fap{\vargs{#1}}}
\newcommand{\sroot}{\mathit{root}}

\newcommand{\vertsi}{\indap{V}}
\newcommand{\svlabi}{\indap{\mathit{lab}}}
\newcommand{\vlabi}[1]{\fap{\svlabi{#1}}}
\newcommand{\svargsi}{\indap{\mathit{args}}}
\newcommand{\vargsi}[1]{\fap{\svargsi{#1}}}

\newcommand{\srooti}{\indap{\mathit{root}}}

\newcommand{\svanc}{\mathit{anc}}
\newcommand{\vanc}{\fap{\svanc}}
\newcommand{\svin}{\mathit{call}}
\newcommand{\vin}{\fap{\svin}}
\newcommand{\svout}{\mathit{return}}
\newcommand{\vout}{\fap{\svout}}

\newcommand{\svanci}{\indap{\mathit{anc}}}
\newcommand{\vanci}[1]{\fap{\svanci{#1}}}
\newcommand{\svini}{\indap{\mathit{call}}}
\newcommand{\vini}[1]{\fap{\svini{#1}}}
\newcommand{\svouti}{\indap{\mathit{return}}}
\newcommand{\vouti}[1]{\fap{\indap{\svout}{#1}}}

\newcommand{\stgsucci}{\indap{\rightarrowtail}}
\newcommand{\tgsucci}[1]{\mathrel{\stgsucci{#1}}}

\newcommand{\srootof}{\textit{rt}}

\newcommand{\soutsof}{\textit{ins}}

\newcommand{\srootofi}{\indap{\srootof}}
\newcommand{\rootofi}[1]{\fap{\srootofi{#1}}}

\newcommand{\soutsofi}{\indap{\soutsof}}
\newcommand{\outsofi}[1]{\fap{\soutsofi{#1}}}

\newcommand{\srecspec}{\mathit{rec}}
\newcommand{\recspec}{\fap{\srecspec}}
\newcommand{\srecspeci}{\indap{\mathit{rec}}}
\newcommand{\recspeci}[1]{\fap{\srecspeci{#1}}}
\newcommand{\srootsym}{\darkcyan{\mathsf{r}}}
\newcommand{\srootsymi}[1]{\indap{\srootsym}{\darkcyan{#1}}}

\newcommand{\graphsrootsym}{R}
\newcommand{\graphsrootsymi}{\indap{\graphsrootsym}}

\newcommand{\srootsymvar}{r}
\newcommand{\srootsymvari}{\indap{\srootsymvar}}

\newcommand{\sntgrootsym}{\darkcyan{\mathsf{n}}}

\newcommand{\graphsntgrootsym}{N}

\newcommand{\subtgat}[2]{{#1}|_{#2}}

\newcommand{\sdependson}{\leftspoon}
\newcommand{\dependson}{\mathrel{\sdependson}}

\newcommand{\dependsonARS}{\leftspoon}
\newcommand{\dependsonARSignoreof}[1]{\dependsonARS}

\newcommand{\stepdependsonARSignoreof}[1]{\mathrel{\dependsonARSignoreof{#1}}}

\newcommand{\prefix}[1]{(\hspace*{-0.75pt}{#1}\hspace*{-1pt})}

\newcommand{\rrgs}{{\cal R}}

\newcommand{\rrgsi}{\indap{\rrgs}}

\newcommand{\preexp}[2]{(#1){#2}}
\newcommand{\emptypreexp}[1]{\preexp{\niks}{#1}}

\newcommand{\avert}{v}
\newcommand{\bvert}{w}
\newcommand{\cvert}{u}
\newcommand{\dvert}{x}
\newcommand{\evert}{y}
\newcommand{\averti}{\indap{\avert}}
\newcommand{\bverti}{\indap{\bvert}}

\newcommand{\avertacc}{\supap{\avert}{\prime}}
\newcommand{\bvertacc}{\supap{\bvert}{\prime}}

\newcommand{\avertacci}[1]{\indap{\avertacc}{\hspace*{-1pt}{#1}}}
\newcommand{\bvertacci}[1]{\indap{\bvertacc}{\hspace*{-1pt}{#1}}}

\newcommand{\classtgsover}{\fap{{\text{\normalfont TG}}}}

\newcommand{\classntgsover}{\fap{\text{$\mathcal{N\hspace*{-1pt}G}$}}}
\newcommand{\classrtgsover}{\fap{\text{\normalfont RG}}}

\newcommand{\sntgstortgs}{I}
\newcommand{\ntgstortgs}{\fap{\sntgstortgs}}
\newcommand{\srtgstontgs}{{R}}

\newcommand{\antg}{{\cal N}}
\newcommand{\antgi}{\indap{\antg}}

\newcommand{\aedntg}{{\cal G}}
\newcommand{\aedntgi}{\indap{\aedntg}}

\newcommand{\atg}{G}
\newcommand{\atgi}{\indap{\atg}}

\newcommand{\srgstontg}{{\cal N}}
\newcommand{\rgstontg}{\fap{\srgstontg}}
\newcommand{\ntgspecby}{\rgstontg}

\newcommand{\sahom}{\phi}
\newcommand{\ahom}{\fap{\sahom}}

\newcommand{\sahomwords}{\wordsover{\sahom}}
\newcommand{\ahomwords}{\fap{\sahomwords}}

\newcommand{\sahomne}{\indap{\sahom}{\text{ne}}}
\newcommand{\ahomne}{\fap{\sahomne}}

\newcommand{\sabisim}{B}
\newcommand{\abisim}{\mathrel{\sabisim}}
\newcommand{\abisimfg}[2]{\mathrel{\abisim_{{#1}%,
                                                {#2}}}}

\newcommand{\abisimspacefg}[2]{\;\abisimfg{}{}\;}

\newcommand{\sabisimne}{B_{\snested}}
\newcommand{\abisimne}{\mathrel{\sabisimne}}
\newcommand{\abisimnespace}{\;\abisimne\;}

\newcommand{\sbisim}{%
    \setbox0=\hbox{\kern-.1ex{$\leftrightarrow$}\kern-.1ex}
    \setbox1=\vbox{\hbox{\raise .1ex \box0}\hrule}%
    \ensuremath{\mathrel{\hbox{\kern.1ex\box1\kern.1ex}}}
  }
\newcommand{\bisim}{\mathrel{\sbisim}}

\newcommand{\sbisimne}{\sbisim^{\snested}}
\newcommand{\bisimne}{\mathrel{\sbisimne}}
\newcommand{\sbisimnei}{\pbap{\sbisim}{\snested}}

\newcommand{\sbisimorbisimne}{\sbisim^{(\snested)}}
\newcommand{\bisimorbisimne}{\mathrel{\sbisimorbisimne}}

\newcommand{\sfunbisimne}{\supap{\sfunbisim}{\snested}}
\newcommand{\funbisimne}{\mathrel{\sfunbisimne}}
\newcommand{\sfunbisimnei}{\pbap{\sfunbisim}{\snested}}
\newcommand{\funbisimnei}[1]{\mathrel{\sfunbisimnei{#1}}}

\newcommand{\sconvfunbisimne}{\supap{\sconvfunbisim}{\snested}}

\newcommand{\sconvfunbisimorfunbisimne}{\supap{\sconvfunbisim}{(\snested)}}
\newcommand{\convfunbisimorfunbisimne}{\mathrel{\sconvfunbisimorfunbisimne}}

\newcommand{\sbisimsubscript}{
    \setbox0=\hbox{\kern-.1ex{$\leftrightarrow$}\kern-.1ex}
    \setbox1=\vbox{\hbox{\raise .1ex \box0}\hrule}%
    \ensuremath{\mathrel{\hbox{\scalebox{0.75}{\box1}}}}
  }

\newcommand{\sfunbisim}{%
    \setbox0=\hbox{\kern-.1ex{$\rightarrow$}\kern-.1ex}
    \setbox1=\vbox{\hbox{\raise .1ex \box0}\hrule}%
    {\hbox{\kern.1ex\box1\kern.1ex}}
  }
\newcommand{\funbisim}{\mathrel{{\sfunbisim}}}

\newcommand{\snotfunbisim}{%
    \setbox0=\hbox{\kern-.1ex{$\not\rightarrow$}\kern-.1ex}
    \setbox1=\vbox{\hbox{\raise .1ex \box0}\hrule}%
    {\hbox{\kern.1ex\box1\kern.1ex}}
  }
\newcommand{\notfunbisim}{\mathrel{{\snotfunbisim}}}

\newcommand{\notfunbisimorfunbisimne}{\mathrel{\supap{\snotfunbisim}{(\snested)}}}

\newcommand{\sfunbisimi}{\indap{\sfunbisim}}
\newcommand{\funbisimi}[1]{\mathrel{\sfunbisimi{#1}}}

\newcommand{\sconvfunbisim}[1][]{%
    \setbox0=\hbox{\kern-.1ex{$\leftarrow$}\kern-.1ex}
    \setbox1=\vbox{\hbox{\raise .1ex \box0}\hrule}%
    \mathrel{\hbox{\kern.1ex\box1\kern.1ex}}
  }
\newcommand{\convfunbisim}{\mathrel{\sconvfunbisim}}

\newcommand{\snotconvfunbisim}[1][]{%
    \setbox0=\hbox{\kern-.1ex{$\not\leftarrow$}\kern-.1ex}
    \setbox1=\vbox{\hbox{\raise .1ex \box0}\hrule}%
    \mathrel{\hbox{\kern.1ex{\box1}\kern.1ex}}
  }

\newcommand{\snotconvfunbisimne}{\snotconvfunbisim^{\snested}}

\newcommand{\sconvfunbisimi}{\indap{\sconvfunbisim}}
\newcommand{\convfunbisimi}[1]{\mathrel{\sconvfunbisimi{#1}}}

\newcommand{\sfunbisimsubscript}{
    \setbox0=\hbox{\kern-.1ex{$\rightarrow$}\kern-.1ex}
    \setbox1=\vbox{\hbox{\raise .1ex \box0}\hrule}%
    \ensuremath{\mathrel{\hbox{\scalebox{0.75}{\box1}}}}
  }

\newcommand{\ARSroot}{a}

\newcommand{\aARS}{{\to}}
\newcommand{\aARStc}{{\aARS^+}}
\newcommand{\aARSrtc}{{\aARS^*}}

\newcommand{\stepaARStc}{\mathrel{\aARStc}}
\newcommand{\stepaARSrtc}{\mathrel{\aARSrtc}}

\newcommand{\antmn}{{S}}
\newcommand{\bntmn}{{T}}
\newcommand{\cntmn}{{U}}
\newcommand{\dntmn}{{V}}
\newcommand{\entmn}{{W}}
\newcommand{\fntmn}{{X}}

\newcommand{\bnfis}{\mathrel{{:}{:}{=}}}

\newcommand{\darkcyan}[1]{#1}

\newcommand\itemizeprefs{
  \setlength\itemsep{-0.2ex}
}
\title{Nested Term Graphs\\[.5ex]
       \normalsize{(Work In Progress)}}
\author{
  Clemens Grabmayer
  \institute{Department of Computer Science\\
             VU University Amsterdam\\
             The Netherlands}
  \email{C.A.Grabmayer@vu.nl}
\and
  Vincent van Oostrom 
  \institute{Philosophy\\
             Utrecht University\\
             The Netherlands}
  \email{V.vanOostrom@uu.nl}
}
\def\titlerunning{Nested Term Graphs}
\def\authorrunning{C. Grabmayer \& V. van Oostrom}

\begin{document}
\maketitle

%-----------------------------------------------------------------------
\begin{abstract}
  We report on work in progress on `nested term graphs'
  %that are intended to formalize 
  for formalizing higher-order terms (e.g.\ finite or infinite \lambdaterms),
  including those expressing recursion (e.g.\ terms in the \lambdacalculus\ with letrec).
  The idea is to represent
  the nested scope structure of a higher-order term by 
  a nested structure of term graphs.
%   The idea is that 
%   the nested scope structure of a higher-order term is represented by 
%   a nested structure of term graphs.
%   The representation of a term starts with an ordinary term graph
%   in which some of the vertices are labelled by `nested' symbols
%   that designate outermost bindings together with their scope.
%   Any such vertex is additionally associated with a usual term graph 
%   that represents the subterm context describing the scope, 
%   where inner scopes are again expressed by nested symbols.  
% %  where inner scopes are again expressed by nested symbols.
%   This association between vertices and term graphs %defining scopes 
%   continues, following the scope nesting structure (possibly infinitely~deep).
%  
  Based on a signature that is partitioned into atomic and nested function symbols,
  we define nested term graphs 
  both in a functional representation,
  as tree-like recursive graph specifications that associate nested symbols with usual term graphs,
  and in a structural representation, as enriched term graph structures.
%   we define nested term graphs both intensionally,
%   as tree-like recursive graph specifications that associate nested symbols with usual term graphs,
%   and extensionally, as enriched term graph structures.
  These definitions induce corresponding notions of bisimulation % (`nested bisimulation')
  between nested term graphs.
  %We explain the close relationship between nested term graphs and higher-order term graphs. 
  %Finally, 
  Our main result states that nested term graphs can be implemented faithfully by first-order~term~graphs. 
\end{abstract}
\section{Introduction}
  \label{sec:intro}
% %----------------------------------------------------------------------- 

As an instance of the general question of how to faithfully
represent structures enriched with a notion of scope using
the same structures without it, we study the question how
to faithfully represent higher-order term graphs using
first-order term graphs.

To set the stage, we first informally recapitulate how to 
faithfully represent first-order terms using strings, and how
to faithfully represent higher-order terms using first-order terms.
The guiding intuition is that the notion of 
scope corresponds to a notion of context-freeness.

First-order terms can be represented using recursive string 
specifications (context-free grammars) such as
$ \{
   \antmn \bnfis \bntmn \times \cntmn 
   \mathrel{|}
   \bntmn \bnfis 2,
   \cntmn \bnfis \dntmn + \entmn,
   \dntmn \bnfis 3,
   \entmn \bnfis 1
   \}
$.
The string $2 \times 3 + 1$ obtained from the specification by 
repeated substitution for variables\footnote{%
Variables can be thought of as named subterms.}
is not a faithful representation of the first-order term though, 
as the nesting structure is lost;
the same string is obtained from the different first-order term 
$ \{
   \antmn \bnfis \bntmn + \cntmn 
   \mathrel{|}
   \bntmn \bnfis \dntmn \times \entmn,
   \dntmn \bnfis 2,
   \entmn \bnfis 3,
   \cntmn \bnfis 1
   \}
$.
A nameless (anonymous) alternative to recursive string specifications is to
introduce a box (scope) construct in the language of strings,
which indeed allows to faithfully represent the first-order terms:
$\fbox{$\fbox{$2$}\times\fbox{$\fbox{$3$}+\fbox{$1$}$}$}$ vs
$\fbox{$\fbox{$\fbox{$2$}\times\fbox{$3$}$}+\fbox{$1$}$}$ or
after unboxing values
$\fbox{$2\times\fbox{$3+1$}$}$ vs
$\fbox{$\fbox{$2\times3$}+1$}$.
However, having the box construct makes this representation 
go beyond a string representation proper (apart from the representation
quickly becoming unwieldy, on paper).
A standard way to overcome this is to split the box
$\Box$ into\footnote{%
For obvious visual reasons we use square brackets here instead of the usual parentheses.
Parentheses are not needed at all when the symbols in the alphabet are enriched with arities.
As shown by (Reverse) Polish Notation, arities are sufficient to capture context-freeness.
} 
symbols $[$ and $]$ that are adjoined to the alphabet 
yielding the proper strings $[2\times[3+1]]$ vs $[[2\times3]+1]$.
This is the common faithful representation of first-order terms as strings.
Note that not just any string represents a first-order term. In particular,
left and right brackets must be matching, the context-freeness aspect 
mentioned above, e.g.\ it would not do to substitute the 
string $3] + [1$ for $\fntmn$ in $[2\times \fntmn]$.

Higher-order terms can be represented by using recursive first-order term 
specifications.\footnote{%
In functional programming recursive first-order term specifications are known
as supercombinators and the transformation of $\lambda$-terms into supercombinators
is known as %$\lambda$-lifting
            \lambdalifting~\cite{hugh:1982}.}
To illustrate this we make use of an example in functional programming 
(Lisp) taken from~\cite{Kohl:etal:86} concerning the unhygienic expansion of the macro
$ (\mathsf{or}\, \langle\mathit{exp}\rangle_1 \langle\mathit{exp}\rangle_2)
  \bnfis
  (\mathsf{let}\, v\,[\,]_{\langle\mathit{exp}\rangle_1} 
   (\mathsf{if}\, v\,v\,[\,]_{\langle\mathit{exp}\rangle_2})) $.
Expanding this $\mathsf{or}$-macro in $(\mathsf{or}\,\mathsf{nil}\,v)$ yields
$(\mathsf{let}\,v\,\mathsf{nil}\,(\mathsf{if}\,v\,v\,v))$
which always yields $\mathsf{nil}$ due to the inadvertent capturing of~$v$.
A representation of the example by means of a recursive first-order
term specification would be
$ \{
   \antmn(v) \bnfis \mathsf{or}(\mathsf{nil},v) 
   \mathrel{|}
   \mathsf{or}(x,y) \bnfis \mathsf{let}(x,\bntmn(y)),
   \bntmn(z) \bnfis \mathsf{if}(v,v,z) 
   \}
$. 
This representation leaves the binding effect of $v$ in the in-part of the $\mathsf{let}$ implicit,
by it not occurring among the arguments to $\bntmn$; 
too implicit, as repeated substitution yields $\mathsf{let}(\mathsf{nil},\mathsf{if}(v,v,v))$.
A nameless alternative to recursive first-order term specifications is to introduce a
box (scope) construct in the language of first-order terms, the idea being that for
every first-order term $t$ over a vector $\vec{x}$ of $n$ variables\footnote{%
Linearity of $t$ in $\vec{x}$ may additionally be imposed.} 
and one additional variable, $\fbox{$t$}_{\vec{x}}$ is an $n$-ary function symbol again,
e.g.\  $\fbox{$\mathsf{if}(v,v,z)$}_z$ allowing to faithfully represent the higher-order term as
$\mathsf{let}(\mathsf{nil},\fbox{$\mathsf{if}(v,v,z)$}_z(z))$.
However, having the box construct makes this representation go beyond a first-order
term representation proper. A standard way to overcome this is to split the
box $\Box$ into\footnote{%
For obvious visual reasons we use these symbols instead of the
usual $\lambda$, $S$ and $0$.
Decomposing a box `vertically' into brackets here instead of `horizontally' as before,
corresponds to the matching of the brackets here being `vertically' 
(along paths in the first-order term tree) whereas before it was `horizontally' (within the string).}  
unary symbols $\sqcap$ and $\sqcup$ (for opening and closing) and a nullary symbol 
$\bullet$ (for using the bound variable) that are adjoined to the alphabet yielding
the proper first-order term 
$\mathsf{let}(\mathsf{nil},{\sqcap}(\mathsf{if}({\bullet},{\bullet},{\sqcup}(z))$.
This is the common faithful representation of higher-order terms as first-order terms,
known for the special case of $\lambda$-terms as
the (extended) De Bruijn representation~\cite{bird:pate:1999}.
Note that not just any first-order term represents a higher-order term.
In particular, open and close brackets must be matching, the context-freeness
aspect mentioned above.

\begin{comment}

An example in programming (Lisp)
is unhygienic expansion~\cite{Kohl:etal:86} of the macro
$ (\mathsf{or}\, \langle\mathit{exp}\rangle_1 \langle\mathit{exp}\rangle_2)
  \bnfis
  (\mathsf{let}\, v\,[\,]_{\langle\mathit{exp}\rangle_1} 
   (\mathsf{if}\, v\,v\,[\,]_{\langle\mathit{exp}\rangle_2})) $.
Expanding this $\mathsf{or}$-macro in $(\mathsf{or}\,\mathsf{nil}\,v)$ yields
$(\mathsf{let}\,v\,\mathsf{nil}\,(\mathsf{if}\,v\,v\,v))$
which always yields $\mathsf{nil}$ due to the inadvertent capturing of $v$.
Again, the issue at stake is the forgetfulness of the mapping
now from the macro language to the term language; the scope structure is lost.
And again the solution is to introduce an artefact into the less-structured
term language in order to represent the extra scope-structure of the macro language: 
In this case the $\adbmal$, a device from~\cite{hend:oost:2003},
may be used to represent scope-endings of binders. Using this the macro becomes
$ (\mathsf{or}\, \langle\mathit{exp}\rangle_1 \langle\mathit{exp}\rangle_2)
  \bnfis
  (\mathsf{let}\, v\,[\,]_{\langle\mathit{exp}\rangle_1} 
   (\mathsf{if}\, v\,v\,\adbmal v.[\,]_{\langle\mathit{exp}\rangle_2})) $ 
yielding    
$(\mathsf{let}\,v\,\mathsf{nil}\,(\mathsf{if}\,v\,v\,\adbmal v.v))$,
avoiding that the substituted $v$ becomes bound by the $\mathsf{let}\,v$ 
of the macro by unbinding the latter by $\adbmal v$, ending that scope.
\end{comment}

%\begin{wrapfigure}[20]{r}{0.85\textwidth}
\begin{figure}[t]  
  \begin{flushleft}
    \hspace*{2ex}
    \scalebox{0.78}{\begin{minipage}{\textwidth}
                     \input{figs/nested-and-implementation-ext}
                   \end{minipage}}
      %{\input{figs/nested-named-and-implementation-1}}
  \end{flushleft}
  \vspace*{-2.5ex}
  \caption{
  \label{fig:nested-and-implementation}%
           Pretty-printed nested term graph representing the \protect\gletrecexpression\ left in Figure~\ref{fig:gletrec:expressions},
           and its interpretation as a first-order term graph
           (back-links from $\sinvert$- and $\sinroot$-labeled vertices are, typically, hinted).
           } 
\end{figure}
%\end{wrapfigure} 

\begin{figure}[t!]
\hspace*{2ex}  
  \begin{center}
    $
    \begin{aligned}[c]
      \begin{array}{lrcl}
        \mathsf{gletrec} 
        & \sntgrootsym() & \bnfis & \lambda x.\afunsymi{1}(x) \afunsymi{2}(x,\bfunsym()) \\
        & \afunsymi{1}(X_1) & \bnfis & \lambda y.\mathsf{letrec}\,\alpha = X_1 {y} \alpha\,\mathsf{in}\,\alpha \\
        & \afunsymi{2}(X_1,X_2) & \bnfis & \lambda z.\mathsf{letrec}\,\beta = X_1 {z} (X_2 {z} \beta)\,\mathsf{in}\,\beta \\
        & \bfunsym() & \bnfis & \lambda w.w \\
        \mathsf{in}       
        & \sntgrootsym()
      \end{array}
    \end{aligned}
    \hspace*{10ex}
    \begin{aligned}   
      \begin{array}{lrcl}
        \mathsf{gletrec} 
        & \afunsym() & \bnfis & \lambda x.\bfunsym(x) \\
        & \bfunsym(X_1) & \bnfis & \lambda y.\bfunsym(y)X_1 \\
        \mathsf{in}       
        & \afunsym()
      \end{array}
    \end{aligned}
    $
  \end{center}
  \vspace*{-1ex}
  \caption{\label{fig:gletrec:expressions}
                  Two CRS-inspired \protect\gletrecexpressions\ that represent infinite \protect\lambdaterms.}
  \vspace*{-1ex}                
\end{figure}

In this paper we are concerned with the same phenomenon 
for `nested term graphs' in relation to
their interpretations as first-order term graphs. 
We describe an interpretation that is faithful
with regard to the respective notions of behavioral (bisimulation) semantics.
As a running example we use %the left one, \eqref{eq1:gletrec},
the \gletrecexpression\ left in Figure~\ref{fig:gletrec:expressions} 
that expresses a cyclic \lambdaterm, and thereby a regular infinite \lambdaterm,
by means of the Combinatory Reduction System (CRS) inspired \gletrecnotation. 
This expression corresponds to the 
pretty printed `recursive graph specification' on the left in Figure~\ref{fig:nested-and-implementation}
(the graph with scopes indicated by dotted lines).
\enlargethispage{3ex}%
Our main result entails that the behavioral semantics of this specification is the same as that
of the first-order term graph obtained from it, 
displayed on the right in Figure~\ref{fig:nested-and-implementation}. %in the same figure.
Note that in this first-order term graph artefacts, additional vertices, and edges
between them have been inserted to delimit scopes appropriately;
they play the same r\^ole as the brackets in the string and term examples.

It is interesting to observe that edges connecting a bound variable
to its binder seem to be forced upon us in this interpretation in order
to preserve the behavorial equivalence of scopes
(and their integrity; partial sharing is prevented).
Interesting, as this allows for a rational
reconstruction of sorts of using such edges to represent binding (instead of using variables for that purpose) as introduced 
in~\cite{Quin:1940,Bour:1954} and common nowadays in the 
implementation of $\lambda$-terms.

%This example 
The example in Figure~\ref{fig:nested-and-implementation} belongs to a particularly well-behaved subclass
of recursive graph specifications that we call nested term graphs,
for which the dependency between the nested symbols 
($\sntgrootsym$, $\afunsymi{1}$, $\afunsymi{2}$, $\bfunsym$ in the example) is tree-like.
The first-order term graph is nearly a `$\lambda$-term graph'~\cite{grab:roch:2013:a:TERMGRAPH},
and it is closely related to a higher-order term graph \cite{blom:2001}.
For defining nested term graphs
we will also consider specifications with arbitrary dependencies, 
allowing for both sharing and cyclicity,
such as the specification left in Figure~\ref{fig:rgs-entangled-ntgs},
which corresponds to the \gletrecexpression\ right in Figure~\ref{fig:gletrec:expressions},
and represents the infinite \lambdaterm\ in Figure~\ref{fig:rgs-entangled-ntgs}.

\myparagraphbf{Overview}
  In Section~\ref{sec:ntgs} we define nested term graphs as such recursive term graph specifications
  in which the dependency `is directly used in the definition of' between occurrences of defined (nested) symbols
  in the specification forms a tree. We also define structural representations of nested term graphs as integral graph structures
  with additional reference links, and an ancestor function that records the nesting of symbols. 
  In Section~\ref{sec:bisim:nested:bisim} we define adequate notions of homomorphism and bisimilarity
  between nested term graphs in two forms:
  a version with a `big-step semantics' condition for dealing with vertices labeled with defined symbols,
  and a `nested' version that is
  based on purely local progression conditions and the use of stacks to record the nesting history. 
  Finally in Section~\ref{sec:interpretation} we explain how nested term graphs can be interpreted
  by first-order term graphs in such a way that homomorphism and bisimilarity are preserved and reflected.

\myparagraphbf{Contribution}
  In its present stage, our contribution is primarily a conceptual one. 
  Inspired by Blom's higher-order term graphs~\cite{blom:2001},
  and by the faithful interpretation of `\lambdahigherordertgs' as first-order `\lambdatgs'
  described by the first author and Rochel in~\cite{grab:roch:2013:a:TERMGRAPH} 
  (which facilitates a maximal-sharing algorithm for the Lambda Calculus with \txtletrec\ \cite{grab:roch:2014:ICFP}),
  we set out to formalize objects with nested attributes
  (e.g.\ \lambdaterms\ with nested `extended scopes') 
  as enriched, and as plain, term graphs. 
  In more detail our contribution is threefold: 
  furnishing term graphs with a concept of nesting,
  developing adequate notions of behavioral semantics (homorphism, bisimulation) for nested term graphs,
  and describing a natural interpretation as first-order term graphs.
  We think that the possibility to implement higher-order features
  in a behavioral-semantics preserving and reflecting manner by first-order means
  can potentially be very~fruitful. 
  
  While for the purpose of this preliminary exploration we deliberately kept to the framework of term graphs due to its simplicity,
  we intend to adapt the results obtained for nested term graphs also
  to other graph formalisms like hypergraphs, jungles, bigraphs, interaction nets, or port graphs.
  Also, we want to compare the concepts developed with well-known formalisms for expressing nested structures
  and reasoning with them,
  for example: bigraphs, proofnets, and Fitch-style natural-deduction proofs in predicate logic. 

%   \todo{%
%   at this stage foremostly  
%   conceptual: extracting a general essence from work in \cite{grab:roch:2013:a:TERMGRAPH,grab:roch:2014:ICFP}
%   for representing \lambdaterms\ by the higher-order \lambdahotgs\ and the first-order \lambdatgs.
%   Trying to base ourselves on the basic concept of first-order term graph.
%   We do not ignore related work, but first try to work out our conceptual ideas in this easy framework.
%   We see our contribution in showing an interpretation of scoped objects
%   by first-order objects, which is potentially very fruitful as it allows
%   to implement higher-order features by first-order means.% 
%   }

%-------------------------------
\myparagraphbf{Preliminaries on term graphs}
  \label{sec:tgs}
%-------------------------------
% For a binary relation $\ssucc$ on a set $\aset$ we denote by
% $\stcsucc$ and $\srtcsucc$ the reflexive closure, and respectively, the reflexive and transitive closure,
% of $\ssucc$. 
% A structure $\triple{\aset}{{\ssucc}}{r}$ consisting of a set $\aset$,
% a binary relation $\ssucc$ on $\aset$, and an element $r\in\aset$
% is called a \emph{tree with root}~$r$
% if $\ssucc$ is acyclic (that is, there is no $x\in\aset$ such that $x \tcsucc x$%
%                         %where $\stcsucc$ the transitive closure of $\ssucc$
%                         ),
% every element $x\in\aset\setminus\setexp{r}$ has precisely one $\ssucc$\nb-predecessor 
%      (that is, a unique $x_0\in\aset$ with $x_0 \succ x$),
% and finally, every element $x\in\aset$ is reachable from $r$ via zero, one, or more $\ssucc$\nb-steps
%      (that is, $r \rtcsucc x$%where $\rtcsucc$ is the reflexive and transitive closure of $\ssucc$
%                             ).
%
By $\nats$ we denote the natural numbers including zero. 
For a set $\Sigma$, $\wordsover{\Sigma}$ stands for the set of words over alphabet $\Sigma$.
We denote the empty word by $\emptyword$, and write ${u}\wcns{v}$ for the concatenation of words $u$ and $v$. 
For a word $w$ and $i\in\nat$, we denote by $\fap{w}{i}$ its $(i+1)$\nb-th letter, % (if that exists),
and $\length{w}$ for the length of $w$. 

Let $\asig$ be a (first-order) signature for function symbols with arity function $\sarity \funin \asig \to \nats$.
For a function symbol $\afunsymvar\in\asig$, we indicate by $\afunsymvar/i$ that $\afunsymvar$ has arity $i$.
A \emph{term graph over $\asig$} (a \emph{$\asig$\nb-term-graph})
is a tuple $\tuple{\verts,\svlab,\svargs,\sroot}$ 
where $\verts$ is a set of \emph{vertices},
$\svlab \funin \verts \to \asig$ the \emph{(vertex) label function},
$\svargs \funin \verts \to \verts^*$ the \emph{argument function} 
  that maps every vertex $\avert$ to the word $\vargs{\avert}$ consisting of the $\arity{\vlab{\avert}}$ successor vertices of $\avert$
  (hence it holds $\length{\vargs{\avert}} = \arity{\vlab{\avert}}$),
and $\sroot\in\verts$ is the \emph{root} of the term graph.
% \begin{itemize}
%   \item[--] $\verts$ is a set of \emph{vertices},
%   \item[--] $\svlab \funin \verts \to \asig$ the \emph{(vertex) label function},
%   \item[--] $\svargs \funin \verts \to \verts^*$ the \emph{argument function} 
%      that maps every vertex $\avert$ to the word $\vargs{\avert}$ consisting of the $\arity{\vlab{\avert}}$ successor vertices of $\avert$
%     (hence it holds $\length{\vargs{\avert}} = \arity{\vlab{\avert}}$),
%   \item[--] $\sroot$, the \emph{root} is a vertex in $\verts$. 
% \end{itemize}
%Note the fact that term graphs may have infinitely many vertices.
% \begin{itemize}
%   \item[--] $\verts$ is a set of \emph{vertices},
%   \item[--] $\svlab \funin \verts \to \asig$ the \emph{(vertex) label function},
%   \item[--] $\svargs \funin \verts \to \verts^*$ the \emph{argument function} 
%      that maps every vertex $\avert$ to the word $\vargs{\avert}$ consisting of the $\arity{\vlab{\avert}}$ successor vertices of $\avert$
%     (hence it holds $\length{\vargs{\avert}} = \arity{\vlab{\avert}}$),
%   \item[--] $\sroot$, the \emph{root} is a vertex in $\verts$. 
% \end{itemize}
A term graph is called \emph{\rootconnected}
if every vertex is reachable from the root by a path that arises by repeatedly going from a vertex to one of its successors.
{By $\classtgsover{\asig}$ we denote the class of all \rootconnected\ term graphs over $\asig$.}
%In this notation we have already anticipated the meaning in which we will use the expression `term graph' from now on, namely as follows.
%\noindent
%\emph{Note:}\label{note:root-connected}
  By a `term graph' we will mean by default a `\rootconnected~term~graph'.
  
For a $\asig$\nb-term-graph $\atg$ and a vertex $\avert$ of $\atg$
we denote by $\subtgat{\atg}{\avert}$ the \emph{sub-term-graph of $\atg$ at $\avert$},
that is, 
the (\rootconnected) term graph with root $\avert$
that consists of all vertices that are reachable from $\avert$ in $\atg$.
As a useful notation for referring to edges in a term graph~$\atg$,
we will write $\avert \tgsucci{i} \bvert$ to indicate that the $(i+1)$-th outgoing edge from vertex $\avert$
leads to vertex $\bvert$
(that is, $\vargsati{\avert}{i} = \bvert$ with $\svargs$ the argument function~of~$\atg$).
  
%   with some exceptions, in which we will explicitly state otherwise, 
%   e.g.\ by referring to a term graph $\atg\in\classtgsminover{\asig}$ over some signature $\asig$. 
  % which then does not need to be root-connected.

% Let $\asig$ be a signature, and $\asigi{0}\subseteq\asig$.
% We say that a term graph $\atg$ over $\asig$ is \emph{$\asigi{0}$\nb-linear}
% if, for every symbol $\afunsym\in\asigi{0}$, $\atg$ contains at most a single vertex with label $\afunsym$. 
  
A \emph{rooted} \ARS\ is the extension of an {abstract rewriting system} (\ARS)~$\aARS$ 
by specifying one of its objects as designated \emph{root}.
%We also use $\aARS$ for the rooted \ARS.
%
A rooted \ARS~$\aARS$ with objects $\aset$ and root $\ARSroot$ is called a \emph{tree}
if $\aARS$ is 
acyclic 
  (there is no $x\in\aset$ such that $x \stepaARStc x$),
co-deterministic
  (for every {$x\in\aset$} %$\setminus\setexp{\ARSroot}$
  there is at most one step of $\aARS$ with target $x$),
and \rootconnected\
(every element $x\in\aset$ is reachable from $\ARSroot$ via a sequence of steps of $\aARS$,
 i.e.\
 % that is, 
 $\ARSroot \stepaARSrtc x$).
\section{Nested term graphs}
  \label{sec:ntgs}
%-----------------------------------------------------------------------  

We will use the words `nested' and `nesting' here in a meaning derived from that of the verb `nest',
which a dictionary%
  \footnote{Merriam-Webster (\href{http://www.merriam-webster.com/dictionary/nest}{http://www.merriam-webster.com/dictionary/nest}),
            visited on March 29, 2015.}
explains as            
`to fit compactly together or within one another',
and as 
`to form a hierarchy, series, or sequence of with each member, element, 
 or set contained in or containing the next $\langle\text{nested subroutines}\rangle$'.

A \emph{signature for nested term graphs} (an \emph{\ntgsig})
is a signature $\asig$ for term graphs that is partitioned 
into a part $\asigat$ for \emph{atomic} symbols,
and a part $\asigne$ for \emph{nested} symbols (cf.\ the terminals and non-terminals of a context-free string grammar.), %
  %\footnote{Cf.\ the terminals and non-terminals of a context-free string grammar.}
that is, $\asig = \asigat \cup \asigne$ and $\asigat \cap \asigne = \emptyset$.  
In addition to a given signature $\asig$ for nested term graphs we always assume additional \emph{interface symbols}
from the set $\OutIns = \Outs \cup \Ins$,
where 
$\Outs = \setexp{\ssout}$ consists of a single unary \emph{output} symbol 
  (symbolizing an edge that can pass on produced output from the root of the term graph definition of a nested symbol),      
and $\Ins = \setexp{ \ssini{1}, \ssini{2}, \ssini{3}, \ldots }$
is a countably infinite set of \emph{input} symbols with arity zero 
  (symbolizing edges to which input can be supplied to leaves of the term graph definition of a nested symbol).
  % where $\Ins = \setexp{\ssin}$ consists of a single unary \emph{input} symbol (symbolizing an input edge into a term graph), 
  % %the symbol set $\Ins = \setexp{\ssin}$, where $\ssin$ is unary,
  % and $\Outs = \setexp{ \ssouti{1}, \ssouti{2}, \ssouti{3}, \ldots }$
  % a countably infinite set of \emph{output} symbols with arity zero (symbolizing a numbered output edge from a term graph).
%We assume that none of these symbols occurs also in $\asig$. 
% Vertices labeled by $\ssin$ will symbolize an `input edge' of a term graph
% from the environment, and vertices labeled by $\ssouti{i}$ will denote the $i$\nb-th `output edge' of a term graph.   
\vspace{2ex}

% We say that a term graph $\atg$ over a signature $\asig$ with $\asig = \asigat \cup \asigne$ 
% for nested term graphs is \emph{nested-symbol linear} if $\atg$ is $\asigne$\nb-linear. 

\begin{definition}[recursive specifications for nested term graphs]\label{def:rgss}
  Let $\asig$ be a signature for nested term graphs.
  A \emph{recursive (nested term) graph specification} (an \emph{\rgs)} over $\asig$
  is a tuple $\pair{\srecspec}{\srootsymvar}$,
  where:
  \begin{itemize}\itemizeprefs
    \item[--]   
      $\srecspec \funin \asigne \to \classtgsover{\asig\cup\OutIns}$ %,
        %$\afunsymvar \mapsto \recspec{\afunsymvar} = \graphafunsym $
      is the \emph{specification function}
      that maps a nested function symbol $\afunsymvar\in\asigne$ with $\arity{\afunsymvar} = m$ 
      to a term graph $\recspec{\afunsymvar} = \graphafunsym \in \classtgsover{\asig\cup\setexp{\ssout,\ssini{1},\ldots,\ssini{m}}}$
      that has precisely one vertex labeled by $\ssout$, the root,
      and that contains precisely one vertex labeled by $\ssini{j}$, for each $j\in\setexp{1,\ldots,m}$;
%       with precisely one vertex labeled by $\ssin$, namely the root,
%       and with precisely one vertex labeled by the $\avarsymi{i}\,$, for each $i\in\setexp{1,\ldots,k}$.
% 
%       $\srecspec \funin \asigne \to \classtgsover{\asig\cup\OutIns}$,
%         $\afunsymvar \mapsto \recspec{\afunsymvar} = \graphafunsym $
%       is the \emph{specification function}
%       that maps every nested function symbol $\afunsymvar$ with arity $\arity{\afunsymvar} = k$ 
%       to a term graph $\recspec{\afunsymvar} = \graphafunsym \in \classtgsover{\asig\cup\setexp{\ssin,\ssouti{1},\ldots,\ssouti{k}}}$
%       that has precisely one vertex labeled by $\ssin$, namely the root,
%       and that contains precisely one vertex labeled by the $\avarsymi{i}$, for each $i\in\setexp{1,\ldots,k}$.
    \item[--]
      $\srootsymvar\in\asigne$, a nullary symbol (that is, $\arity{\srootsymvar} = 0$), is the \emph{root symbol}.
      % $\atg$ is a term graph over $\asig$,
  \end{itemize} 
  % We use the letters $\rrgs$ and $\srgs$ possibly with super- and subscripts, to designate \rgss. 
  %
  For such an \rgs\ $\rrgs = \pair{\srecspec}{\srootsymvar}$ over $\asig$,
  the rooted \emph{dependency}\/ \ARS~$\dependsonARSignoreof{\rrgs}$ of $\rrgs$ 
  has as objects the nested symbols in $\asigne$, it has root $\srootsymvar$,
  and the following steps:  
  for all $\afunsymvar,\bfunsymvar\in\asigne$ such that 
  a vertex labeled by $\bfunsymvar$ occurs in the term graph $\recspec{\afunsymvar}$  
  at position $p$ 
  there is a step $ p \funin \afunsymvar \stepdependsonARSignoreof{\rrgs} \bfunsymvar$. 
  We say that an~\rgs~$\rrgs$ is \emph{\rootconnected} 
  if every nested symbol is reachable from the root symbol of $\rrgs$ via steps
  of the dependency \ARS~$\dependsonARSignoreof{\rrgs}$ of $\rrgs$.
  Analogously as for term graphs, by an `\rgs' we will by default mean a `\rootconnected\ ARS'. 
\end{definition}  

%\todo{root-connectedness van \rgss} (is no issue for \ntgs\ due to their definition)

\begin{figure}[t]
    \vspace*{-1.5ex}
    \begin{center}
      \scalebox{0.925}{\begin{tikzpicture} 
  %\draw[help lines] (-8,-2) grid (8,2); 
   
  \node[anchor=center,draw,thick,circle,scale=0.75,minimum size=0.6cm,inner sep=0pt,densely dotted] (r0) at (-4.6,2.5) {$\srootsymi{0}$};
    \draw[<-,thick,>=latex](r0) -- ++ (90:0.6cm);
    %\draw[<-,thick,>=latex]($(r0.north) + (0pt,2.5pt)$) -- ($(r0.north) + (0pt,2.5pt) + (0pt,0.35cm)$); 
    
  \node[anchor=center,draw,thick,circle,scale=0.75,minimum size=0.6cm,inner sep=0pt,densely dotted] (r) at (6.25,2.5) {$\sntgrootsym$};  
    \draw[<-,thick,>=latex](r) -- ++ (90:0.6cm);
    %\draw[<-,thick,>=latex]($(r.north) + (0pt,2.5pt)$) -- ($(r.north) + (0pt,2.5pt) + (0pt,0.35cm)$);   

  \matrix[anchor=center,row sep=0.3cm,column sep=0.4cm,every node/.style={draw,thick,circle,scale=0.75,minimum size=0.6cm,inner sep=0pt}] at (-5.9,0) {
     & & \node(root_R0){$\ssout$};
     \\
     & & \node(0_R0){$\sslabs$}; 
     \\
     & & \node(00_R0){$\sslapp$};
     \\
     \node(000_R0)[densely dotted]{$\afunsymi{2}$}; & & &[-2.5ex] &[-1ex] \node(001_R0)[densely dotted]{$\afunsymi{2}$};
     \\
     \node(0000_R0){$\snlvar$}; & & \node(helper_R0)[draw=none]{}; & \node(0010_R0){$\snlvar$}; & &[-1ex] \node(0011_R0)[densely dotted]{$\bfunsym$};
     \\
     };
   %\draw[<-,thick,>=latex](root_R0) -- ++ (90:0.6cm);
   \draw[<-,thick,>=latex]($(root_R0.west)$) -- ($(root_R0.west) + (-0.35cm,0pt)$);  
   \draw[->] (root_R0) to (0_R0);
   \draw[->] (0_R0) to (00_R0);
   \draw[->] (00_R0) to (000_R0);
   \draw[->] (00_R0) to (001_R0);
   \draw[->,bend right] (000_R0.240) to (0000_R0);
   \draw[->,bend left]  (000_R0.310) to (0000_R0);
   \draw[->] (001_R0) to (0010_R0);
   \draw[->] (001_R0) to (0011_R0);
   \draw[dotted,thick] 
     (root_R0.north) -- ($ (root_R0.north) + (-1.75cm,0cm) $)
                     -- ($ (helper_R0)     + (-1.75cm,-0.4cm) $)
                     -- ($ (helper_R0)     + (2.3cm,-0.4cm) $)
                     -- ($ (root_R0.north) + (2.3cm,0cm) $)
                     -- (root_R0.north);
   \path (root_R0) ++ (-1cm,0.5cm) node[minimum size=0.6cm,inner sep=0pt] (R0_label) {$\graphsrootsymi{0}$};            
   \draw[|->,thick,shorten <=3pt,shorten >=0pt,bend right,distance=0.4cm]
      (r0) to node[above]{$\scriptstyle\srecspeci{0}$} (R0_label);

  \matrix[anchor=center,row sep=0.3cm,column sep=0.4cm,every node/.style={draw,thick,circle,scale=0.75,minimum size=0.6cm,inner sep=0pt}] at (-2.94,-0.9) {
    \node(root_G){$\ssout$};
    \\
    \node(0_G){$\sslabs$};
    \\
    \node(00_G){$\snlvar$};
    \\
    };
  %\draw[<-,thick,>=latex](root_G) -- ++ (90:0.6cm); 
  \draw[<-,thick,>=latex]($(root_G.west)$) -- ($(root_G.west) + (-0.35cm,0pt)$);  
  \draw[->] (root_G) to (0_G);  
  \draw[->] (0_G) to (00_G);
  \draw[dotted,thick]
    (root_G.north) -- ($ (root_G.north) + (0.7cm,0cm) $)
                   -- ($ (00_G)         + (0.7cm,-0.4cm) $)
                   -- ($ (00_G)         + (-0.7cm,-0.4cm) $)
                   -- ($ (root_G.north) + (-0.7cm,0cm) $)
                   -- (root_G.north);
   \path (00_G) ++ (0cm,-0.7cm) node (G_label) {$\graphbfunsym$};         
   \draw[|->,thick,shorten <=3pt,shorten >=0pt,bend right,distance=0.25cm]
      (0011_R0) to node[below]{$\scriptstyle\srecspeci{0}\hspace*{1.5ex}$} (G_label);

   \matrix[anchor=center,row sep=0.3cm,column sep=0.18cm,every node/.style={draw,thick,circle,minimum size=0.6cm,scale=0.75,inner sep=0pt}] at (0.15,0) { 
      \node[draw=none](helper_F2){}; 
      & & & \node(root_F2){$\ssout$}; & &
      \\
      & & & \node(0_F2){$\sslabs$};  
      \\
      & & & \node(00_F2){$\sslapp$}; & &
      \\
      & \node(000_F2){$\sslapp$}; & & & & \node(001_F2){$\sslapp$};
      \\
      & & \node(0001_F2){$\snlvar$}; & & \node(0010_F2){$\sslapp$};
      \\  
      & & & & & \node(00101_F2){$\snlvar$};
      \\[-2.5ex]
      \node(0000_F2){$\ssini{1}$}; & & & \node(00100_F2){$\ssini{2}$}; 
      \\
    };
   %\draw[<-,thick,>=latex](root_F2) -- ++ (90:0.6cm);
   \draw[<-,thick,>=latex]($(root_F2.west)$) -- ($(root_F2.west) + (-0.35cm,0pt)$); 
   \draw[->](root_F2) to (0_F2);
   \draw[->](0_F2) to (00_F2);
   \draw[->](00_F2) to (000_F2);
   \draw[->](00_F2) to (001_F2);
   \draw[->](000_F2) to (0000_F2);
   \draw[->](000_F2) to (0001_F2); 
   \draw[->](001_F2) to (0010_F2);
   \draw[->](001_F2) to[out=-30,in=60,distance=1.25cm] (00_F2);
   \draw[->](0010_F2) to (00100_F2);
   \draw[->](0010_F2) to (00101_F2);
   \draw[dotted,thick]
     (0000_F2.south)   -- (00100_F2.south);
   \draw[dotted,thick]
     (0000_F2.south)   -- ($ (0000_F2.south)    + (-0.45cm,0cm) $)
                       -- ($ (helper_F2.north) + (-0.45cm,0cm) $)
                       -- (root_F2.north);
   \draw[dotted,thick]        
     (root_F2.north) -- ($ (root_F2.north)  + (2.1cm,0cm) $)
                     -- ($ (00100_F2.south)  + (2.1cm,0cm) $)
                     -- (00100_F2.south);
   \path (root_F2) ++ (-0.8cm,0.51cm) node (F2_label) {$\graphafunsymi{2}$}; 
   \draw[|->,thick,near end,shorten <=3pt,shorten >=5pt,bend left,distance=1.4cm]
      (000_R0) to node[above]{$\scriptstyle\srecspeci{0}$} (F2_label); 
   \draw[|->,thick,shorten <=3pt,shorten >=-2pt,bend left,near start,distance=0.8cm]
      (001_R0) to node[above]{$\scriptstyle\srecspeci{0}\;\;\;\;\;$} (F2_label);

  \matrix[anchor=center,row sep=0.3cm,column sep=0.1cm,every node/.style={draw,thick,circle,minimum size=0.6cm,scale=0.75,inner sep=0pt}] at (3.95,-0.3) { 
      \node[draw=none](helper1_F1){}; 
      & & \node(root_F1){$\ssout$}; & 
      \\
      & & \node(0_F1){$\sslabs$};  
      \\
      & & \node(00_F1){$\sslapp$}; 
      \\
      & \node(000_F1){$\sslapp$};
      \\ 
      & & \node(0001_F1){$\snlvar$};
      \\[-2ex]
      \node(0000_F1){$\ssini{1}$}; & & \node[draw=none](helper2_F1){};
      \\
      }; 
    %\draw[<-,thick,>=latex](root_F1) -- ++ (90:0.6cm);
    \draw[<-,thick,>=latex]($(root_F1.west)$) -- ($(root_F1.west) + (-0.35cm,0pt)$); 
    \draw[->](root_F1) to (0_F1);
    \draw[->](0_F1) to (00_F1);
    \draw[->](00_F1) to (000_F1);
    \draw[->](00_F1) to[out=-50,in=50,distance=1.25cm] (00_F1);
    \draw[->](000_F1) to (0000_F1);
    \draw[->](000_F1) to (0001_F1);
    \draw[dotted,thick]
      (0000_F1.south)   -- ($ (0000_F1.south)     + (-0.45cm,0) $)
                        -- ($ (helper1_F1.north) + (-0.45cm,0) $)
                        -- (root_F1.north);
    \draw[dotted,thick]        
      (root_F1.north) -- ($ (root_F1.north)    + (0.95cm,0) $)
                      -- ($ (helper2_F1.south) + (0.95cm,0) $)
                      -- (0000_F1.south);   
    \path (root_F1) ++ (0.45cm,0.5cm) node (F1_label) {$\graphafunsymi{1}$};            
  
  \matrix[anchor=center,row sep=0.3cm,column sep=0.4cm,every node/.style={draw,thick,circle,scale=0.75,minimum size=0.6cm,inner sep=0pt}] at (7.75,0) {
     & & \node(root_R){$\ssout$};
     \\
     & & \node(0_R){$\sslabs$}; 
     \\
     & & \node(00_R){$\sslapp$};
     \\
     \node(000_R)[densely dotted]{$\afunsymi{1}$}; & & &[-2.5ex] &[-1ex] \node(001_R)[densely dotted]{$\afunsymi{2}$};
     \\
     \node(0000_R){$\snlvar$}; & & \node(helper_R)[draw=none]{}; & \node(0010_R){$\snlvar$}; & &[-1ex] \node(0011_R)[densely dotted]{$\bfunsym$};
     \\
     };
   %\draw[<-,thick,>=latex](root_R) -- ++ (90:0.6cm); 
   \draw[<-,thick,>=latex]($(root_R.west)$) -- ($(root_R.west) + (-0.35cm,0pt)$);  
   \draw[->] (root_R) to (0_R);
   \draw[->] (0_R) to (00_R);
   \draw[->] (00_R) to (000_R);
   \draw[->] (00_R) to (001_R);
   \draw[->] (000_R) to (0000_R);
   \draw[->] (001_R) to (0010_R);
   \draw[->] (001_R) to (0011_R);
    \draw[dotted,thick] 
     (root_R.north) -- ($ (root_R.north)   + (-1.75cm,0cm) $)
                    -- ($ (helper_R.south) + (-1.75cm,-0.2cm) $)
                    -- ($ (helper_R.south) + (2.3cm,-0.2cm) $)
                    -- ($ (root_R.north)   + (2.3cm,0cm) $)
                    -- (root_R.north);
   \path (root_R) ++ (1.15cm,0.5cm) node (R_label) {$\graphsntgrootsym$}; 
   \draw[|->,thick,shorten <=3pt,shorten >=0pt,bend left,distance=0.75cm]
      (r) to node[above]{$\scriptstyle\srecspec$} (R_label);
   \draw[|->,thick,shorten <=3pt,shorten >=-3pt,bend right,distance=0.65cm]
      (000_R.90) to node[above,near start]{$\scriptstyle\hspace*{2.5ex}\srecspec$} (F1_label);    
   \draw[|->,thick,shorten <=3pt,shorten >=0pt,bend right,distance=1.95cm]
      (001_R.65) to node[above,near end]{$\scriptstyle\hspace*{2.5ex}\srecspec$} (F2_label);   
   \draw[|->,thick,shorten <=3pt,bend left,distance=1.3cm]
      (0011_R.240) to node[below]{$\scriptstyle\hspace*{2.5ex}\srecspec$} (G_label.350);        
     
\end{tikzpicture} }  
    \end{center}  
    \vspace*{-3ex}
    \caption{\label{fig:ex:rgs:ntg}
      Definitions of a recursive graph specification~$\rrgsi{0}$~(Ex.~\ref{ex:rgs}),
      and a nested term graph~$\antg$~(Ex.~\ref{ex:ntg}).}
\end{figure}

\begin{example}\label{ex:rgs}
  We choose a signature part %
    %\footnote{For a function symbol $\afunsymvar$ we indicate by $\afunsymvar/i$ that $\afunsymvar$ has arity $i$.}
  $\asigat = \setexp{\sslabs/1,\, \sslapp/2,\, \snlvar/0 }$ for expressing \lambdaterms\ as term graphs.
  \begin{enumerate}[(i)]\itemizeprefs%\renewcommand{\labelenumi}{(\roman{enumi})}\itemizeprefs
    \item{}\label{ex:rgs:item:1} 
      %Let $\asigat = \setexp{\sslabs/1,\, \sslapp/2,\, \snlvar/0 }$,
      Let $\asigine{0} = \setexp{\srootsymi{0}/0,\, \afunsymi{2}/2,\, \bfunsym/0}$.
      Then $\rrgsi{0} = \pair{\srecspeci{0}}{\srootsymi{0}}$,
      where $\srecspeci{0} \funin \asigine{0} \to \classtgsover{\asig\cup\OutIns}$
      is defined by
      $\srootsymi{0} \mapsto \graphsrootsymi{0}$, 
      $\afunsymi{2} \mapsto \graphafunsymi{2}$, and
      $\bfunsym \mapsto \graphbfunsym$
      as shown in Figure~\ref{fig:ex:rgs:ntg} (starting from $\srootsymi{0}$ on the left),
      is an \rgs.
%     Note that the term graph $\graphsrootsymi{0}$ has two occurrences of the nested symbol~$\afunsymi{2}$,
%     and hence it is not nested-symbol linear.
%     As a consequence also $\rrgsi{0}$ is not nested-symbol linear. 
% 
    \item{}\label{ex:rgs:item:2} 
      Let $\asigne = \setexp{ \sntgrootsym/0,\, \afunsymi{1}/1,\, \afunsymi{2}/2,\, \bfunsym/0}$.
      Then $\pair{\srecspec}{\sntgrootsym}$,
      where $\srecspec \funin \asigne \to \classtgsover{\asig\cup\OutIns}$ is defined by
      $\sntgrootsym \mapsto \graphsntgrootsym$,  
      $\afunsymi{1} \mapsto \graphafunsymi{1}$,
      $\afunsymi{2} \mapsto \graphafunsymi{2}$, and
      $\bfunsym \mapsto \graphbfunsym$
      as shown in Figure~\ref{fig:ex:rgs:ntg} (starting from $\sntgrootsym$ on the right),
      is an \rgs.
      It is an \rgsrepresentation\ of the \gletrecexpression\ left in Figure~\ref{fig:gletrec:expressions}.
      
   \item{}\label{ex:rgs:item:3}    
      Let $\asigine{1} = \setexp{ \afunsym/0, \bfunsym/1 }$.
      Then $\rrgsi{1} = \pair{\srecspeci{1}}{\afunsym}$,
      where $\srecspeci{1} \funin \asigine{1} \to \classtgsover{\asig\cup\OutIns}$
      is defined by
      $\afunsym \mapsto \graphafunsym$,
      $\bfunsym \mapsto \graphbfunsym$ 
      as shown left in Figure~\ref{fig:rgs-entangled-ntgs} is an \rgs.
      It represents the \gletrecexpression\ right in Figure~\ref{fig:gletrec:expressions}. 
      
    \item{}\label{ex:rgs:item:4} 
      Let $\asigine{2} = \setexp{ \afunsym } \cup \descsetexp{ \bfunsymi{i}/1 }{i\in\nat,\, i\ge 1 }$.
      Then $\pair{\srecspeci{2}}{\afunsym}$,
      where $\srecspeci{2} \funin \asigine{1} \to \classtgsover{\asig\cup\OutIns}$
      is defined by
      $\afunsym \mapsto \graphafunsym$, 
      $\bfunsymi{1} \mapsto \graphbfunsymi{1}$, 
      $\bfunsymi{2} \mapsto \graphbfunsymi{2}$, 
      $\bfunsymi{3} \mapsto \graphbfunsymi{3}$, \ldots\
      as shown right in Figure~\ref{fig:rgs-entangled-ntgs} is an \rgs.
      It represents the infinite \lambdaterm\ to the left of it in Figure~\ref{fig:rgs-entangled-ntgs}.
\end{enumerate}  
\end{example}
  
% \begin{definition}
%   Let $\rrgs = \pair{\srecspec}{\srootsymvar}$ be an \rgs\ over $\asig$.
%   The rooted \emph{dependency}\/ \ARS~$\dependsonARSignoreof{\rrgs}$ of $\rrgs$ 
%   has as objects the nested symbols in $\asigne$, it has root $\srootsymvar$,
% %   and as steps those of the form $%\astep \funin 
% %                                   \afunsymvar \dependsonARSignoreof{\rrgs} \bfunsymvar$
% %   for every $\afunsymvar\in\asigne$, and for
% %   every occurrence of $\bfunsymvar\in\asigne$ in the term graph $\recspec{\afunsymvar}$
% %   (note the possibility of parallel steps from $\afunsymvar$ to $\bfunsymvar$). 
%   and the following steps:  
%   for all $\afunsymvar,\bfunsymvar\in\asigne$ such that $\bfunsymvar$ occurs in the term graph $\recspec{\afunsymvar}$  
%   at position $p$ 
%   % and every occurrence $o_{\afunsymvar}$ of a $\bfunsymvar\in\asigne$ in the term graph $\recspec{\afunsymvar}$
%   there is a step $ p \funin \afunsymvar \stepdependsonARSignoreof{\rrgs} \bfunsymvar$.
% %   every occurrence of $\bfunsymvar$ in the term graph $\recspec{\afunsymvar}$
% %   induces a step $ \afunsymvar \stepdependsonARSignoreof{\rrgs} \bfunsymvar$ 
% %  (so note the possibility of parallel steps from $\afunsymvar$ to $\bfunsymvar$). 
%   
%   
% \end{definition}  

\begin{definition}[nested term graphs]
  Let $\asig$ be an \ntgsig. %signature for nested term graphs. 
  A \emph{nested term graph} (an \emph{\ntg}) 
                              over $\asig$ 
  is an \rgs~$\antg = \pair{\srecspec}{\srootsymvar}$ 
  such that the rooted dependency \ARS~$\dependsonARSignoreof{\antg}$ is a tree.
%   $\srecspec$ is nested-symbol linear,
%   and 
%   $\triple{\asigne}{\sdependsonof{\antg}}{\sntgrootsym}$ is a tree with root $\sntgrootsym$.
  %Here and later we use the letter $\antg$, possibly with super- and subscripts, to designate \ntgs.
  %
  By $\classntgsover{\asig}$ we denote the class of all nested term graphs over $\asig$. 
\end{definition}  
  
\begin{example}\label{ex:ntg}
  We first consider the \rgs~$\rrgsi{0} = \pair{\srecspeci{0}}{\srootsymi{0}}$ from Example~\ref{ex:rgs}, \eqref{ex:rgs:item:1}.
  Its rooted dependency \ARS~$\dependsonARSignoreof{\rrgsi{0}}$ is not a tree, 
  because there are two steps that witness $\srootsymi{0} \dependsonARSignoreof{\rrgsi{0}} \afunsymi{2}$,
  namely those that are induced by the two occurrences of $\afunsymi{2}$ 
  in the term graph $\graphsrootsymi{0} =\recspeci{0}{\srootsymi{0}}$.
  As a consequence, $\rrgsi{0}$ is not a nested term graph.
  
  Similarly, the \rgs~$\rrgsi{1}$ from Example~\ref{ex:rgs}, \eqref{ex:rgs:item:3},
  is not a nested term graph, because its dependency \ARS~$\dependsonARSignoreof{\rrgsi{1}}$
  contains the cycle $\bfunsym \dependson \bfunsym$, and hence is not a tree.  
  
  But for the \rgs~$\antg = \pair{\srecspec}{\sntgrootsym}$  
  from Example~\ref{ex:rgs}, \eqref{ex:rgs:item:2}, %\ref{ex:rgs:item:1}
  we find that the rooted dependency \ARS~$\dependsonARSignoreof{\antg}$ is a tree with root $\sntgrootsym$. 
  Hence $\antg$ is a nested term graph.
  For a `pretty print' of $\antg$, see the left graph in Figure~\pageref{fig:nested-and-implementation}.
  
  Also for the \rgs~$\antgi{2} = \pair{\srecspeci{2}}{\afunsym}$ 
  from Example~\ref{ex:rgs}, \eqref{ex:rgs:item:4}, 
  we find that the rooted dependency \ARS~$\dependsonARSignoreof{\antg}$ is a tree with root $\afunsym$,
  since it is of the form:
  $\afunsym \dependson \bfunsymi{1} \dependson \bfunsymi{2} \dependson \bfunsymi{3} \dependson \ldots\,$.
  Hence $\antgi{2}$ is a nested term graph with infinitely deep nesting.
  It represents the infinite \lambdaterm\ with infinitely deep nesting of its `extended scopes'
  (minimal extensions of bound variable scopes in order to obtain nestedness)
  to the left of it in Figure~\ref{fig:rgs-entangled-ntgs},
  which has the $\textsf{gletrec}$\nb-re\-pre\-sen\-ta\-tion on the right in Figure~\ref{fig:gletrec:expressions}.  
\end{example}  
  
\begin{figure}[t!]
\begin{flushleft}
  \hspace*{-8ex}
  $  
  \begin{gathered}[c]
    \phantom{\text{infinite nesting}}
    \\
    \scalebox{0.7}{\begin{tikzpicture} 
  %\draw[help lines] (-8,-6) grid (8,0); 

\matrix[anchor=north,row sep=0.5cm,column sep=0.4cm,every node/.style={draw,thick,circle,scale=0.75,minimum size=0.6cm,inner sep=0pt}] at (-7,0)%
{  
  \node (f0)[densely dotted]{$\afunsym$}; 
  \\
};    
\draw[<-,thick,>=latex]($(f0.north)$) -- ($(f0.north) + (0pt,0.35cm)$);   
  
\matrix[anchor=north,row sep=0.5cm,column sep=0.4cm,every node/.style={draw,thick,circle,scale=0.75,minimum size=0.6cm,inner sep=0pt}] at (-5,0)%
{
  \node(root_F){$\ssout$}; 
  \\
  \node (0_F){$\sslabs$};
  \\
  \node(00_F)[densely dotted]{$\bfunsym$};
  \\
  \node(000_F){$\snlvar$};
  \\  
}; 
\path (root_F) ++ (0cm,0.6cm) node (F_label) {$\graphafunsym$}; 
%\path (root_F) ++ (1cm,0.5cm) node[minimum size=0.6cm,inner sep=0pt] (R0_label) {$\graphsrootsymi{0}$};

\draw[<-,thick,>=latex]($(root_F.west)$) -- ($(root_F.west) + (-0.35cm,0pt)$); 
\draw[->,thick](root_F) to (0_F);
\draw[->,thick](0_F) to (00_F);
\draw[->,thick](00_F) to (000_F);
\draw[dotted,thick]
  (root_F.north) -- ($ (root_F.north) + (-1cm,0cm) $)
                  -- ($ (000_F.south) + (-1cm,-0.3cm) $)
                  -- ($ (000_F.south) + (1cm,-0.3cm) $)
                  -- ($ (root_F.north) + (1cm,0cm) $)
                  -- (root_F.north);  
                  
\draw[|->,thick,shorten <=3pt,shorten >=0pt,bend left,distance=0.25cm]
      (f0.north east) to node[above]{$\scriptstyle\;\;\srecspeci{1}$} 
                                    (F_label);

\matrix[anchor=north,row sep=0.5cm,column sep=0.4cm,every node/.style={draw,thick,circle,scale=0.75,minimum size=0.6cm,inner sep=0pt}] at (-2.25,0)%
{
  \node(root_G){$\ssout$}; 
  \\
  \node (0_G){$\sslabs$};
  \\
  \node(00_G){$\sslapp$};
  \\
  \node(000_G)[densely dotted]{$\bfunsym$};
  \\
  \node(0000_G){$\snlvar$};
  \\[1ex]
  \node(00000_G){$\ssini{1}$};
  \\  
}; 
\draw[<-,thick,>=latex]($(root_G.west)$) -- ($(root_G.west) + (-0.35cm,0pt)$);
\path (root_G) ++ (0cm,0.6cm) node (G_label) {$\graphbfunsym$};  
\draw[->,thick](root_G) to (0_G);
\draw[->,thick](0_G) to (00_G);
\draw[->,thick](00_G) to (000_G);
\draw[->,thick](000_G) to (0000_G); 
% \draw[->,thick]
%   (0000_G.east) -- ($ (0000_G.east) + (0.6cm,0cm) $)
%                  -- ($ (0_G.east) + (0.6cm,0cm) $)
%                  -- (0_G.east); 
\draw[->,thick]
  (00_G.east) -- ($ (00_G.east) + (0.7cm,0cm) $)
               -- ($ (00000_G.north) + (0.925cm,0.35cm) $)   
               -- ($ (00000_G.north) + (0cm,0.35cm) $)                  
               -- (00000_G.north);
\draw[dotted,thick]
  (root_G.north) -- ($ (root_G.north) + (-1cm,0cm) $)
                  -- ($ (00000_G.south) + (-1cm,0cm) $)
                  -- ($ (00000_G.south) + (1.4cm,0cm) $)
                  -- ($ (root_G.north) + (1.4cm,0cm) $)
                  -- (root_G.north);  
                                 
\draw[|->,thick,shorten <=5pt,shorten >=0pt,bend left,distance=0.75cm]
      (00_F.45) to node[above,near end]{$\scriptstyle\srecspeci{1}\hspace*{2ex}\mbox{}$}
                                        (G_label);                
               
\draw[|->,thick,shorten <=5pt,shorten >=-3pt,bend right,distance=1.75cm]
      (000_G.45) to node[right,near end]{$\scriptstyle\srecspeci{1}$}
                                        (G_label);                                             
                                                   
\end{tikzpicture}}
  \end{gathered}%
  \hspace*{3ex}
  \begin{aligned}[c]
    & \scalebox{0.7}{\begin{tikzpicture}
\matrix[anchor=center,row sep=0.5cm,column sep=0.5cm]%,every node/.style={draw,scale=0.75,minimum size=0.6cm,inner sep=0pt}]
{  
  \node(root){$\slabsmindot{\avari{0}}$}; &
  \\
  \node(0){$\slabsmindot{\avari{1}}$}; &
  \\
  \node(00){$\sslapp$}; & \node(001){$\avari{0}$};
  \\
  \node(000){$\slabsmindot{\avari{2}}$}; &
  \\
  \node(0000){$\sslapp$}; & \node(00001){$\avari{1}$};
  \\
  \node(00000){$\slabsmindot{\avari{3}}$}; &
  \\
  \node(000000){$\sslapp$}; & \node(0000001){$\avari{2}$};
  \\  
  %\node(0000000){$\vdots$}; &
  \node(0000000){$\hspace*{7pt}\vdots\hspace*{7pt}$}; &
  \\
%   \node(00000000){$\sslapp$}; & \node(000000001){$\avari{3}$};
%   \\  
%   \node(000000000){$\vdots$}; &
%   \\
};  
\draw[->,thick](root) to (0);
\draw[->,thick](0) to (00);            \draw[->,thick](00) to (001);
\draw[->,thick](00) to (000);
\draw[->,thick](000) to (0000);        \draw[->,thick](0000) to (00001);
\draw[->,thick](0000) to (00000);
\draw[->,thick](00000) to (000000);    \draw[->,thick](000000) to (0000001);
\draw[->,thick](000000) to (0000000);
% \draw[->,thick](0000000) to (00000000);    \draw[->,thick](00000000) to (000000001);
% \draw[->,thick](00000000) to (000000000);

\path (root) ++ (-0.85cm,0.3cm) node (F0_label) {$\afunsym$}; 
\draw[dotted,thick]
  (root.east) -- ($(root.east) + (1.7cm,0cm)$)
              -- ($(0000000.east) + (1.7cm,-0.5cm)$);
\draw[dotted,thick]
  (root.west) -- ($(root.west) + (-0.9cm,0cm)$)
              -- ($(0000000.west) + (-0.9cm,-0.5cm)$);

\path (0) ++ (-0.75cm,0.3cm) node (F0_label) {$\bfunsymi{1}$}; 
\draw[dotted,thick]
  (0.east) -- ($(0.east) + (0.2cm,0cm)$)
           -- ($(000.east) + (0.2cm,0.5cm)$)
           -- ($(000.east) + (1.5cm,0.5cm)$)
           -- ($(0000000.east) + (1.5cm,-0.5cm)$);
\draw[dotted,thick]
  (0.west) -- ($(0.west) + (-0.7cm,0cm)$)
           -- ($(0000000.west) + (-0.7cm,-0.5cm)$);

\path (000) ++ (-0.65cm,0.3cm) node (F0_label) {$\bfunsymi{2}$};
\draw[dotted,thick]
  (000.east) -- ($(000.east) + (0.2cm,0cm)$)
             -- ($(00000.east) + (0.2cm,0.5cm)$)
             -- ($(00000.east) + (1.3cm,0.5cm)$)
             -- ($(0000000.east) + (1.3cm,-0.5cm)$);
\draw[dotted,thick]
  (000.west) -- ($(000.west) + (-0.5cm,0cm)$)
             -- ($(0000000.west) + (-0.5cm,-0.5cm)$);

\path (00000) ++ (-0.55cm,0.3cm) node (F0_label) {$\bfunsymi{3}$};
\draw[dotted,thick]
  (00000.east) -- ($(00000.east) + (0.2cm,0cm)$)
               -- ($(0000000.east) + (0.2cm,0.5cm)$)
               -- ($(0000000.east) + (1.1cm,0.5cm)$)
               -- ($(0000000.east) + (1.1cm,-0.5cm)$);
\draw[dotted,thick]
  (00000.west) -- ($(00000.west) + (-0.3cm,0cm)$)
               -- ($(0000000.west) + (-0.3cm,-0.5cm)$);  

% %\path (0000000) ++ (-0.55cm,0.3cm) node (F0_label) {$\afunsymi{4}$};
% \draw[dotted,thick]
%   (0000000.east) -- ($(0000000.east) + (0.2cm,0cm)$)
%                  -- ($(0000000.east) + (0.2cm,-1.5cm)$)
%                  -- ($(0000000.east) + (0.9cm,-1.5cm)$)
%                  -- ($(0000000.east) + (0.9cm,-2.5cm)$);
% \draw[dotted,thick]
%   (0000000.west) -- ($(0000000.west) + (-0.1cm,0cm)$)
%                  -- ($(0000000.west) + (-0.1cm,-2.5cm)$); 
%  
\end{tikzpicture}}
  \end{aligned}  
  \hspace*{0ex}
  \begin{gathered}[c]
    \phantom{\text{infinite nesting}}
    \\
    \scalebox{0.7}{\begin{tikzpicture} 
  %\draw[help lines] (-8,-6) grid (8,0); 

\matrix[anchor=north,row sep=0.5cm,column sep=0.4cm,every node/.style={draw,thick,circle,scale=0.75,minimum size=0.6cm,inner sep=0pt}] at (-7,0)%
{  
  \node (f)[densely dotted]{$\afunsym$}; 
  \\
};    
\draw[<-,thick,>=latex]($(f.north)$) -- ($(f.north) + (0pt,0.35cm)$);   
  
\matrix[anchor=north,row sep=0.5cm,column sep=0.4cm,every node/.style={draw,thick,circle,scale=0.75,minimum size=0.6cm,inner sep=0pt}] at (-5,0)%
{
  \node(root_F){$\ssout$}; 
  \\
  \node (0_F){$\sslabs$};
  \\
  \node(00_F)[densely dotted]{$\bfunsymi{1}$};
  \\
  \node(000_F){$\snlvar$};
  \\  
}; 
\path (root_F) ++ (0cm,0.6cm) node (F_label) {$\graphafunsym$}; 
%\path (root_F) ++ (1cm,0.5cm) node[minimum size=0.6cm,inner sep=0pt] (R0_label) {$\graphsrootsymi{0}$};

\draw[<-,thick,>=latex]($(root_F.west)$) -- ($(root_F.west) + (-0.35cm,0pt)$); 
\draw[->,thick](root_F) to (0_F);
\draw[->,thick](0_F) to (00_F);
\draw[->,thick](00_F) to (000_F);
\draw[dotted,thick]
  (root_F.north) -- ($ (root_F.north) + (-1cm,0cm) $)
                  -- ($ (000_F.south) + (-1cm,-0.3cm) $)
                  -- ($ (000_F.south) + (1cm,-0.3cm) $)
                  -- ($ (root_F.north) + (1cm,0cm) $)
                  -- (root_F.north);  
                  
\draw[|->,thick,shorten <=3pt,shorten >=0pt,bend left,distance=0.25cm]
      (f.north east) to node[above]{$\scriptstyle\;\;\srecspeci{1}$} 
                                    (F_label);

\matrix[anchor=north,row sep=0.5cm,column sep=0.4cm,every node/.style={draw,thick,circle,scale=0.75,minimum size=0.6cm,inner sep=0pt}] at (-2.25,0)%
{
  \node(root_G1){$\ssout$}; 
  \\
  \node (0_G1){$\sslabs$};
  \\
  \node(00_G1){$\sslapp$};
  \\
  \node(000_G1)[densely dotted]{$\bfunsymi{2}$};
  \\
  \node(0000_G1){$\snlvar$};
  \\[1ex]
  \node(00000_G1){$\ssini{1}$};
  \\  
}; 
\draw[<-,thick,>=latex]($(root_G1.west)$) -- ($(root_G1.west) + (-0.35cm,0pt)$);
\path (root_G1) ++ (0cm,0.6cm) node (G1_label) {$\graphbfunsymi{1}$};  
\draw[->,thick](root_G1) to (0_G1);
\draw[->,thick](0_G1) to (00_G1);
\draw[->,thick](00_G1) to (000_G1);
\draw[->,thick](000_G1) to (0000_G1); 
% \draw[->,thick]
%   (0000_G1.east) -- ($ (0000_G1.east) + (0.6cm,0cm) $)
%                  -- ($ (0_G1.east) + (0.6cm,0cm) $)
%                  -- (0_G1.east); 
\draw[->,thick]
  (00_G1.east) -- ($ (00_G1.east) + (0.7cm,0cm) $)
               -- ($ (00000_G1.north) + (0.925cm,0.35cm) $)   
               -- ($ (00000_G1.north) + (0cm,0.35cm) $)                  
               -- (00000_G1.north);
\draw[dotted,thick]
  (root_G1.north) -- ($ (root_G1.north) + (-1cm,0cm) $)
                  -- ($ (00000_G1.south) + (-1cm,0cm) $)
                  -- ($ (00000_G1.south) + (1.4cm,0cm) $)
                  -- ($ (root_G1.north) + (1.4cm,0cm) $)
                  -- (root_G1.north);  
                                 
\draw[|->,thick,shorten <=5pt,shorten >=0pt,bend left,distance=0.75cm]
      (00_F.45) to node[above,near end]{$\scriptstyle\srecspeci{1}\hspace*{2ex}\mbox{}$}
                                        (G1_label);

\matrix[anchor=north,row sep=0.5cm,column sep=0.4cm,every node/.style={draw,thick,circle,scale=0.75,minimum size=0.6cm,inner sep=0pt}] at (0.9,0)%
{
  \node(root_G2){$\ssout$}; 
  \\
  \node (0_G2){$\sslabs$};
  \\
  \node(00_G2){$\sslapp$};
  \\
  \node(000_G2)[densely dotted]{$\bfunsymi{3}$};
  \\
  \node(0000_G2){$\snlvar$};
  \\[1ex]
  \node(00000_G2){$\ssini{1}$};
  \\  
}; 
\draw[<-,thick,>=latex]($(root_G2.west)$) -- ($(root_G2.west) + (-0.35cm,0pt)$); 
\path (root_G2) ++ (0cm,0.6cm) node (G2_label) {$\graphbfunsymi{2}$}; 
\draw[->,thick](root_G2) to (0_G2);
\draw[->,thick](0_G2) to (00_G2);
\draw[->,thick](00_G2) to (000_G2);
\draw[->,thick](000_G2) to (0000_G2); 
% \draw[->,thick]
%   (0000_G2.east) -- ($ (0000_G2.east) + (0.6cm,0cm) $)
%                  -- ($ (0_G2.east) + (0.6cm,0cm) $)
%                  -- (0_G2.east); 
\draw[->,thick]
  (00_G2.east) -- ($ (00_G2.east) + (0.7cm,0cm) $)
               -- ($ (00000_G2.north) + (0.925cm,0.35cm) $)   
               -- ($ (00000_G2.north) + (0cm,0.35cm) $)                  
               -- (00000_G2.north);
\draw[dotted,thick]
  (root_G2.north) -- ($ (root_G2.north) + (-1cm,0cm) $)
                  -- ($ (00000_G2.south) + (-1cm,0cm) $)
                  -- ($ (00000_G2.south) + (1.4cm,0cm) $)
                  -- ($ (root_G2.north) + (1.4cm,0cm) $)
                  -- (root_G2.north);

\draw[|->,thick,shorten <=5pt,shorten >=0pt,bend left,distance=1cm]
      (000_G1.45) to node[above]{$\scriptstyle\srecspeci{1}\hspace*{2.5ex}\mbox{}$} 
                                 (G2_label);

\matrix[anchor=north,row sep=0.5cm,column sep=0.4cm,every node/.style={draw,thick,circle,scale=0.75,minimum size=0.6cm,inner sep=0pt}] at (4.05,0)%
{
  \node(root_G3){$\ssout$}; 
  \\
  \node(0_G3){$\sslabs$};
  \\
  \node(00_G3){$\sslapp$};
  \\
  \node(000_G3)[densely dotted]{$\bfunsymi{4}$};
  \\
  \node(0000_G3){$\snlvar$};
  \\[1ex]
  \node(00000_G3){$\ssini{1}$};
  \\  
}; 
\draw[<-,thick,>=latex,draw=none]($(root_G3.west)$) -- ($(root_G3.west) + (-0.35cm,0pt)$); 
\path (root_G3) ++ (0cm,0.6cm) node (G3_label) {$\graphbfunsymi{3}$}; 
\draw[->,thick](root_G3) to (0_G3);
\draw[->,thick](0_G3) to (00_G3);
\draw[->,thick](00_G3) to (000_G3);
\draw[->,thick](000_G3) to (0000_G3); 
% \draw[->,thick,draw=none]
%   (0000_G3.east) -- ($ (0000_G3.east) + (0.6cm,0cm) $)
%                  -- ($ (0_G3.east) + (0.6cm,0cm) $)
%                 -- (0_G3.east); 
\draw[->,thick]
  (00_G3.east) -- ($ (00_G3.east) + (0.7cm,0cm) $)
               -- ($ (00000_G3.north) + (0.925cm,0.35cm) $)   
               -- ($ (00000_G3.north) + (0cm,0.35cm) $)                  
               -- (00000_G3.north);
\draw[dotted,thick]
  (root_G3.north) -- ($ (root_G3.north) + (-1cm,0cm) $)
                  -- ($ (00000_G3.south) + (-1cm,0cm) $)
                  -- ($ (00000_G3.south) + (1.4cm,0cm) $)
                  -- ($ (root_G3.north) + (1.4cm,0cm) $)
                  -- (root_G3.north);                               
                              
\draw[|->,thick,shorten <=5pt,shorten >=0pt,bend left,distance=1cm]
      (000_G2.45) to node[above]{$\scriptstyle\;\;\srecspeci{1}\hspace*{2.5ex}\mbox{}$} 
                                 (G3_label);                              
                     
\path (G3_label.east) ++ (2.9cm,0cm) node (F4_label) {\phantom{$\graphbfunsymi{3}$}};                                                      
\draw[|->,thick,shorten <=5pt,shorten >=0pt,bend left,distance=1cm]
      (000_G3.45) to node[above]{$\scriptstyle\srecspeci{1}\hspace*{2.5ex}\mbox{}$} 
                                  (F4_label);              
                                                   
\end{tikzpicture}}
  \end{gathered}%
  \hspace*{6ex}
  $
\end{flushleft} 
  \vspace*{-2.5ex}
  \caption{\label{fig:rgs-entangled-ntgs}%
           Illustrations of a recursive graph specification~$\rrgsi{1}$ (left, see Example~\ref{ex:rgs}, \eqref{ex:rgs:item:3})
           with cyclic dependency \ARS,
           and a nested term graph~$\antgi{1}$ (right, see Example~\ref{ex:rgs}, \eqref{ex:rgs:item:4})
           with infinite dependency ARS.
           Both represent the 
           infinite \protect\lambdaterm\ (in between them) with infinite nesting of extended scopes.} 
\end{figure}

Next to nested term graphs as functional representations,
we also introduce corresponding representations of \ntgs\ as enrichments of ordinary term graphs. 
The reason is threefold.
We obtain a characterization of nested term graphs
as integral graph structures with functional dependencies represented by explicit links
(see Proposition~\ref{prop:ntgs:entgs}). 
Furthermore, such structural representations directly induce
a behavioral semantics via the associated notions of homomorphism and bisimulation
(see Section~\ref{sec:bisim:nested:bisim}).
And finally, they will be instrumental in defining the interpretation of nested term graphs
as first-order term graphs (in Section~\ref{sec:interpretation}).

In `structural representations' of nested term graphs as defined below, the device of the
`ancestor function' records, and---due to appropriate conditions on it---guarantees, 
the nesting structure of vertices by assigning to every vertex $\avert$
the word $\vanc{\avert} = \averti{1}\worddots\averti{n}$ made up of the vertices in which $\avert$ is nested.

\begin{definition}[nested term graphs, 
                   %extensionally
                   as structures]\label{eq:ergs}
  Let $\asig$ be a signature for nested term graphs. 
  A \emph{structural representation}
    %An \emph{extensional description} 
  of a nested term graph (an \emph{\entg}) over $\asig$
  is a tuple
  $\tuple{\verts,\svlab,\svargs,\svin,\svout,\svanc,\sroot}$,
  where $\atgi{0} = \tuple{\verts,\svlab,\svargs,\sroot}$ is a (typically not root-connected) term graph over $\asig\cup\OutIns$,
  and additionally:
\begin{itemize}\itemizeprefs
  \item[--]
    $\svin \funin \verts \rightharpoonup \verts$ 
      is the \emph{call} (or \emph{step-into}) partial function  
      that assigns to every vertex $\avert$ labeled with a nested symbol 
        %$\avert\in\vertsofnested$ 
      the root of the term graph nested into $\avert$
      (this root is an output vertex); 
  \item[--]
    $\svout \funin \verts \rightharpoonup \verts$ 
      is the \emph{return} (or \emph{step-out}) partial function
      that to every input vertex $\avert$ labeled by $\ssini{j}$ assigns the $j$\nb-th successor of the vertex 
      into which the term graph containing $\avert$ is nested;
\item[--]%[$\svanc \funin \verts \to \verts^*$]
    $\svanc \funin \verts \to \verts^*$ 
      is the \emph{ancestor function} 
      that to every vertex $\avert$ assigns the word $\vanc{\avert} = \averti{1}\worddots\averti{n}$
      made up of the vertices in which $\avert$ is nested:
      $\avert$ is nested in $\averti{n}$, 
      $\averti{n}$ is nested in $\averti{n-1}$, \ldots, $\averti{2}$ is nested in $\averti{1}$;      
\end{itemize}
that satisfy, more precisely,
the following conditions, for all $i,k\in\nats$, and all $\bvert,\bverti{i},\averti{1},\ldots,\averti{k}\in\verts$:\vspace*{-0.7ex}
\begin{align*}
  %\begin{aligned} 
        (\text{root})_{\svlab,\svanc} & \hspace*{3ex} 
          \vlab{\sroot} \in \asigne 
            \;\logand\;
          \vanc{\sroot} = \emptyword   
        \\   
        (\text{nested})_{\svanc} & \hspace*{3ex}     
          \vanc{\bvert} = \averti{1}\worddots\averti{n} 
            \;\;\Longrightarrow\;\;
          \text{$\averti{1}$, \ldots, $\averti{n}$, and $\bvert$ are distinct}
        \displaybreak[0]\\
        (\text{arguments})_{\svanc} & \hspace*{3ex}  
          %\bvert \tgsucci{i} \bverti{i}
          \bvert \tgsucci{i} \bverti{i}
          %\bverti{i} = \vargsati{\bvert}{i}
            \;\;\Longrightarrow\;\;
          \vanc{\bverti{i}} = \vanc{\bvert}  
        \displaybreak[0]\\
        (\text{defined})_{\svin,\svout} & \hspace*{3ex}    
          (\,\defd{\vin{\bvert}}
               \;\Longleftrightarrow\;
             \vlab{\bvert}\in\asigne\,)
          \,\logand\,
          (\,\defd{\vout{\bvert}}
               \;\Longleftrightarrow\;
           \vlab{\bvert}\in\Ins\,)  
        \displaybreak[0]\\  
        (\text{step-into})_{\svin} & \hspace*{3ex}    
          %\defd{\vin{\bvert}}
          \vlab{\bvert} \in \asigne
            \;\Longrightarrow\;
          \left\{\,   
          \parbox{265pt}{$\vlab{\vin{\bvert}} = \ssout \in\Outs 
                            \;\;\wedge\;\;
                          \vanc{\vin{\bvert}} = \vanc{\bvert}\wcns\bvert$
                         \\   
                         $\;\wedge\;\;\vin{\bvert}$ is the single vertex with label $\ssout$ in  
                         the 
                         \\
                         $\phantom{\;\wedge\;}$%
                         sub-term-graph $\subtgat{\atgi{0}}{\vin{\bvert}}$ of $\atgi{0}$ at vertex $\vin{\bvert}$}
          \,\right.
        \displaybreak[0]\\   
        \hspace*{3ex}(\text{step-out})_{\svout} & \hspace*{3ex} 
          %\defd{\vin{\bvert}}
          \vlab{\bvert} %= \afunsym 
                        \in \asigne 
            \;\Longrightarrow\;
          \left\{\,   
          \parbox{265pt}{for all $j\in\setexp{1,\ldots,\arity{\vlab{\bvert}}}$,
                         %the sub-term-graph 
                         $\subtgat{\atgi{0}}{\vin{\bvert}}$ %of $\atgi{0}$ at $\vin{\bvert}$
                         contains
                         precisely one vertex $\bvertacci{\!j}$ 
                         with label $\ssini{j}\in\Ins$,
                         and it holds:
                         $\bvert \tgsucci{j} \vout{\bvertacci{j}}$; $\;$
                         % $\vout{\bvertacci{\!j}} = \vargsati{\bvert}{j}$; $\;$
                         $\subtgat{\atgi{0}}{\vin{\bvert}}$ has no other vertices with labels in~$\Ins$}
          \,\right.
          % \displaybreak[0]\\
          % (\text{nested-into})_{\svanc,\svin} & \hspace*{3ex}    
          %   %\defd{\vin{\bvert}}
          %   \vlab{\bvert}\in\asigne 
          %     \;\;\Longrightarrow\;\;
          %   \vanc{\vin{\bvert}} = \vanc{\bvert}\wcns\bvert
            %\end{aligned}
\end{align*}   
\end{definition}

\begin{example}
  An \entg\ that corresponds to the nested term graph~$\antg$ in Example~\ref{ex:ntg}
  is depicted in Figure~\ref{fig:edntg}. 
    % The illustration in Figure~\ref{fig:edntg} depicts an \entg\ that corresponds to 
    % the nested term graph~$\antg$ in Example~\ref{ex:ntg}.
    % %with names for vertices with nested symbols (right of such vertices), and the values of the ancestor function indicated in brackets 
    % % (left of the vertices).
\begin{figure}[tb]
  \begin{center}   
      \scalebox{0.93}{\begin{tikzpicture}[scale=0.75,
                    node/.style={%
                      draw,
                      circle,
                      inner sep=0,
                      outer sep=0,
                      minimum size=0,
                      node distance=0}]
\matrix[row sep=0.4cm,column sep=0.4cm,every node/.style={draw,thick,circle,scale=0.75,minimum size=0.6cm,inner sep=0pt}%,outer sep=.5mm}
                                                                                                                      ]{ 
  & & \node(root_F1){$\ssout$}; & &[5ex] & &[1ex] \node(root_R){$\ssout$}; & & \node(root){$\sntgrootsym$}; & &[5ex] \node(root_G){$\ssout$}; &[5ex] & &[2.5ex] \node(root_F2){$\ssout$};
  \\
  & & \node(0_F1){$\sslabs$}; & & & & \node(0_R){$\sslabs$}; & & & &     \node(0_G){$\sslabs$};  & & & \node(0_F2){$\sslabs$};
  \\
  & & \node(00_F1){$\sslapp$}; & & & & \node(00_R){$\sslapp$};  & & & &  \node(00_G){$\snlvar$}; & & & \node(00_F2){$\sslapp$}; 
  \\
  & \node(000_F1){$\sslapp$}; & & & & \node(000_R){$\afunsymi{1}$}; & & \node(001_R){$\afunsymi{2}$}; & & & & & \node(000_F2){$\sslapp$}; & & & & \node(001_F2){$\sslapp$}; 
  \\
  & & \node(0001_F1){$\snlvar$}; & & & \node(0000_R){$\snlvar$}; & \node(0010_R){$\snlvar$}; & & \node(0011_R){$\bfunsym$}; & & & & & \node(0001_F2){$\snlvar$}; & & \node(0010_F2){$\sslapp$};    
  \\  
  \node(0000_F1){$\ssini{1}$}; & & & & & & & & & & & & & & & & \node(00101_F2){$\snlvar$};
  \\[-2ex]
  & & & & & & & & & & & \node(0000_F2){$\ssini{1}$}; & & & \node(00100_F2){$\ssini{2}$};
  \\
  };
\path (root) ++ (+0.4cm,-0.35cm) node {$\scriptstyle\averti{0}$};
\path (root) ++ (-0.5cm,-0.35cm) node {$\scriptstyle\prefix{\emptyword}$};

\path (root_R) ++ (-0.6cm,-0.2cm) node {$\scriptstyle\prefix{\averti{0}}$};
\path (0_R) ++ (-0.6cm,-0.2cm) node {$\scriptstyle\prefix{\averti{0}}$};
\path (00_R) ++ (-0.6cm,-0.2cm) node {$\scriptstyle\prefix{\averti{0}}$};
\path (000_R) ++ (-0.6cm,-0.2cm) node {$\scriptstyle\prefix{\averti{0}}$};
\path (001_R) ++ (-0.6cm,-0.2cm) node {$\scriptstyle\prefix{\averti{0}}$};
\path (0000_R) ++ (-0.6cm,-0.2cm) node {$\scriptstyle\prefix{\averti{0}}$};
\path (0010_R) ++ (-0.6cm,-0.2cm) node {$\scriptstyle\prefix{\averti{0}}$};
\path (0011_R) ++ (-0.6cm,-0.2cm) node {$\scriptstyle\prefix{\averti{0}}$};

\path (root_F1) ++ (-0.75cm,-0.2cm) node {$\scriptstyle\prefix{\averti{0}\averti{1}}$};
\path (0_F1) ++ (-0.75cm,-0.2cm) node {$\scriptstyle\prefix{\averti{0}\averti{1}}$};
\path (00_F1) ++ (-0.75cm,-0.2cm) node {$\scriptstyle\prefix{\averti{0}\averti{1}}$};
\path (000_F1) ++ (-0.75cm,-0.2cm) node {$\scriptstyle\prefix{\averti{0}\averti{1}}$};
\path (0000_F1) ++ (-0.75cm,-0.2cm) node {$\scriptstyle\prefix{\averti{0}\averti{1}}$};
\path (0001_F1) ++ (-0.75cm,-0.2cm) node {$\scriptstyle\prefix{\averti{0}\averti{1}}$};

\path (root_G) ++ (-0.75cm,-0.2cm) node {$\scriptstyle\prefix{\averti{0}\averti{3}}$};
\path (0_G) ++ (-0.75cm,-0.2cm) node {$\scriptstyle\prefix{\averti{0}\averti{3}}$};
\path (00_G) ++ (-0.75cm,-0.2cm) node {$\scriptstyle\prefix{\averti{0}\averti{3}}$};

\path (root_F2) ++ (-0.75cm,-0.2cm) node {$\scriptstyle\prefix{\averti{0}\averti{2}}$};
\path (0_F2) ++ (-0.75cm,-0.2cm) node {$\scriptstyle\prefix{\averti{0}\averti{2}}$};
\path (00_F2) ++ (-0.75cm,-0.2cm) node {$\scriptstyle\prefix{\averti{0}\averti{2}}$};
\path (000_F2) ++ (-0.75cm,-0.2cm) node {$\scriptstyle\prefix{\averti{0}\averti{2}}$};
\path (0000_F2) ++ (-0.75cm,-0.2cm) node {$\scriptstyle\prefix{\averti{0}\averti{2}}$};
\path (0001_F2) ++ (-0.75cm,-0.2cm) node {$\scriptstyle\prefix{\averti{0}\averti{2}}$};
\path (001_F2) ++ (-0.75cm,-0.2cm) node {$\scriptstyle\prefix{\averti{0}\averti{2}}$};
\path (0010_F2) ++ (-0.75cm,-0.2cm) node {$\scriptstyle\prefix{\averti{0}\averti{2}}$};
\path (00100_F2) ++ (-0.75cm,-0.2cm) node {$\scriptstyle\prefix{\averti{0}\averti{2}}$};
\path (00101_F2) ++ (-0.75cm,-0.2cm) node {$\scriptstyle\prefix{\averti{0}\averti{2}}$};

\path (000_R) ++ (+0.4cm,-0.35cm) node {$\scriptstyle\averti{1}$};
\path (001_R) ++ (+0.5cm,-0.2cm) node {$\scriptstyle\averti{2}$};
\path (0011_R) ++ (+0.5cm,-0.2cm) node {$\scriptstyle\averti{3}$};

\draw[<-,thick,>=latex](root) -- ++ (90:0.7cm);
\draw[->](root_R) to (0_R);
\draw[->](0_R) to (00_R);
\draw[->](00_R) to  (000_R);  \draw[->](00_R) to  (001_R); 
\draw[->](000_R) to (0000_R); \draw[->](001_R) to (0010_R); \draw[->](001_R) to (0011_R); 
\draw[->](root_F1) to (0_F1); 
\draw[->](0_F1) to (00_F1);
\draw[->](00_F1) to (000_F1); \draw[->](00_F1) to[out=-50,in=50,distance=1.75cm] (00_F1);
\draw[->](000_F1) to (0000_F1); \draw[->](000_F1) to (0001_F1);
\draw[->](root_F2) to (0_F2);
\draw[->](0_F2) to (00_F2);
\draw[->](00_F2) to (001_F2);
\draw[->](00_F2) to (000_F2);
\draw[->](000_F2) to (0000_F2);
\draw[->](000_F2) to (0001_F2);
\draw[->](001_F2) to (0010_F2);
\draw[->](001_F2) to[out=-10,in=50,distance=1.5cm] (00_F2);
\draw[->](0010_F2) to (00100_F2);
\draw[->](0010_F2) to (00101_F2);
\draw[->](root_G) to (0_G);
\draw[->](0_G) to (00_G);
\begin{scope}[|->,shorten <=3pt,shorten >=3pt]  
\draw[|->,thick,densely dashed]
  (root) to node[above]{$\svin$} (root_R);
\draw[|->,thick,densely dashed] 
  (000_R) to node[above]{$\;\;\;\;\;\;\;\svin$}  (root_F1); 
\draw[|->,thick,densely dashed,bend right,distance=2cm,shorten <=10pt]
  (001_R) to node[below]{$\svin$}  (root_F2); 
\draw[|->,thick,densely dashed,bend left,distance=1.25cm,shorten >=13pt] 
  (0011_R) to node[above]{$\svin\hspace*{4ex}$} (root_G);  
\draw[|->,thick,densely dashed,shorten >=16pt]
  (0000_F1) to node[below]{$\svout$} (0000_R); 
\draw[|->,thick,densely dashed,shorten <=5pt]
  (0000_F2) to node[below]{$\svout$} (0010_R);
\draw[|->,thick,densely dashed,bend right,distance=1.5cm]
  (00100_F2) to node[above]{$\svout$} (0011_R);
\end{scope}          
\end{tikzpicture} }
    \end{center}
    \vspace*{-3ex}
    \caption{\label{fig:edntg}%
             Illustration of a structural representation of the nested term graph~$\antg$ from Ex.~\ref{ex:ntg},
             with names for vertices with nested symbols (right of such vertices), and the ancestor function values indicated in brackets.}  
\end{figure}
\end{example}

\begin{proposition}\label{prop:ntgs:entgs}
  Every nested term graph has a structural representation.
  And for every structural representation $\mathcal{G}$ of a nested term graph 
  there is a nested term graph for which $\mathcal{G}$ is the structural representation.  
\end{proposition}

% \begin{remark}
%   \todo{%
%   This proposition can be viewed as a term-graph analogue to the 
%   standard result for string languages that context-free grammars 
%   are equivalent to push-down automata.}
% \end{remark}

%-----------------------------------------------------------------------
\section{Bisimulation and nested bisimulation}
  \label{sec:bisim:nested:bisim}
%-----------------------------------------------------------------------   
  
In order to motivate appropriate definitions of behavioral semantics for nested term graphs and 
recursive graph specifications, we start with the rather clear behavioral semantics for \entgs. 
Then we adapt these definitions to nested term graphs, and yield corresponding concepts. 
Subsequently we develop a definition of homomorphism and bisimilarity that also applies
to recursive graph specifications, and is based on purely local progression rules 
together with stacks that record the nesting history. We call these further concepts
`nested homomorphism' and `nested bisimilarity'. 
Finally we gather statements that relate bisimilarity and nested bisimilarity.

%-------
\myparagraphbf{Homomorphisms and bisimulations between \entgs} %{\rgss\ and \ntgs}
  \label{subsec:hom:bisim:entgs}
%-------
%
Since structural representations of nested term graphs can be viewed as coalgebras, 
they carry clear associated notions of homomorphism and bisimilarity.
To see this, let $\aedntgi{1}$ and $\aedntgi{2}$ be \entgs\ over signatures $\asigi{1}$ and $\asigi{2}$
with the same part $\asigat$ for atomic symbols.
A homomorphism from $\aedntgi{1}$ to $\aedntgi{2}$ (indicated by $\aedntgi{1} \funbisim \aedntgi{2}$)
is a function $\sahom \funin \vertsi{1} \to \vertsi{2}$ between their vertex sets
that preserves the property of being root, preserves atomic, nested, and interface labels, %$\OutIns$ labels,
commutes with the partial functions $\svin$ and $\svout$,
commutes with the (individual) argument function on vertices with atomic labels,
and preserves the ancestor function. % (see also Appendix~B). %(\ref{app:hom:bisim}). 
A bisimulation between \entgs~$\aedntgi{1}$ and $\aedntgi{2}$ % over $\asig$
  %(indicated by $\aedntgi{1} \bisim \aedntgi{2}$) 
can then be defined as an \entg~$\aedntg$ with the property
$\aedntgi{1} \convfunbisim \aedntg \funbisim \aedntgi{2}$.

\begin{definition}[homomorphism, bisimulation between \entgs]
  Let $\asigi{1} = \asigat \cup \asigine{1}$ 
  and $\asigi{2} = \asigat \cup \asigine{2}$
  be \ntgsigs\ with the same signature $\asigat$ for atomic symbols.
  Furthermore, let for each of $i\in\setexp{1,2}$,
  $\aedntgi{i} = \tuple{\vertsi{i},\svlabi{i},\svargsi{i},\svanci{i},\svini{i},\svouti{i},\srooti{i}}$
  be an \entg\ over signature $\asig$.
  
  A \emph{homomorphism} (functional bisimulation) between $\aedntgi{1}$ and $\aedntgi{2}$ is
  a $\quadruple{\asigne}{\asigat}{\Outs}{\Ins}$\nb-respecting morphism 
  $\sahom \funin \vertsi{1} \to \vertsi{2}$
  between the structures $\aedntgi{1}$ and $\aedntgi{2}$,
    % in the sense that 
  that is, 
  for all $\bvert\in\vertsi{1}$~the following conditions hold:
  \begin{align*}
    & \Longrightarrow\;\;
      \ahom{\srooti{1}} = \srooti{2} 
        \;\,\logand\,\;
      \ahomwords{\vanci{1}{\bvert}} = \vanci{2}{\ahom{\bvert}}
    & & (\sroot,\,\svanc)
    \displaybreak[0]\\
    \vlabi{1}{\bvert}\in\asigat 
      \;\; & \Longrightarrow\;\;
        \vlabi{2}{\ahom{\bvert}} = \vlabi{1}{\bvert} \in\asigat
          \;\,\logand\,\;
        \ahomwords{\vargsi{1}{\bvert}} = \vargsi{2}{\ahom{\bvert}} 
    & & (\svlab,\,\svargs)_{{\asigat}}     
    \displaybreak[0]\\
    \vlabi{1}{\bvert}\in\asigine{1} 
      \;\; & \Longrightarrow\;\;
        \vlabi{2}{\ahom{\bvert}}\in\asigine{2}
          \;\,\logand\,\;
        \ahom{\vini{1}{\bvert}} = \vini{2}{\ahom{\bvert}}
    & & (\svlab,\,\svin)_{{\asigne}}               
    \displaybreak[0]\\
    \vlabi{1}{\bvert}\in\Outs
      \;\; & \Longrightarrow\;\;
        \vlabi{2}{\ahom{\bvert}}\in\Outs
    & & (\svlab)_{{\Outs}}               
    \displaybreak[0]\\
    \vlabi{1}{\bvert}\in\Ins
      \;\; & \Longrightarrow\;\;
        \vlabi{2}{\ahom{\bvert}}\in\Ins
          \;\,\logand\,\;
        \ahom{\vouti{1}{\bvert}} = \vouti{2}{\ahom{\bvert}} 
    & & (\svlab,\,\svout)_{{\Ins}} 
  \end{align*}
  where $\sahomwords$ is the homomorphic extension of $\sahom$ 
  to a function from $\wordsover{\vertsi{1}}$ to $\wordsover{\vertsi{2}}$. 
  If there is a homomorphism $\sahom$ from $\aedntgi{1}$ to $\aedntgi{2}$,
  we write 
  $\aedntgi{1} \funbisimi{\sahom} \aedntgi{2}$
  and $\aedntgi{2} \convfunbisimi{\sahom} \aedntgi{1}$,
  or, dropping $\sahom$ as subscript, 
  $\aedntgi{1} \funbisim \aedntgi{2}$ and $\aedntgi{2} \convfunbisim \aedntgi{1}$. 
  
  A \emph{bisimulation} between $\aedntgi{1}$ and $\aedntgi{2}$ 
  is an 
  \entg~$\aedntg = \tuple{\sabisim,\svlab,\svargs,\svin,\svout,\svanc,\sroot}$
  %\todo{not root-connected}
  where
  $\sabisim \subseteq \vertsi{1}\times\vertsi{2}$ 
  and $\sroot = \pair{\srooti{1}}{\srooti{2}}$
  such that
  $ \aedntgi{1} \convfunbisimi{\sproji{1}} \aedntg \funbisimi{\sproji{2}} \aedntgi{2}$
  where $\sproji{1}$ and $\sproji{2}$ are projection functions that are defined, for $i\in\setexp{1,2}$,
  by $ \sproji{i} \funin \vertsi{1}\times\vertsi{2} \to \vertsi{i} $,
  $\pair{\averti{1}}{\averti{2}} \mapsto \averti{i}$.
  If there exists a bisimulation $\sabisim$ between $\aedntgi{1}$ and $\aedntgi{2}$,
  then we write %$\aedntgi{1} \bisimi{\sabisim} \aedntgi{2}$, 
                %or just 
  $\aedntgi{1} \bisim \aedntgi{2}$, and say that $\aedntgi{1}$ is \emph{bisimilar to} $\aedntgi{2}$.
\end{definition}

%-------
\myparagraphbf{Homomorphisms and bisimulations between nested term graphs}
  \label{subsec:hom:bisim:rgss:ntgs}
%-------
%
The definitions for \entgs\ above can motivate similar definitions for \ntgs.
Let $\antgi{1} = \pair{\srecspeci{1}}{\srootsymvari{1}}$ and $\antgi{2} = \pair{\srecspeci{2}}{\srootsymvari{2}}$ be \ntgs\
over signatures with the same atomic symbols.
A homomorphism between $\antgi{1}$ and $\antgi{2}$ %(denoted by $\rrgsi{1} \funbisim \rrgsi{2}$)
will be defined as a function $\sahom \funin \vertsi{1} \to \vertsi{2}$ 
between the vertex sets of the disjoint unions %$\atgi{1}$ and $\atgi{2}$ 
of the term graphs 
in the image of $\srecspeci{1}$ and $\srecspeci{2}$, respectively;
on vertices %of $\atgi{1}$ and $\atgi{2}$ 
labeled with atomic, or interface labels, 
$\sahom$ behaves like an ordinary term graph homomorphism;
and on   vertices labeled with nested symbols %$\afunsymvari{1}$ and $\afunsymvari{2}$, 
an `interface' clause applies.
This condition, illustrated in Figure~\ref{fig:nested-bisimulation}, 
demands that via $\sahom$ related vertices $\avert$ and $\bvert$ with nested symbols
entail, following $\svin$\nb-links of the corresponding \entgs, that the roots of the symbol definition are related via $\sahom$,
and, following the $\svout$\nb-links of the corresponding \entgs, that respective successors of $\avert$ and $\bvert$
are related via $\sahom$. 

\begin{figure}[t]
\begin{center} 
  \begin{tikzpicture}  
  %\draw[help lines] (-8,-2) grid (8,2); 

\pgfdeclarelayer{foreground}
\pgfdeclarelayer{main}
\pgfdeclarelayer{background}
\pgfsetlayers{background,main,foreground}

\begin{pgfonlayer}{main}   
%\begin{pgfonlayer}{background}
  \draw [draw=white,rounded corners=1mm,fill=blue!10] (-8,-1.6) rectangle (-0.6,1.7);
  
  \draw [draw=white,rounded corners=1mm,fill=green!10] (0,-1.6) rectangle (7.5,1.7);
%\end{pgfonlayer}  

%\begin{pgfonlayer}{foreground}   
  \matrix[anchor=center,row sep=0.8cm,column sep=0.5cm,
           every node/.style={draw,circle,fill=white,scale=0.65,minimum size=0.6cm,inner sep=0pt}] 
          at (-6.5,0) {
    & \node(f1)[thick,densely dotted]{$\afunsymvari{1}$}; 
        %\node(f1_label)[draw=white]{$\scriptstyle\averti{1}$};    
    \\
    \node(f1_succ_1){}; & \node(f1_succ_i){}; & \node(f1_succ_n1){};
    \\     
    };
  \path (f1) ++ (+0.31cm,-0.15cm) node {$\scriptstyle\bvert$};  
    
  \draw[<-](f1) -- ++ (90:0.6cm);\draw[<-](f1) -- ++ (60:0.6cm);\draw[<-](f1) -- ++ (120:0.6cm);    
  \draw[->](f1) to node[left]{$\scriptstyle 1$} (f1_succ_1);
  \draw[->](f1) to node[left]{$\scriptstyle i$} (f1_succ_i);
  \draw[->](f1) to node[right]{$\scriptstyle n_1$} (f1_succ_n1);
    
  \draw[dotted,shorten >=3pt,shorten <=3pt](f1_succ_1) to (f1_succ_i); 
  \draw[dotted,shorten >=3pt,shorten <=3pt](f1_succ_i) to (f1_succ_n1); 
  
  \matrix[anchor=center,row sep=1.4cm,column sep=0.6cm,
           every node/.style={draw,circle,fill=white,scale=0.65,minimum size=0.6cm,inner sep=0pt}] 
          at (-2.5,0) {
    & \node(F1_root)[thick]{$\ssout$};   
    \\
    \node[thick](F1_x1){$\ssini{1}$}; & \node[thick](F1_xi){$\ssini{i}$}; & \node[thick](F1_xn1){$\ssini{n_1}$};
    \\     
    };
    \path (F1_root) ++ (-1.2cm,+0.45cm) node (F1_label) {$\graphafunsymi{1}$};
  
    \draw[|->,thick,shorten <=5pt]
      (f1) to node[below]{$\scriptstyle\srecspeci{1}$} (F1_label);
      
    \draw[|->,densely dotted,shorten <=8pt,shorten >=5pt,bend right,distance=0.6cm]
      (f1) to node[above]{$\scriptstyle\svin$} (F1_root);   
    \draw[|->,densely dotted,shorten <=3pt,shorten >=3pt,bend right,distance=1.35cm]
      (F1_xi) to node[above]{$\scriptstyle\svout$} (f1_succ_i);

    %\draw[<-,thick,>=latex](F1_root) -- ++ (90:0.6cm);
    \draw[<-,thick,>=latex]($(F1_root.west)$) -- ($(F1_root.west) + (-0.3cm,0pt)$);
    %\draw[->](F1_x1) -- ++ (90:-0.6cm); 
      \draw[<-](F1_x1) -- ++ (80:+0.8cm);
    %\draw[->](F1_xi) -- ++ (90:-0.6cm); 
      \draw[<-](F1_xi) -- ++ (90:+0.8cm);
    %\draw[->](F1_xn1) -- ++ (90:-0.6cm); 
      \draw[<-](F1_xn1) -- ++ (110:+0.8cm);
  
    \draw[dotted,shorten >=0.4cm,shorten <=0.4cm]($(F1_x1) + (0,0.15cm)$) to ($ (F1_xi) + (0,0.15cm) $); 
    \draw[dotted,shorten >=0.4cm,shorten <=0.4cm]($(F1_xi) + (0,0.15cm)$) to ($ (F1_xn1) + (0,0.15cm) $);

  \matrix[anchor=center,row sep=0.8cm,column sep=0.5cm,
           every node/.style={draw,circle,fill=white,scale=0.65,minimum size=0.6cm,inner sep=0pt}] 
          at (1.6,0) {
    & \node(f2)[densely dotted,thick]{$\afunsymvari{2}$};   
    \\
    \node(f2_succ_1){}; & \node(f2_succ_j){}; & \node(f2_succ_n2){};
    \\     
    };
  \path (f2) ++ (+0.55cm,-0.15cm) node {$\scriptstyle\ahom{\bvert}$};  
  
  \draw[<-](f2) -- ++ (110:0.6cm);\draw[<-](f2) -- ++ (70:0.6cm);    
  \draw[->](f2) to node[left]{$\scriptstyle 1$} (f2_succ_1);
  \draw[->](f2) to node[left]{$\scriptstyle j$} (f2_succ_j);
  \draw[->](f2) to node[right]{$\scriptstyle n_2$} (f2_succ_n2);
  
  \draw[dotted,shorten >=3pt,shorten <=3pt] (f2_succ_1) to (f2_succ_j); 
  \draw[dotted,shorten >=3pt,shorten <=3pt] (f2_succ_j) to (f2_succ_n2);

  \matrix[anchor=center,row sep=1.4cm,column sep=0.6cm,
           every node/.style={draw,circle,fill=white,scale=0.65,minimum size=0.6cm,inner sep=0pt}] 
          at (5.6,0) {
    & \node(F2_root)[thick]{$\ssout$};   
    \\
    \node(F2_x1)[thick]{$\ssini{1}$}; & \node(F2_xj)[thick]{$\ssini{j}$}; & \node(F2_xn2)[thick]{$\ssini{n_2}$};
    \\     
    };
    \path (F2_root) ++ (-1.65cm,+0.3cm) node (F2_label) {$\graphafunsymi{2}$};
  
    \draw[|->,thick,shorten <=5pt,shorten >=0pt]
      (f2) to node[above]{$\scriptstyle\srecspeci{2}$} (F2_label);
      
    \draw[|->,densely dotted,shorten <=15pt,shorten >=3pt,bend right,distance=0.9cm]
      (f2) to node[above]{$\scriptstyle\svin$} (F2_root);          
    \draw[|->,densely dotted,shorten <=3pt,shorten >=3pt,bend right,distance=1.25cm]
      (F2_xj) to node[above]{$\scriptstyle\svout$} (f2_succ_j);    
  
    %\draw[<-,thick,>=latex](F2_root) -- ++ (90:0.6cm);
    \draw[<-,thick,>=latex]($(F2_root.west)$) -- ($(F2_root.west) + (-0.3cm,0pt)$);
    %\draw[->](F2_x1) -- ++ (90:-0.6cm); 
      \draw[<-](F2_x1) -- ++ (80:+0.8cm);
    %\draw[->](F2_xj) -- ++ (90:-0.6cm); 
      \draw[<-](F2_xj) -- ++ (90:+0.8cm);
    %\draw[->](F2_xn2) -- ++ (90:-0.6cm); 
      \draw[<-](F2_xn2) -- ++ (110:+0.8cm);
  
    \draw[dotted,shorten >=0.4cm,shorten <=0.4cm] ($(F2_x1) + (0,0.15cm)$) to ($ (F2_xj) + (0,0.15cm) $); 
    \draw[dotted,shorten >=0.4cm,shorten <=0.4cm] ($(F2_xj) + (0,0.15cm)$) to ($ (F2_xn2) + (0,0.15cm) $);

  %\draw (nodeA) -- (nodeB) node [midway, above, sloped] (TextNode) {path text};
  
  \draw[|->,shorten <= 3pt,shorten >=3pt,thick,densely dashed,bend left,distance=3cm] (f1) to node[above,near start] {$\sahom$} node [above,midway,sloped] (f1f2) {} (f2);
  \draw[|->,shorten <= 3pt,shorten >=3pt,thick,densely dashed,bend left,distance=2.4cm] (F1_root) to node[above,near end] {$\sahom$} node [above,sloped] (F1_root_F2_root){} (F2_root) ; 
  
  \draw[|->,shorten <= 3pt,shorten >=3pt,thick,densely dashed,bend right,distance=1.75cm] (F1_xi) to node[below,near end] {$\sahom$} node[below,midway,sloped] (F1_xi_F2_xj){} (F2_xj); 
  \draw[|->,shorten <= 3pt,shorten >=3pt,thick,densely dashed,bend right,distance=2.6cm] (f1_succ_i) to node[below,near start] {$\sahom$} node [below,midway,sloped] (f1_succ_i_f2_succ_j){} (f2_succ_j);  
 
    \draw[double,->,thick,bend left,distance=0.6cm] (f1f2) to (F1_root_F2_root); 
    \draw[double,->,thick,bend left,distance=0.6cm] (F1_xi_F2_xj) to node [above,midway,sloped] (point) {} (f1_succ_i_f2_succ_j);
    \draw[double,->,thick,bend left,distance=2.85cm,shorten <=0.3cm] (f1f2.90) to (point); 

%\end{pgfonlayer}{foreground} 

%\begin{pgfonlayer}{main}    
  
  \node[draw=none,fill=none] (recspec1_label) at ($ (F1_root.north)   + (-4.5cm,+0.8cm) $) {$\atgi{1}$}; % {$\rrgsi{1}$};
  
  \node[draw=none,fill=none] (recspec1_label) at ($ (F2_root.north)   + (+1cm,+0.8cm) $) {$\atgi{2}$};   % {$\rrgsi{2}$};
  
  \draw [thick,dotted%,fill=green!18
                     ] 
    ($ (F2_xj.south)   + (-1.4cm,0cm) $) rectangle ($ (F2_root.north) + (+1.4cm,0cm) $);
         
  \draw[dotted,thick%,fill=blue!18
                    ] 
    ($ (F1_xi.south)   + (-1.4cm,0cm) $) rectangle ($ (F1_root.north) + (+1.4cm,0cm) $);
      
\end{pgfonlayer}{main}

\end{tikzpicture}    
\end{center}  
  \vspace*{-2ex}
  \caption{\label{fig:nested-bisimulation} 
           Bisimulation `interface clause' for homorphisms between \ntgs\ in case of related vertices with nested symbols. % $\afunsymvari{1}$ and $\afunsymvari{2}$.
           Its motivation consists in following the dotted $\svin$- and $\svout$-links of the corresponding \entgs:
           if $\sahom$ maps a vertex $\bvert$ with nested symbol $\afunsymvari{1}$ to vertex $\ahom{\bvert}$ with nested symbol $\afunsymvari{2}$,
           then $\sahom$ must also map the root of the definition $\graphafunsymi{1}$ of $\afunsymvari{1}$ 
           to the root of the definition $\graphafunsymi{2}$ of $\afunsymvari{2}$;
           and if $\sahom$ maps the input vertex~$\ssini{i}$ of $\graphafunsymi{1}$
           to an input vertex $\ssini{j}$ of $\graphafunsymi{2}$,
           then $\sahom$ must also map the $i$-th successor of $\bvert$ to the $j$-th successor~of~$\ahom{\bvert}$.}
  \vspace*{-0.75ex}         
\end{figure}

% (see the picture in Figure~\ref{fig:nested-bisimulation}).
% A bisimulation between $\rrgsi{1}$ and $\rrgsi{2}$ (denoted by $\rrgsi{1} \bisim \rrgsi{2}$) will then be defined as 
% an \rgs~$\rrgs$ such that $\rrgsi{1} \convfunbisim \rrgs \funbisim \rrgsi{2}$.
% More precisely, the following definition applies.

\begin{definition}[homomorphism, bisimulation between \ntgs]\label{def:hom:bisim:rgss}
  Let $\asigi{1} = \asigat \cup \asigine{1}$ 
  and $\asigi{2} = \asigat \cup \asigine{2}$
  be \ntgsig{s} with the same signature $\asigat$ for atomic symbols. 
  Let $\antgi{1} = \pair{\srecspeci{2}}{\srootsymvari{1}}$ and $\antgi{2} = \pair{\srecspeci{2}}{\srootsymvari{2}}$
  be nested term graphs over $\asigi{1}$ and $\asigi{2}$, respectively.  
  Let
  $\atgi{i} = \tuple{\vertsi{i},\svlabi{i},\svargsi{i},\srootofi{i},\soutsofi{i}}$
  for $i\in\setexp{1,2}$ be 
  the enriched (not necessarily root-connected) term graphs 
  that arise as the disjoint union of the term graphs $\recspeci{i}{\afunsymvar}$ for $\afunsymvar\in\asigine{i}$
  together with functions $\srootofi{i} \funin \asigine{i} \to \vertsi{i}$ 
                and       $\soutsofi{i} \funin \asigine{i} \to \powersetof{\vertsi{i}}$                                 
  that map a nested function symbol $\afunsymvar$ to the root $\rootofi{i}{\afunsymvar}$,
  and to the set $\outsofi{i}{\afunsymvar}$ of input vertices, 
  of the definition of $\afunsymvar$ in $\atgi{i}$,
  respectively.                                
    % that map a nested function symbol $\afunsymvar$ to its root $\rootofi{i}{\afunsymvar}$ in $\atgi{i}$,
    % and to its set $\outsofi{i}{\afunsymvar}$ of input vertices in $\atgi{i}$, respectively. 
  
  A \emph{homomorphism} %(functional bisimulation) 
                        between $\antgi{1}$ and $\antgi{2}$ 
  is a function $\sahom \funin \vertsi{1} \to \vertsi{2}$
  such that for all $\bvert\in\vertsi{1}$ it holds
  (a condition in brackets $[\ldots]$ has been added for clarity, but is redundant,
   see Remark~\ref{rem:def:hom:bisim:rgss}):
  \begin{align*}
    \ahom{\rootofi{1}{\srootsymvari{1}}} & = \rootofi{2}{\srootsymvari{2}}
    & & (\text{roots of \rgss})
    %\displaybreak[0]
    \\
        % \ahomne{\afunsymvari{1}} = \afunsymvari{2}
        %   \;\; & \Longrightarrow\;\;
        %     \ahom{\rootofi{1}{\afunsymvari{1}}} = \rootofi{2}{\afunsymvari{2}}
        % & & (\text{roots of spec's})
        % \displaybreak[0]\\
     \vlabi{1}{\bvert}\in\asigat 
      \;\; & \Longrightarrow\;\;
        \vlabi{2}{\ahom{\bvert}} = \vlabi{1}{\bvert} \in\asigat
          \;\,\logand\,\;
        \ahomwords{\vargsi{1}{\bvert}} = \vargsi{2}{\ahom{\bvert}} 
    & & (\svlab,\,\svargs)_{{\asigat}} 
    \displaybreak[0]\\
    \vlabi{1}{\bvert}\in\Outs 
      \;\; & \Longrightarrow\;\;
        \vlabi{2}{\ahom{\bvert}}\in\Outs
    & & (\svlab)_{{\Outs}}    
    \displaybreak[0]\\
    \vlabi{1}{\bvert}\in\Ins
      \;\; & \Longrightarrow\;\;
        \vlabi{2}{\ahom{\bvert}}\in\Ins
    & & (\svlab)_{{\Ins}}    
    \displaybreak[0]\\
    \vlabi{1}{\bvert} % = \afunsymvari{1}
                        \in\asigine{1} 
      \;\; & \Longrightarrow\;\;
        \left\{\,
        \begin{aligned}
          &
          \vlabi{2}{\ahom{\bvert}} \in \asigine{2}
            \;\logand\;
          \ahom{\rootofi{1}{\vlabi{1}{\bvert}}} = \rootofi{2}{\vlabi{2}{\ahom{\bvert}}}  
          \\%[-0.5ex] 
          &
          [\, 
          \,\logand\;
          \forall \cvert\in\outsofi{1}{\vlabi{1}{\bvert}}. \;\;
            \ahom{\cvert}\in\outsofi{2}{\vlabi{2}{\ahom{\bvert}}}
          \,]  
          \\
          &
          \,\logand\;
          \forall \cvert\in\outsofi{1}{\vlabi{1}{\bvert}}. \,
            \forall i,j\in\nat. \,
              \forall \dvert\in\vertsi{1}.\, \forall \evert\in\vertsi{2}.\,
          \\[-0.5ex]
          & \hspace*{5ex}
          \vlabi{1}{\cvert} = \ssini{i} 
            \;\logand\;
          \vlabi{2}{\ahom{\cvert}} = \ssini{j}
            \;\logand\;
          \bvert \tgsucci{i} \dvert
            \;\,\logand\,\;
          \ahom{\bvert} \tgsucci{j} \evert    
          \\[-0.5ex]
          & \hspace*{10ex}
            \Rightarrow\;\;
              %\ahom{\vargsiatj{1}{\bvert}{i}} = \vargsiatj{2}{\ahom{\bvert}}{j} 
              \ahom{\dvert} = \evert    
        \end{aligned}
        \right.
    & & (\svlab,\,\svargs)_{{\asigne}} 
  \end{align*}
  hold, where $\sahomwords$ is the homomorphic extension of $\sahom$ 
  to a function from $\wordsover{\vertsi{1}}$ to $\wordsover{\vertsi{2}}$. 
  See  Figure~\ref{fig:nested-bisimulation} 
  for an illustration of the `interface clause' $(\svlab,\,\svargs)_{{\asigne}}$.
  If there is a homomorphism $\sahom$ from $\antgi{1}$ to $\antgi{2}$,
  we write $\antgi{1} \funbisimi{\sahom} \antgi{2}$
  and $\antgi{2} \convfunbisimi{\sahom} \antgi{1}$,
  or, dropping $\sahom$ as subscript, 
  $\antgi{1} \funbisim \antgi{2}$ and $\antgi{2} \convfunbisim \antgi{1}$.
  
  A \emph{bisimulation} between $\antgi{1}$ and $\antgi{2}$ is
  an \ntg~$\antg$ over signature $\asig = \asigat \cup \asigne$
  with $\asigne \subseteq \asigine{1} \times \asigine{2}$ 
  such that 
  $ \antgi{1} \convfunbisimi{\pair{\sproji{1}}{\sahom}} \antg \funbisimi{\pair{\sproji{2}}{\sahom}} \antgi{2}$
  where $\sproji{1}$ and $\sproji{2}$ are projection functions, defined, for $i\in\setexp{1,2}$,
  by $ \sproji{i} \funin \asigine{1}\times\asigine{2} \to \asigine{i} $,
  $\pair{\afunsymvari{1}}{\afunsymvari{2}} \mapsto \afunsymvari{i}$.
\end{definition}

\begin{remark}\label{rem:def:hom:bisim:rgss}
  In condition~$(\svlab,\,\svargs)_{{\asigne}}$ 
  for a homomorphism between nested term graphs in Definition~\ref{def:hom:bisim:rgss}
  the part
  $\forall \cvert\in\outsofi{1}{\vlabi{1}{\bvert}}. \;\;
            \ahom{\cvert}\in\outsofi{2}{\vlabi{2}{\ahom{\bvert}}}$
  is redundant.  
  It expresses that if a homomorphism~$\sahom$ 
  maps the root $\rootofi{1}{\afunsymi{1}}$ in $\atgi{1}$ of the definition a nested function symbol $\afunsymi{1}$
  to the root $\rootofi{2}{\afunsymi{2}}$ in $\atgi{2}$ of the definition of a nested function symbol $\afunsymi{2}$ 
  (by Definition~\ref{def:rgss}, $\rootofi{1}{\afunsymi{1}}$ and $\rootofi{2}{\afunsymi{2}}$ must be output vertices),  
  then $\sahom$ maps input vertices of the definition of $\afunsymi{1}$ in $\atgi{1}$
  to input vertices of the definition of $\afunsymi{2}$ in $\atgi{2}$. 
  This, and additionally also the fact that
  $\forall \dvert\in\outsofi{2}{\vlabi{2}{\ahom{\bvert}}} \,
     \exists \cvert\in\outsofi{1}{\vlabi{1}{\bvert}}. \;\; 
       \ahom{\cvert} = \dvert$ holds,
  follow from the other conditions since a homomorphism is a function, 
  and importantly, since definitions of nested symbols are term graphs that were assumed to be root-connected
  by default.
  By the latter, input vertices of the definition of a nested symbol are always reachable from 
  the output vertex at the root of the definition, which facilitates a proof of these properties
  using induction on the length of paths from output to input vertices. 
\end{remark}

%$\supap{\indap{\bvert}{\!\!j}}{\prime}$,  $w_j^\prime$, $w_j'$

% In the case of nested term graphs, one can rely on a definition for \rgss. 
%   
% \begin{definition}[homomorphism, bisimulation between \ntgs]
%   Let $\antgi{1}$ and $\antgi{2}$ be nested term graphs over signature
%   $\asigi{1} = \asigat \cup \asigine{1}\,$, 
%   and $\asigi{1} = \asigat \cup \asigine{2}\,$, respectively. 
%   A \emph{homomorphism} from $\antgi{1}$ to $\antgi{2}$
%   is a homomorphism from the \rgs~$\antgi{1}$ to the \rgs~$\antgi{2}$. 
%   A \emph{bisimulation} between $\antgi{1}$ and $\antgi{2}$
%   is a bisimulation between the \rgss~$\antgi{1}$ and $\antgi{2}$.
% \end{definition}  

\begin{example}  
  See Figure~\ref{fig:2-vs-1-exits} for four nested term graphs that are related by homomorphisms,
  and hence are bisimilar.  
  Note that homomorphisms can map a nested symbol to one of smaller arity (here from arity~2 to arity~1). 
  For the nested term graphs $\antg$ and $\ntgspecby{\rrgs}$ in Figure~\ref{fig:ex-nested-bisim}
  (the notation $\ntgspecby{\rrgs}$ will become clear later in Definition~\ref{def:rgs-specified-ntg})
  it holds that $\ntgspecby{\rrgs} \funbisim \antg$, and hence that they are bisimilar;
  but there is no homomorphism from $\antg$ to $\ntgspecby{\rrgs}$, and hence $\antg \notfunbisim \ntgspecby{\rrgs}$.
  % We note that $\ntgspecby{\rrgs}$ is the nested term graph that is `specified by' the \rgs~$\rrgs$ in Figure~\ref{fig:ex-nested-bisim}
  % as defined later in Definition~\ref{def:rgs-specified-ntg}.
\end{example}

\begin{proposition}
  The notions of homomorphism and bisimilarity for \ntgs\
  correspond to the notions of homomorphism and bisimilarity for \entgs,
  via the mappings between these concepts stated in Proposition~\ref{prop:ntgs:entgs}.  
\end{proposition}

\begin{figure}[t]
\vspace*{-2ex} 
\begin{center} 
  {\begin{tikzpicture}[scale=0.6,
    node/.style={%
      draw,
      circle,
      inner sep=0,
      outer sep=0,
      minimum size=0,
      node distance=0,
    }]
      
\matrix[row sep=0.3cm,column sep=0.1cm,every node/.style={draw,thick,circle,scale=0.6,minimum size=0.6cm,inner sep=0pt}] at (-0.1,0) { 
  \node[draw=none,fill=none](111_helper1){}; & \node(111_root){$\ssout$}; &
  \\
                                         & \node(111_0){$\sslabs$}; & 
  \\
                                         & \node(111_00){$\sslapp$}; &
  \\ 
  \node(111_000){$\ssini{1}$}; & \node[draw=none,fill=none](111_helper2){};
  \\[1ex]
  \node(111_0000){$\aconstsym$}; & 
  \\
  }; 
\draw[<-,thick,>=latex]($(111_root.north) + (0pt,2.5pt)$) -- ($(111_root.north) + (0pt,2.5pt) + (0pt,0.5cm)$);  
\draw[->](111_root) to (111_0);
\draw[->](111_0) to (111_00);
\draw[->](111_00) to (111_000);
\draw[->,shorten <= 2pt](111_000) to (111_0000);
\draw[->](111_00) to[out=-40,in=60,distance=1.75cm] (111_00);
\draw[dotted,thick]
  (111_000.south)   -- ($ (111_000.south) + (-0.7cm,0) $)
                -- ($ (111_helper1.north) + (-0.7cm,0) $)
                -- (111_root.north);
\draw[dotted,thick]        
  (111_root.north)  -- ($ (111_root.north) + (1.5cm,0) $)
                -- ($ (111_helper2.south) + (1.5cm,0) $)
                -- (111_000.south);

\matrix[row sep=0.3cm,column sep=0.1cm,every node/.style={draw,fill=white,thick,circle,minimum size=0.6cm,scale=0.6,inner sep=0pt}] at (5,0) { 
  \node[draw=none,fill=none](211_helper1){}; & \node(211_root){$\ssout$};
  \\
                                         & \node(211_0){$\sslabs$};
  \\                                       
                                         & \node(211_00){$\sslapp$}; 
  \\  
  &                                      & \node(211_001){$\sslapp$};
  \\
  \node(211_000){$\ssini{1}$}; & \node[draw=none,fill=none](211_helper2){}; 
  \\[1ex]
  \node(211_0000){$\aconstsym$}; 
  \\
  };
%\draw[<-,thick,>=latex](211_root) -- ++ (90:0.7cm);
\draw[<-,thick,>=latex]($(211_root.north) + (0pt,2.5pt)$) -- ($(211_root.north) + (0pt,2.5pt) + (0pt,0.5cm)$);  
\draw[->](211_root) to (211_0);
\draw[->](211_0) to (211_00);
\draw[->](211_00) to (211_000);
\draw[->,shorten <= 2pt](211_000) to (211_0000);
\draw[->](211_001) to (211_000);
\draw[->](211_00) to (211_001);
\draw[->](211_001) to[out=-40,in=60,distance=1.6cm] (211_00);

\draw[dotted,thick%,red
                  ]
  (211_000.south) -- ($ (211_000.south) + (-0.7cm,0) $)
              -- ($ (211_helper1.north) + (-0.7cm,0) $)
              -- (211_root.north);
\draw[dotted,thick]        
  (211_root.north) -- ($ (211_root.north) + (1.9cm,0) $)
               -- ($ (211_helper2.south) + (1.9cm,0) $)
               -- (211_000.south);

% homomorphism from 211 to 111
% 
\draw[|->,densely dotted,shorten >= 2pt,shorten <= 2pt]
  (211_root) to (111_root);                
\draw[|->,densely dotted,shorten >= 2pt,shorten <= 2pt]
  (211_0) to (111_0);    
\draw[|->,densely dotted,shorten >= 2pt,shorten <= 2pt]
  (211_00) to (111_00);                      
\draw[|->,densely dotted,shorten >= 2pt,shorten <= 2pt]
  (211_001) to (111_00);                          
\draw[|->,densely dotted,shorten >= 2pt,shorten <= 2pt]
  (211_000) to (111_000);                     
\draw[|->,densely dotted,shorten >= 2pt,shorten <= 2pt]
  (211_0000) to (111_0000);

\matrix[row sep=0.3cm,column sep=0.18cm,every node/.style={draw,thick,circle,minimum size=0.6cm,scale=0.6,inner sep=0pt}] at (10.85,0) { 
  \node[draw=none](221_helper){}; & & & \node(221_root){$\ssout$}; & & 
  \\
                              & & & \node(221_0){$\sslabs$}; 
  \\
                              & & & \node(221_00){$\sslapp$}; & &
  \\  
                              & & & & \node(221_001){$\sslapp$};
  \\
  \node(221_000){$\ssini{1}$}; & & & \node(221_0010){$\ssini{2}$}; 
  \\[1ex]
  \node(221_0000){$\aconstsym$};
  \\
  };
%\draw[<-,thick,>=latex](221_root) -- ++ (90:0.7cm); 
\draw[<-,thick,>=latex]($(221_root.north) + (0pt,2.5pt)$) -- ($(221_root.north) + (0pt,2.5pt) + (0pt,0.5cm)$);   
\draw[->](221_root) to (221_0);
\draw[->](221_0) to (221_00);
\draw[->,shorten <= 2pt](221_000.south) to (221_0000);
\draw[->](221_00) to (221_001);
\draw[->,shorten <= 4pt](221_0010.south) to (221_0000);
\draw[->](221_001) to[out=-40,in=60,distance=1.6cm] (221_00);
\draw[->](221_00) to (221_000);
\draw[->](221_001) to (221_0010);
\draw[dotted,thick]
  (221_000.south)  -- (221_0010.south);
\draw[dotted,thick]
  (221_000.south)  -- ($ (221_000.south) + (-0.7cm,0) $)
                   -- ($ (221_helper.north) + (-0.7cm,0) $)
                   -- (221_root.north);
\draw[dotted,thick]        
  (221_root.north) -- ($ (221_root.north) + (2.2cm,0) $)
                   -- ($ (221_0010.south) + (2.2cm,0) $)
                   -- (221_0010.south);
               
% homomorphism from 221 to 211
% 
\draw[|->,densely dotted,shorten >= 2pt,shorten <= 2pt]
  (221_root) to (211_root);                
\draw[|->,densely dotted,shorten >= 2pt,shorten <= 2pt]
  (221_0) to (211_0);    
\draw[|->,densely dotted,shorten >= 2pt,shorten <= 2pt]
  (221_00) to (211_00);                      
\draw[|->,densely dotted,shorten >= 2pt,shorten <= 2pt]
  (221_001) to (211_001);                          
\draw[|->,densely dotted,shorten >= 2pt,shorten <= 2pt]
  (221_000) to (211_000);                            
\draw[|->,densely dotted,shorten >= 4pt,shorten <= 2pt,bend right,distance=1.2cm]
  (221_0010) to (211_000);                     
\draw[|->,densely dotted,shorten >= 2pt,shorten <= 2pt]
  (221_0000) to (211_0000);

\matrix[row sep=0.3cm,column sep=0.18cm,every node/.style={draw,thick,circle,minimum size=0.6cm,scale=0.6,inner sep=0pt}] at (17.1,0) { 
  \node[draw=none](222_helper){}; & & & \node(222_root){$\ssout$};
  \\ 
                              & & & \node(222_0){$\sslabs$}; 
  \\
                              & & & \node(222_00){$\sslapp$}; & &
  \\  
                              & & & & \node(222_001){$\sslapp$};
  \\
  \node(222_000){$\ssini{1}$}; & & & \node(222_0010){$\ssini{2}$}; 
  \\[1ex]
  \node(222_0000){$\aconstsym$};  & & & \node(222_00100){$\aconstsym$};
  \\
  };
%\draw[<-,thick,>=latex](222_root) -- ++ (90:0.7cm); 
\draw[<-,thick,>=latex]($(222_root.north) + (0pt,2.5pt)$) -- ($(222_root.north) + (0pt,2.5pt) + (0pt,0.5cm)$);   
\draw[->](222_root) to (222_0);
\draw[->](222_0) to (222_00);
\draw[->,shorten <= 2pt](222_000) to (222_0000);
\draw[->](222_00) to (222_001);
\draw[->,shorten <= 2pt](222_0010) to (222_00100);
\draw[->](222_001) to[out=-40,in=60,distance=1.6cm] (222_00);
\draw[->](222_00) to (222_000);
\draw[->](222_001) to (222_0010);
\draw[dotted,thick]
  (222_000.south)  -- (222_0010.south);
\draw[dotted,thick]
  (222_000.south)  -- ($ (222_000.south) + (-0.7cm,0) $)
               -- ($ (222_helper.north) + (-0.7cm,0) $)
               -- (222_root.north);
\draw[dotted,thick]        
  (222_root.north) -- ($ (222_root.north) + (2.2cm,0) $)
               -- ($ (222_0010.south)  + (2.2cm,0) $)
               -- (222_0010.south);

% homomorphism from 221 to 222
% 
\draw[|->,densely dotted,shorten >= 2pt,shorten <= 2pt]
  (222_root) to (221_root);                
\draw[|->,densely dotted,shorten >= 2pt,shorten <= 2pt]
  (222_0) to (221_0);    
\draw[|->,densely dotted,shorten >= 2pt,shorten <= 2pt]
  (222_00) to (221_00);                      
\draw[|->,densely dotted,shorten >= 2pt,shorten <= 2pt]
  (222_001) to (221_001);                          
\draw[|->,densely dotted,shorten >= 2pt,shorten <= 2pt,bend left,distance=1.2cm]
  (222_000) to (221_000);                            
\draw[|->,densely dotted,shorten >= 4pt,shorten <= 2pt,bend right,distance=1.2cm]
  (222_0010) to (221_0010);                     
\draw[|->,densely dotted,shorten >= 2pt,shorten <= 2pt]
  (222_0000) to (221_0000);      
\draw[|->,densely dotted,shorten >= 2pt,shorten <= 2pt,bend left  ,distance=1.2cm]
  (222_00100) to (221_0000);      

\end{tikzpicture}}
\end{center}  
\vspace*{-3.5ex}
\caption{\label{fig:2-vs-1-exits}
         Four simple nested term graphs that are related by converse functional bisimilarity $\sconvfunbisim$ (and hence also by bisimilarity $\sbisim$)
         via homorphisms that are indicated as dotted assignments.}
\end{figure}        
 
%-------
\myparagraphbf{Nested bisimulation and nested homomorphism between \rgss\ and \ntgs}
  \label{app:nested-hom:nested-bisim:rgss:ntgs}
%-------
%
A `nested bisimulation' %between \rgss~$\rrgsi{1}$ and $\rrgsi{2}$
compares \ntgs, or for that matter also \rgss, 
by keeping track, along any chosen path, of the nesting history by means of stacks of nested vertices. 
% in a way in which the nesting history of comparing the \ntgs\ or \rgss\ is recorded by means of stacks of vertices.
It is defined between prefixed expressions $\preexp{\averti{1}\worddots\averti{k}}{\avert}$ and
$\preexp{\bverti{1}\worddots\bverti{k}}{\bvert}$ that describe a visit of the vertices $\avert$ and $\bvert$
in the context of histories of visits to vertices $\averti{i}$ and $\bverti{i}$ as recorded by the stacks
$\averti{1}\worddots\averti{k}$ and $\bverti{1}\worddots\bverti{k}$ 
of the nested vertices %of $\atgi{1}$ and $\atgi{2}$ 
in the nesting hierarchy above $\avert$ and $\bvert$, respectively.
  % The topmost stack element always indicates the parent nested vertex,
  % enabling a definition by local progression clauses. 
  % %
  % A definition by local progression rules is possible, since
  % %Convenient for a definition is that
  % the topmost stack element always indicates the parent nested vertex. 
These stacks facilitate the definition of nested bisimulation by purely local progression rules,
since the immediate nesting ancestor of a vertex can always be found on top of the stack.
% the topmost element always indicates the immediate ancestor in the nesting.
% We denote bisimilarity by $\bisimne$ and functional bisimilarity by $\funbisimne$. 

\begin{definition}[nested bisimulation and nested homomorphism between \rgss\ and \ntgs]
  Let $\asigi{1} = \asigat \cup \asigine{1}$ 
  and $\asigi{2} = \asigat \cup \asigine{2}$
  be \ntgsigs\ with the same signature $\asigat$ for atomic symbols. 
  Let $\rrgsi{1} = \pair{\srecspeci{2}}{\srootsymvari{1}}$ and $\rrgsi{2} = \pair{\srecspeci{2}}{\srootsymvari{2}}$
  be \rgss\ over $\asigi{1}$ and $\asigi{2}$, respectively.  
  Let
  $\atgi{i} = \tuple{\vertsi{i},\svlabi{i},\svargsi{i},\srooti{i},\srootofi{i}}$
  for $i\in\setexp{1,2}$ be 
  the enriched (typically not root-connected) term graph 
  that arises as the disjoint union of the term graphs $\recspeci{i}{\afunsymvar}$ for $\afunsymvar\in\asigine{i}$
  such that its root $\srooti{i}\in\vertsi{i}$ is the root of $\recspeci{i}{\srootsymvari{i}}$,
  and with as enrichment the function $\srootofi{i} \funin \asigine{i} \to \vertsi{i}$                                
  that maps a nested function symbol $\afunsymvar\in\asigine{i}$ to its root $\rootofi{i}{\afunsymvar}$ in $\atgi{i}$
  (hence $\srooti{i} = \rootofi{i}{\srootsymvari{i}}$).
  
  %\vspace{0.75ex}
  A \emph{nested bisimulation} between $\rrgsi{1}$ and $\rrgsi{2}$ 
  is a relation 
    $\sabisimne \subseteq \wordsover{\vertsi{1}} \times \vertsi{1} \times \wordsover{\vertsi{2}} \times \vertsi{2}$,
  for which we will indicate elements
  $\tuple{\averti{1}\worddots\averti{k},\avert,\bverti{1}\worddots\bverti{k},\bvert}\in\sabisimne$
  as
  $\preexp{\averti{1}\worddots\averti{k}}{\avert} \abisimnespace \preexp{\bverti{1}\worddots\bverti{k}}{\bvert}$, 
  with the following properties,
  for all $i,j,k\in\nats$, $\avert,\averti{1},\ldots,\averti{k},\avertacci{i}\in\vertsi{1}$,
  $\bvert,\bverti{1},\ldots,\bverti{k},\bvertacci{i},\bvertacci{j}\in\vertsi{2}\,$,
  $\afunsymvar\in\asigat$,
  $\afunsymvari{1}\in\asigine{1}$, and $\afunsymvari{2}\in\asigine{2}$:
  \begin{align*}
    (\sroot)^{\snested} & \hspace{3ex}
      \emptypreexp{\srooti{1}} 
        \abisimnespace
      \emptypreexp{\srooti{2}}  
        \hspace*{3ex}
      (\text{equivalently: }
       \emptypreexp{\rootofi{1}{\srootsymvari{1}}} 
         \abisimnespace
       \emptypreexp{\rootofi{2}{\srootsymvari{2}}}) 
    \displaybreak[0]\\
    (\svlab)^{\snested} & \hspace*{3ex}
      \preexp{\averti{1}\worddots\averti{k}}{\avert} 
        \abisimnespace
      \preexp{\bverti{1}\worddots\bverti{k}}{\bvert}
        \;\;\Longrightarrow\;\;
      \begin{aligned}[t]  
        &
        (\,\vlab{\avert} = \vlab{\bvert} \in \asigat\,)
          \logor
        (\,\vlab{\avert}\in\asigine{1} \logand \vlab{\bvert}\in\asigine{2}\,)
        \\
        &
          \logor
        (\, \vlab{\avert} = \vlab{\bvert} = \ssout \in\Outs \,)  
        % (\, \vlab{\avert}\in\Ins \logand \vlab{\bvert}\in\Ins \,)
          \logor
        (\, \vlab{\avert}\in\Ins \logand \vlab{\bvert}\in\Ins \,)
      \end{aligned} 
    \displaybreak[0]\\
    (\svargs)_{\asigat}^{\snested} & \hspace*{3ex}
      \begin{aligned}[t]
        &
        \preexp{\averti{1}\worddots\averti{k}}{\avert} 
          \abisimnespace
        \preexp{\bverti{1}\worddots\bverti{k}}{\bvert}
          \;\logand\;
        \vlabi{1}{\avert} = \vlabi{2}{\bvert} = \afunsymvar \in\asigat
          \;\logand\;
        \avert \tgsucci{i} \avertacci{i}
          \;\logand\;
        \bvert \tgsucci{i} \bvertacci{i}   
        \\
        & 
        \hspace*{3ex} \;\;\Longrightarrow\;\;
        \preexp{\averti{1}\worddots\averti{k}}{\avertacci{i}} 
            \abisimnespace
          \preexp{\bverti{1}\worddots\bverti{k}}{\bvertacci{i}}
          % \forall i\in\setexp{1,\ldots,\arity{\afunsymvar}}. \;\,
          %   \preexp{\averti{1}\worddots\averti{k}}{\vargsiatj{1}{\avert}{i}} 
          %     \abisimnespace
          %   \preexp{\bverti{1}\worddots\bverti{k}}{\vargsiatj{2}{\bvert}{i}}
      \end{aligned} 
    \displaybreak[0]\\
    (\svargs)_{\asigne}^{\snested} & \hspace*{3ex}
      \begin{aligned}[t]
        &
        \preexp{\averti{1}\worddots\averti{k}}{\avert} 
          \abisimnespace
        \preexp{\bverti{1}\worddots\bverti{k}}{\bvert}
          \;\logand\;
        \vlabi{1}{\avert} = \afunsymvari{1}\in\asigine{1}
          \;\logand\;
        \vlabi{2}{\bvert} = \afunsymvari{2} \in\asigine{2}
        \\
        & 
        \hspace*{3ex} \;\;\Longrightarrow\;\;
          \preexp{\averti{1}\worddots\averti{k}\avert}{\rootofi{1}{\afunsymvari{1}}} 
            \abisimnespace
          \preexp{\bverti{1}\worddots\bverti{k}\bvert}{\rootofi{2}{\afunsymvari{2}}} 
      \end{aligned}  
    \displaybreak[0]\\
    (\svargs)_{\Outs}^{\snested} & \hspace*{3ex}
      \begin{aligned}[t]
        &
        \preexp{\averti{1}\worddots\averti{k}}{\avert} 
          \abisimnespace
        \preexp{\bverti{1}\worddots\bverti{k}}{\bvert}
          \;\logand\;
        \vlabi{1}{\avert} = \vlabi{2}{\bvert} = \ssout\in\Outs
          \;\logand\;
        \avert \tgsucci{0} \avertacci{0}
          \;\logand\;
        \bvert \tgsucci{0} \bvertacci{0}    
        \\
        & 
        \hspace*{3ex} \;\;\Longrightarrow\;\;
        \preexp{\averti{1}\worddots\averti{k}}{\avertacci{0}} 
          \abisimnespace
        \preexp{\bverti{1}\worddots\bverti{k}}{\bvertacci{0}}
          % \preexp{\averti{1}\worddots\averti{k}}{\vargsiatj{1}{\avert}{1}} 
          %   \abisimnespace
          % \preexp{\bverti{1}\worddots\bverti{k}}{\vargsiatj{2}{\bvert}{1}}
      \end{aligned}    
    \displaybreak[0]\\
    (\svargs)_{\Ins}^{\snested} & \hspace*{3ex}
      \begin{aligned}[t]
        &
        \preexp{\averti{1}\worddots\averti{k}}{\avert} 
          \abisimnespace
        \preexp{\bverti{1}\worddots\bverti{k}}{\bvert}
          \;\logand\;
        \vlabi{1}{\avert} = \ssini{i}\in\Ins \;\logand\; \vlabi{2}{\bvert} = \ssini{j} \in\Ins
        \\
        & 
        \hspace*{3ex} \;\;\Longrightarrow\;\;
        \exists \avertacci{i}\in\vertsi{1}. 
          \, \exists \bvertacci{j}\in\vertsi{2}. \,
            \bigl(\,
              \averti{k} \tgsucci{i} \avertacci{i}
                \;\logand\;
              \bverti{k} \tgsucci{j} \bvertacci{j} 
                \;\logand\; 
              \preexp{\averti{1}\worddots\averti{k -1}}{\avertacci{i}} 
                \abisimnespace
              \preexp{\bverti{1}\worddots\bverti{k -1}}{\bvertacci{j}}  
            \,\bigr)  
          % k,k \ge 1 
          %   \;\logand\;
          % \preexp{\averti{1}\worddots\averti{k -1}}{\vargsiatj{1}{\averti{k}}{i}} 
          %   \abisimnespace
          % \preexp{\bverti{1}\worddots\bverti{k -1}}{\vargsiatj{2}{\bverti{k}}{j}}
      \end{aligned}      
  \end{align*}
  If there is a nested bisimulation $\sabisimne$ between $\rrgsi{1}$ and $\rrgsi{2}$,
  then we write $\rrgsi{1} \sbisimnei{\sabisimne} \rrgsi{2}$, 
  or just $\rrgsi{1} \bisimne \rrgsi{2}$.
  
  %\vspace{0.75ex}
  A \emph{nested homomorphism} from $\rrgsi{1}$ to $\rrgsi{2}$ 
  is a partial function 
    $ \sahomne \funin  \wordsover{\vertsi{1}} \times \vertsi{1}   \rightharpoonup   \wordsover{\vertsi{2}} \times \vertsi{2}$
  such that the relation
    $ \descsetexp{ \tuple{\averti{1}\worddots\averti{n},\avert,\bverti{1}\worddots\bverti{n},\bvert} 
                      \in \wordsover{\vertsi{1}} \times \vertsi{1} \times  \wordsover{\vertsi{2}} \times \vertsi{2} }
                 { \defd{\ahomne{\pair{\averti{1}\worddots\averti{n}}{\avert}}} = \pair{\bverti{1}\worddots\bverti{n}}{\bvert} }$
  is a nested bisimulation between $\rrgsi{1}$ and $\rrgsi{2}$. 
  If there is such a function $\sahomne$, % between $\rrgsi{1}$ and $\rrgsi{2}$,                               
  % If there exists a nested homomorphism between $\rrgsi{1}$ and $\rrgsi{2}$,    
  %then 
  we write $\rrgsi{1} \funbisimnei{\sahomne} \rrgsi{2}$, or just $\rrgsi{1} \funbisimne \rrgsi{2}$. 
    % A nested homomorphism $\sahomne$ between $\rrgsi{1}$ and $\rrgsi{2}$ is called a \emph{nested isomorphism}
    % if $\sahomne$ is injective. 
    % If there is a nested isomorphism between $\rrgsi{1}$ and $\rrgsi{2}$,    
    % we write $\rrgsi{1} \isonei{\sahom} \rrgsi{2}$, 
    % or just $\rrgsi{1} \isone \rrgsi{2}$. 
\end{definition}

\begin{example}
  The set $\abisimne$ defined in Figure~\ref{fig:ex-nested-bisim} 
  is a nested homomorphism from the rgs~$\rrgs$ to the nested term graph $\ntgspecby{\rrgs}$
  (the notation $\ntgspecby{\rrgs}$ is explained in Definition~\ref{def:rgs-specified-ntg} below).
  Hence it witnesses $\rrgs \funbisimnei{\abisimne} \ntgspecby{\rrgs}$. 
  Note that its converse also is a nested homomorphism, and hence that $\ntgspecby{\rrgs} \funbisimne \rrgs$ holds, too.
%   \begin{align*}
%      \abisimne =
%      \{\;
%      %&
%      \pair{\preexp{}{\averti{0}}}{\preexp{}{\bverti{0}}},\,
%      %\\
%      & 
%      \pair{\preexp{\averti{0}}{\averti{1}}}{\preexp{\bverti{0}}{\bverti{1}}}, \,
%      \pair{\preexp{\averti{0}}{\averti{2}}}{\preexp{\bverti{0}}{\bverti{2}}}, \,
%      \pair{\preexp{\averti{0}}{\averti{3}}}{\preexp{\bverti{0}}{\bverti{3}}}, \,
%      \ldots,\,
%      \pair{\preexp{\averti{0}}{\averti{7}}}{\preexp{\bverti{0}}{\bverti{7}}},  
%      \\
%      &
%      \pair{\preexp{\averti{0}\averti{3}}{\averti{8}}}{\preexp{\bverti{0}\bverti{3}}{\bverti{8}'}}, \,
%      \pair{\preexp{\averti{0}\averti{3}}{\averti{9}}}{\preexp{\bverti{0}\bverti{3}}{\bverti{9}'}}, \,
%      %\pair{\preexp{\averti{0}\averti{3}}{\averti{10}}}{\preexp{\bverti{0}\bverti{3}}{\bverti{10}'}}, \,
%      \ldots,\,
%      \pair{\preexp{\averti{0}\averti{3}}{\averti{12}}}{\preexp{\bverti{0}\bverti{3}}{\bverti{12}'}}, \,
%      \\
%      &
%      \pair{\preexp{\averti{0}\averti{3}}{\averti{8}}}{\preexp{\bverti{0}\bverti{3}}{\bverti{8}''}}, \, 
%      \pair{\preexp{\averti{0}\averti{3}}{\averti{9}}}{\preexp{\bverti{0}\bverti{3}}{\bverti{9}''}}, \,
%      %\pair{\preexp{\averti{0}\averti{3}}{\averti{10}}}{\preexp{\bverti{0}\bverti{3}}{\bverti{10}''}}, \,
%      \ldots,\,
%      \pair{\preexp{\averti{0}\averti{3}}{\averti{12}}}{\preexp{\bverti{0}\bverti{3}}{\bverti{12}''}} 
%      \;\}
%   \end{align*}
  There is also an obvious nested homomorphism from~$\ntgspecby{\rrgs}$ to the nested term graph $\antg$ in Figure~\ref{fig:ex-nested-bisim} right,
  but not the other way round, that is, $\antg \funbisimne \ntgspecby{\rrgs}$ does not hold.
 
  In the example concerning the four nested term graphs in Figure~\ref{fig:2-vs-1-exits},
  the indicated homomorphisms induce obvious corresponding nested homomorphisms. 
\end{example}

\begin{figure}
  \vspace*{-1.5ex}
\begin{flushleft}
  \scalebox{0.79}{\input{figs/ex-nested-bisim.tex}}
  \\[2.5ex]
  \hspace*{3ex}
  $
  \begin{small}
  \begin{aligned}
     \abisimne =
     \{\;
     %&
     \pair{\preexp{}{\averti{0}}}{\preexp{}{\bverti{0}}},\,
     %\\
     & 
     \pair{\preexp{\averti{0}}{\averti{1}}}{\preexp{\bverti{0}}{\bverti{1}}}, \,
     \pair{\preexp{\averti{0}}{\averti{2}}}{\preexp{\bverti{0}}{\bverti{2}}}, \,
     \pair{\preexp{\averti{0}}{\averti{3}}}{\preexp{\bverti{0}}{\bverti{3}}}, \,
     \ldots,\,
     \pair{\preexp{\averti{0}}{\averti{7}}}{\preexp{\bverti{0}}{\bverti{7}}},  
     \\
     &
     \pair{\preexp{\averti{0}\averti{3}}{\averti{8}}}{\preexp{\bverti{0}\bverti{3}}{\bverti{8}'}}, \,
     \pair{\preexp{\averti{0}\averti{3}}{\averti{9}}}{\preexp{\bverti{0}\bverti{3}}{\bverti{9}'}}, \,
     %\pair{\preexp{\averti{0}\averti{3}}{\averti{10}}}{\preexp{\bverti{0}\bverti{3}}{\bverti{10}'}}, \,
     \ldots,\,
     \pair{\preexp{\averti{0}\averti{3}}{\averti{12}}}{\preexp{\bverti{0}\bverti{3}}{\bverti{12}'}}, \,
     \\
     &
     \pair{\preexp{\averti{0}\averti{4}}{\averti{8}}}{\preexp{\bverti{0}\bverti{4}}{\bverti{8}''}}, \, 
     \pair{\preexp{\averti{0}\averti{4}}{\averti{9}}}{\preexp{\bverti{0}\bverti{4}}{\bverti{9}''}}, \,
     %\pair{\preexp{\averti{0}\averti{3}}{\averti{10}}}{\preexp{\bverti{0}\bverti{3}}{\bverti{10}''}}, \,
     \ldots,\,
     \pair{\preexp{\averti{0}\averti{4}}{\averti{12}}}{\preexp{\bverti{0}\bverti{4}}{\bverti{12}''}} 
     \;\}
  \end{aligned}
  \end{small}
  $
\end{flushleft}  
  \vspace*{-2.5ex}   
  \caption{\label{fig:ex-nested-bisim}%
           Example of a recursive graph specifications~$\rrgs$ (middle),
           the nested term graph $\ntgspecby{\rrgs}$ specified by $\rrgs$ (left),
           and a bisimilar nested term graph $\antg$ (right)
           that are related by $\sfunbisimne$ and $\bisimne$
           (i.p., $\rrgs \funbisimnei{\abisimne} \ntgspecby{\rrgs}$).}
\end{figure}

Every nested bisimulation $\abisimne$ between \rgss\ gives rise to an \rgs~$\rrgsi{\abisimne}$ in a straightforward
manner by forming, for every pair $\pair{\preexp{\averti{1}\ldots\averti{n}}{\avert}}{\preexp{\bverti{1}\ldots\bverti{n}}{\bvert}}$
a nested function symbol $\afunsymvari{\averti{1}\ldots\averti{n},\bverti{1}\ldots\bverti{n}}$,
and by letting the pair be a vertex with label $\vlabi{1}{\avert} = \vlabi{2}{\bvert}$
in the term graph specifying $\afunsymvari{\averti{1}\ldots\averti{n},\bverti{1}\ldots\bverti{n}}$.
As nesting is recorded in $\abisimne$, the \rgs~$\rrgsi{\abisimne}$ turns out to be a nested term graph.
Of particular interest are nested self-bisimulations on an \rgs, that is, bisimulations between an \rgs\ and itself.

\begin{definition}[nested term graph specified by an \rgs]\label{def:rgs-specified-ntg}
  Let $\rrgs$ be an \rgs\ over \ntgsig~$\asig$, and $\abisimne$ the minimal nested bisimulation
  between $\rrgs$ and itself.
  Together with $\rrgs$, $\abisimne$ specifies an \rgs~$\rrgsi{\abisimne}$
  with a tree as dependency~\ARS, and hence an \ntg.
  This is the nested term graph \emph{specified} by $\rrgs$,
  denoted by $\ntgspecby{\rrgs}$.
\end{definition}

\begin{example}\label{ex:rgs-specified-by}
  In Figure~\ref{fig:ex-nested-bisim},
  the nested term graph $\ntgspecby{\rrgs}$ on the left is specified by the \rgs~$\rrgs$ in the middle.
  In Figure~\ref{fig:rgs-entangled-ntgs},
  the \rgs~$\rrgsi{1}$ on the left specifies the nested term graph $\antgi{1}$ on the right,
  that is, $\antgi{1} = \ntgspecby{\rrgsi{1}}$. 
\end{example}

\begin{proposition}\label{lem:rgs-specified-ntg}
  For every \rgs~$\rrgs$ it holds:
  $\ntgspecby{\rrgs} \funbisimne \rrgs$ and $\rrgs \funbisimne \ntgspecby{\rrgs}$. 
\end{proposition}

%----------------------------------------------------------------------------------------------------
\myparagraphbf{Relationships between homomorphism/bisimilarity, and nested homomorphism/nested bisimilarity}
  \label{app:relation:hom:nested-hom:bisim:nested-bisim}
%----------------------------------------------------------------------------------------------------
%
We conclude this section with two statements
that relate $\sfunbisim$ and $\sbisim$ with $\sfunbisimne$ and $\sbisimne$ on nested term graphs,
and nested bisimilarity of \rgss\ with bisimilarity of the specified nested term graphs. 

\begin{theorem}\label{thm:bisim:bisimne:ntgs}
  For nested term graphs, functional bisimilarity $\sfunbisim$ coincides with nested functional bisimilarity $\sfunbisimne$, 
  and bisimilarity $\bisim$ coincides with nested bisimilarity $\bisimne$. 
\end{theorem}

The intuition behind this statement is as follows. 
In building up a bisimulation $\abisim$ between two bisimilar nested term graphs $\antgi{1}$ and $\antgi{2}$,
the tree structure of the dependency~\ARS{s}~$\dependsonARS$ together with the interface clause for bisimulation
guarantees that a vertex $\avert$ with nesting ancestors $\averti{1}\ldots\averti{k_1}$ is only related to a vertex
$\bvert$ with nesting ancestors $\bverti{1}\ldots\bverti{k_2}$ 
if $k_1 = k_2$.
  % if also $\averti{1}$, \ldots, $\averti{k}$ and $\bverti{1}$, \ldots, $\bverti{k}$ are related, respectively.
And furthermore, that by adding the nesting ancestors of vertices as prefixes
the bisimulation $\abisim$ gives rise to a nested bisimulation $\abisimne$ between $\antgi{1}$ and $\antgi{2}$.
Vice versa, again due to tree structure of the dependency~\ARS{s}~$\dependsonARS$,
the contextual information in a nested bisimulation can be ignored
to obtain a bisimulation.\enlargethispage{2ex}
Formally, Theorem~\ref{thm:bisim:bisimne:ntgs} and Theorem~\ref{thm:recspecs:bisimne:ntgsspecby:bisim} below,
can be proved by using induction on the length of `access paths' (acyclic paths from the root to a vertex).

% Both the tree structure of $\dependsonARS$ 
% and the interface clause for bisimilarity guarantee that
% no confusion can arise in the bisimulation over where to `jump back' from the input vertices
% to successors of the defined symbols. For every vertex there is a unique nested-symbol parent vertex,
% and therefore the contextual information recorded by a nested bisimulation can safely be ignored. 
%   % There is a unique (occurrence of) a nested symbol associated to the scope, 
%   % so that contextual information can safely be forgotten.
% Formally, Theorem~\ref{thm:bisim:bisimne:ntgs}, and Theorem~\ref{thm:recspecs:bisimne:ntgsspecby:bisim} below,
% can be proved by using induction on the length of `access paths' (acyclic paths from the root to a node).

\begin{theorem}\label{thm:recspecs:bisimne:ntgsspecby:bisim}
  Two recursive graph specifications $\rrgsi{1}$ and $\rrgsi{2}$ are nested bisimilar
  (i.e.\ $\rrgsi{1} \bisimne \rrgsi{2}$)
  if and only if
  the nested term graphs specified by $\rrgsi{1}$ and $\rrgsi{2}$, respectively,
  are bisimilar (i.e.\ $\ntgspecby{\rrgsi{1}} \bisim \ntgspecby{\rrgsi{2}}$).
\end{theorem}

\section{Interpretation as first-order term graphs}
  \label{sec:interpretation}
%----------------------------------------------------------------------- 

Nested term graphs can be interpreted in a faithful, and rather natural way as first-order term graphs.
By `faithful' we mean that the interpretation mapping is a retraction that preserves and reflects homomorphisms,
and by `natural' that it can be defined inductively on the nesting structure. % of nested term graphs.
The basic idea is analogous to the interpretation of \lambdahigherordertgs\
                                                     %\lambdatgs\ 
                                                     as first-order \lambdatgs\ developed in \cite{grab:roch:2013:a:TERMGRAPH,grab:roch:2014:ICFP}.

For a nested term graph $\antg = \pair{\srecspec}{\srootsymvar}$
we define the first-order term graph interpretation $\ntgstortgs{\antg}$ of $\antg$
by a stepwise procedure that starts on an \entg\ representation of $\antg$ as input.
The example of the \ntg~$\antg$ and the resulting interpretation $\ntgstortgs{\antg}$ in Figure~\ref{fig:nested-and-implementation} 
may help to provide some guiding intuition.

\begin{definition}\label{def:ntg:to:fotg}
  Let $\antg$ be a nested term graph over $\asig = \asigat \cup \asigne$, and 
  $\aedntg = \tuple{\verts,\svlab,\svargs,\svin,\svout,\svanc,\sroot}$ 
  be an \entg\ representation of $\antg$. 
  With $\asigat = \asigatconst \uplus \asigatfun$ a partitioning of $\asigat$ into constant
  and non-constant symbols,
  let $\asigatconstprime = \descsetexp{\aconstsym^{\prime}/1}{\aconstsym\in\asigatconst}$
          (that is, the constants in $\asigat$ are turned into corresponding unary symbols in $\asigatconstprime$),
  and let $\asigacc = \asigatconstprime \cup \asigatfun \cup \setexp{\ssout/1,\,\ssin/2,\,\ssoutroot/1,\,\ssinroot/1}$.
%   $\asigatacc = \descsetexp{\aconstsym^{\prime}/1}{\aconstsym\in\asigat\text{ a constant}}
%                   \cup
%                 \descsetexp{\cfunsym}{\cfunsym\in\asigat\text{ not a constant}}$
%   (that is, the constants in $\asigat$ are turned into corresponding unary symbols in $\asigatacc$).               
  %
  The \emph{first-order term graph interpretation $\ntgstortgs{\antg}$ of $\antg$}
  is a term graph over $\asigacc$ that is obtained from $\aedntg$ by the following steps:
  \begin{enumerate}[(i)]\itemizeprefs
    \item 
      Remove every vertex $\avert$ with a nested symbol, 
      redirect incoming edges at $\avert$ to the vertex $\vin{\avert}\,$.
    \item
      Relabel every input vertex $\bvert$ with nullary label $\ssini{k}$ 
      by the binary label $\ssin$ (thereby dropping the index $k$),
      directing the first edge (which becomes a back-link) from $\bvert$ to $\vout{\bvert}$,
      and the second edge from $\bvert$ to the vertex $\vin{\averti{n}}$,
      where $\vanc{\bvert} = \averti{1}\ldots\averti{n}$
      (note that $\vin{\averti{n}}$ has label $\ssout$).
    \item 
      Relabel the output vertex (with label $\ssout$) at the root by the special unary symbol $\ssoutroot\,$.
    \item
      Change every vertex with a nullary symbol~$\aconstsym$ into a vertex labeled with a corresponding unary symbol~$\aconstsym'$
      whose outgoing edge targets a chain of new binary input vertices whose back-links
      target respective output vertices of the nesting structure; the outermost input vertex
      gets label $\ssinroot$ and a backlink to $\ssoutroot\,$.
  \end{enumerate}
\end{definition}

\begin{example}
  See Appendix~A on pages~\pageref{app:start}--\pageref{app:end}
  for an application of this procedure on the \entg\ in Example~\ref{ex:ntg}.
\end{example}  

The statements that show that this interpretation is indeed faithful 
are closely analogous to the statements that establish this fact for the interpretation of 
$\lambda$-higher-order-term-graphs by the first-order `$\lambda$-term-graphs' introduced in \cite{grab:roch:2013:a:TERMGRAPH}. 
Here we only describe the most important steps and their underlying intuition. 

The first step is as follows. In analogy with the class $\lambda$-term-graphs in \cite{grab:roch:2013:a:TERMGRAPH},
those first-order term graphs that arise as interpretations of nested term graphs
belong to a class of term graphs that can be defined via the existence of
an ancestor function with appropriate properties, see the definition below. The name of this class
already anticipates the fact that all of its members do indeed represent nested term graphs.

\begin{definition}[term graphs that represent nested term graphs]\label{def:classrtgs}
  Let $\asig$ be an \ntgsig, and let $\asigacc$ be defined as in Definition~\ref{def:ntg:to:fotg}.
  Let $\atg = \tuple{\verts,\svlab,\svargs,\sroot}$ be a term graph,
  and $\svanc \funin \verts \to \wordsover{\verts}$ be a function.
  We say that $\atg$ is \emph{correct with respect to ancestor function $\svanc$} 
  if for all $\bvert,\bverti{0},\bverti{1}\in\verts$ and all $i\in\nat$  
  the following conditions hold (conditions in brackets $[\ldots]$ have been added for readability, but are redundant): 
  \begin{align*}
      \;\; & \Rightarrow \;\;
    \vlab{\sroot} = \soutroot 
      \logand
    \vanc{\sroot} = \emptyword      
    \\
    \vlab{\bvert} \in \asigatconstprime
      \logand
    \bvert \tgsucci{0} \bverti{0}
      \;\; & \Rightarrow \;\;
    \vanc{\bverti{0}} = \vanc{\bvert}
      \logand  
    \begin{aligned}[t]
      &
      \exists n\in\nat.\, 
        \exists\bverti{1},\ldots,\bverti{n}\in\verts.\, 
          \\[-0.5ex]
          & \hspace*{1.5ex} 
          \bigl(\,
          \bverti{0} \tgsucci{0} \bverti{1} \tgsucci{0} \ldots \tgsucci{0} \bverti{n}
          \\[-0.5ex]
          & \hspace*{1.5ex}\;\;
            \logand
            \vlab{\bverti{0}} = \ldots = \vlab{\bverti{n-1}} = \ssin
              \logand
            \vlab{\bverti{n}} = \sinroot  
          \,\bigr)   
    \end{aligned}  
    \displaybreak[0]\\
    \vlab{\bvert} \in \asigatfun 
      \logand
    \bvert \tgsucci{i} \bverti{i}
      \;\; & \Rightarrow \;\;
    \vanc{\bverti{i}} = \vanc{\bvert} 
    \displaybreak[0]\\
    \vlab{\bvert} \in \setexp{\soutroot,\ssout}
      \logand
    \bvert \tgsucci{0} \bverti{0}
      \;\; & \Rightarrow \;\;
    \vanc{\bverti{0}} = \vanc{\bvert} \wcns \bvert
    \displaybreak[0]\\          
    \vlab{\bvert} = \sinroot
      \logand
    \bvert \tgsucci{0} \bverti{0}
      \;\; & \Rightarrow \;\;
    [\, \sroot = \bverti{0} = \,] \;\, \vanc{\bverti{0}} \wcns \bverti{0} = \vanc{\bvert} 
    \;\;\; [\, \logand \vlab{\bverti{0}} = \soutroot\,] 
    \displaybreak[0]\\  
    \vlab{\bvert} = \ssin
      \logand
    \bvert \tgsucci{0} \bverti{0}
      \;\; & \Rightarrow \;\;
    \vanc{\bverti{0}} \wcns \avert = \vanc{\bvert} \text{ for some $\avert\in\verts$}
    \\          
    \vlab{\bvert} = \ssin
      \logand
    \bvert \tgsucci{1} \bverti{1}
      \;\; & \Rightarrow \;\;
    \vanc{\bverti{1}} \wcns \bverti{1} = \vanc{\bvert}
      \;\;\; [\, \logand \vlab{\bverti{1}} = \ssout \,] 
  \end{align*}
  By $\classrtgsover{\asigacc}$ we denote the class of term graphs over $\asigacc$ 
  that are correct with respect to some ancestor function. % $\svanc \funin \verts \to \wordsover{\verts}$.
  We call $\classrtgsover{\asigacc}$ the class of term graphs that \emph{represent} nested term graphs. 
\end{definition}

\begin{proposition}
  %Let $\asig$ be an \ntgsig, and let $\asigacc$ be defined as in Definition~\ref{def:ntg:to:fotg}. 
  The transformation $\sntgstortgs$ as introduced in Definition~\ref{def:ntg:to:fotg} 
  gives rise to a well-defined function 
  $\sntgstortgs \funin \classntgsover{\asig} \to \classrtgsover{\asigacc}$,
  $\antg \mapsto \ntgstortgs{\antg}$
  from $\classntgsover{\asig}$ into $\classrtgsover{\asigacc}$,
  which preserves $\sfunbisimne$ as $\sfunbisim$, and $\sbisimne$ as $\sbisim$. 
\end{proposition}

This can be proved by keeping the ancestor function of the \entg~$\aedntg$ on which the procedure starts,
and by extending it appropriately for the vertices in chains of added input vertices below vertices with constant symbols in $\aedntg$.
In this way an ancestor function is obtained with respect to which the resulting term graph is correct.
Preservation of $\sfunbisimne$ as $\sfunbisim$ along $\sntgstortgs$ can be established
by arguments using induction on the length of `access paths' 
(acyclic paths from the root to a node) in $\aedntg$ and in $\ntgstortgs{\antg}$, respectively. 

We note that the image of $\sntgstortgs$ is not all of $\classrtgsover{\asigacc}$:
e.g.\ the term graph that results from the term graph in Figure~\ref{fig:nested-and-implementation} right
by a homomorphism that identifies all vertices labeled by $\sinroot$ 
is still correct with respect to an ancestor function (compare Lemma~\ref{lem:closed:under:hom}),
but it does not arise as the interpretation of a nested term graph.

% \begin{theorem}[implementation of \ntgs\ by first-order term graphs]\label{thm:implementation}
%   Let $\asig$ be an \ntgsig, and $\asigacc = \asig \cup \Ins \cup \setexp{%\ssin/1,
%                                                                           \ssout/2,\ssinroot/1,\ssoutroot/1}$. 
%   %
%   There are functions %(homomorphism preserving functions)
%   $\sntgstortgs \funin \classntgsover{\asig} \to \classtgsover{\asigacc}$
%   and 
%   $\stgstontgs \funin \classtgsover{\asigacc} \to \classntgsover{\asig}$
%   between the classes of \ntgs\ over $\asig$ and term graphs over $\asigacc$
%   such that $\scompfuns{\stgstontgs}{\sntgstortgs} = \sidfunon{\classntgsover{\asig}}$,
%   (i.e.\ $\stgstontgs$ is a retraction of $\sntgstortgs$, and $\sntgstortgs$ is a section of $\stgstontgs$)
%   that are efficiently computable, and preserve and reflect functional bisimilarity~$\sfunbisim$.  
% \end{theorem}

As the occurrences of matching output and input vertices in the example of the term graph~$\ntgstortgs{\antg}$
in Figure~\ref{fig:nested-and-implementation} indicate, the nesting structure of a nested term graph is
preserved in its term graph interpretation. More importantly, the matching of output and input vertices
is guaranteed by the ancestor function of term graphs in $\classrtgsover{\asigacc}$.
This facilitates the definition of a representation function $\srtgstontgs$ from $\classrtgsover{\asigacc}$ back 
to $\classntgsover{\asig}$ that is the inverse of $\sntgstortgs$.
Similar as for $\sntgstortgs$, also preservation of $\sfunbisim$ and $\sbisim$ along $\srtgstontgs$ can be shown. 

\begin{theorem}%[correspondence between \ntgs\ and first-order term graphs]\label{thm:implementation}
  Let $\asig$ be an \ntgsig, and let $\asigacc$ be defined as in Definition~\ref{def:ntg:to:fotg}. 
  There is a representation function $\srtgstontgs \funin \classrtgsover{\asigacc} \to \classntgsover{\asig}$
  such that $\scompfuns{\srtgstontgs}{\sntgstortgs} = \sidfunon{\classntgsover{\asig}}$ holds
  (that is, $\srtgstontgs$ is a retraction of $\sntgstortgs$, and $\sntgstortgs$ is a section of $\srtgstontgs$).
  Furthermore,
  both of the mappings $\sntgstortgs$ and $\srtgstontgs$ 
  are efficiently computable.
  Along $\sntgstortgs$, $\sfunbisimne$ and $\sbisimne$ are preserved as $\sfunbisim$ and $\sbisim\,$, respectively;
  and along $\srtgstontgs\,$, $\sfunbisim$ and $\sbisim$ are preserved as $\sfunbisimne$ and $\sbisimne$.  
\end{theorem}

This correspondence opens up the possibility to transfer various well-known results for
term graphs to nested term graphs, such as the fact that bisimulation equivalence classes 
are, modulo isomorphism, complete lattices with respect to homomorphism. 
Another example is the existence of unique nested term graph collapses,
a result whose transfer from $\classrtgsover{\asigacc}$ to $\classntgsover{\asig}$
depends on the following lemma. 

\begin{lemma}\label{lem:closed:under:hom}
  The class $\classrtgsover{\asigacc}$ of \ntgs-representing term graphs is closed under homorphism. That is, 
  if $\atgi{1} \funbisim \atgi{2}$ holds for $\atgi{1},\atgi{2}\in\classtgsover{\asigacc}$,
  then $\atgi{1}\in\classrtgsover{\asigacc}$ implies $\atgi{2}\in\classrtgsover{\asigacc}$.
\end{lemma}

This proof of this lemma exploits the fact that the term graphs in $\classrtgsover{\asigacc}$ 
are `fully back-linked' in the following sense: for every vertex $\bvert$ and every output vertex $\avert$
that in the nesting structure resides above $\bvert$, there is a path (in forward direction) from $\bvert$ to $\avert$.
(Since in particular the root output vertex is reachable, this entails that in fact all other vertices
 are reachable by paths from $\bvert$.)
It follows that if a homomorphism $\sahom$ from an \ntg\nb-re\-pre\-sen\-ting term graph $\atgi{1}\in\classrtgsover{\asigacc}$ to 
a term graph $\atgi{2}\in\classtgsover{\asigacc}$ 
identifies two vertices $\bverti{1}$ and $\bverti{2}$, then, due to the local progression clauses of the homomorphism and due the ancestor function on $\atgi{1}$,
$\sahom$ also identifies all corresponding output vertices in the nesting hierarchy above $\bverti{1}$ and $\bverti{2}$, respectively. 
This fact makes it possible to define an ancestor function on $\atgi{2}$ for which $\atgi{2}$ is correct.
Hence $\atgi{2}\in\classrtgsover{\asigacc}$.  

% As a rather direct consequence of Theorem~\ref{thm:implementation}, 
% various well-known results for term graphs can be transferred to nested term graphs.
% As an example, we formulate the following corollary.

% \begin{lemma}%[Reflection of complete lattices under order homomorphisms with left-inverses]
%   \label{lem:reflect:complete:lattice}
%   Let $\pair{\aset}{\spoi{\aset}}$ and $\pair{\bset}{\spoi{\bset}}$ be partial orders. 
%   Suppose that 
%   $\saohom \funin \aset \to \bset$, 
%   and 
%   $\sbohom \funin \bset \to \aset$
%   are order homomorphisms such that 
%   $\sbohom$ is a left-inverse of $\saohom$, that is,
%   $\scompfuns{\sbohom}{\saohom} = \sidfunon{\aset}$ holds.
%   %
%   Then if $\pair{\bset}{\spoi{\bset}}$ is a complete lattice, then so is $\pair{\aset}{\spoi{\aset}}$.
% \end{lemma}

\begin{theorem}\label{cor:thm:implementation}
  Every nested term graph $\antg$ has, up to isomorphism, a unique nested term graph collapse.
%   Let $\antg$ be a nested term graph. 
%   $\antg$ has, up to isomorphism, a unique nested term graph collapse. 
%   The bisimulation equivalence class of $\antg$ (up to isomorphism)
%   forms a complete lattice w.r.t.\ $\sfunbisim$. 
\end{theorem}

\section{Further aims}
  \label{sec:aims}
%----------------------------------------------------------------------- 
%
We are interested in, and have started to investigate, the following further topics:

\renewcommand{\descriptionlabel}[1]%
      {\hspace{\labelsep}{\emph{#1}$\,$}}
\begin{description}
  \item[Context-free graph grammars]
    We want to view \rgss\ as context-free graph grammars in order to
    recognize \rgs-generated nested term graphs as context-free graphs.
    We expect to find a close connection. 

  \item[Proofnets]  
    Formulas containing existential or universal quantifiers, 
    mathematical expressions containing integrals or derivatives, 
    and more generally any language with binding constructs, 
    can be represented as $\lambda$-terms over a simply typed signature.
    Since proofnets refine the latter, it follows that such languages
    can be represented as proofnets over a signature typed with (MELL) formulas from linear logic.
    Such a representation should tie in with the boxed representations for
    first- and higher-order terms of the introduction, but now for nested term graphs.
    On the one hand, we expect that the development of nested term graphs 
    can profit, via this route, from the detailed studies of the fine structure of proofnets, e.g.\ 
    various notions of explicit substitution, that have been carried out in the literature 
    (for example, see Accattoli and Guerrini~\cite{acca:guer:2009}).  
    On the other hand, it is conceivable that the theory of proofnets could benefit
    from work on nested term graphs for what concerns the natural formalization 
    of infinite nesting, the concepts of bisimilarity and nested bisimilarity,  
    and the faithful representation of nested term graphs as first-order term graphs. 
    
  \item[Boxes]    
    Extending the previous item, we want to investigate the connection between 
    nested term graphs and the way how boxes that symbolize scopes are employed in 
    various settings. Apart from the box of linear logic proofnets,
    we are thinking of monads in category theory.
    These have been introduced and studied to express nested first-order signatures by
     L\"{u}th~\cite{luth:1997}, and L\"{u}th and Ghani~\cite{luth:ghan:1997},
    leading to categorical proofs of modularity results in term rewriting. 
        
    We would like to obtain a categorical semantics via algebras and coalgebras
    by viewing nested term graphs as monads over some signature, 
    analogous to how this has been done for first-order term graphs 
    by Ghani, L\"{u}th and de~Marchi \cite{ghan:luth:marc:2005}.
    Moreover, we would like to understand whether, and if so how,
    the respective monadic views can be related via our representation
    of nested term graphs as first-order term graphs.
    In this respect,  the decomposition of boxes into the opening and 
    closing `brackets' that are used in optimal graph reduction techniques 
    for the \lambdacalculus, as studied from a categorical perspective
    by Asperti~\cite{aspe:1995}, should be relevant.
     
   \item[Rewrite theory]
     In the above we have only addressed the \emph{static} aspects,
     how to represent structures with a notion of scope.
     Ultimately, our interest is in \emph{dynamic} aspects, 
     in rewriting systems for such structures.
     In particular,  since higher-order terms have a natural interpretation 
     as nested term graphs,  it is desirable to investigate implementations 
     of higher-order rewriting by nested term graph rewriting,
     and eventually, via the correspondence explained in Section~\ref{sec:interpretation}, 
     by first-order term graph rewriting. 
     Several preliminary investigations into this have been carried out, but
     from different perspectives (corresponding to the different perspectives
     on boxes as in the previous item):
     Lafont presents proofnet reduction as reduction on 
     `nested interaction nets' in~\cite{lafo:1990},
     Van Raamsdonk  defined rewriting modulo proofnets in~\cite{raam:1996},
     and L\"uth and Ghani defined and studied monadic rewriting in their cited papers.
     We first want to develop a notion of rewriting on nested terms graphs that   
     is adequate with respect to higher-order term rewriting (HRSs see e.g.~\cite{terese:2003}),
     analogous to the adequacy of first-order term graph rewriting for first-order  
     term rewriting \cite{kenn:klop:sleep:vrie:1994}.
     To that end, a notion of equivalence on nested term graphs has to be developed 
     that represents $\alpha\beta\eta$-equivalence on \HRSterms.\footnote{%
 Analogous to proofnet reduction. Rewriting should `interact nicely' with boxes.}
     This also gives rise to other questions, cf.\  \cite{slee:plas:eeke:1993}, 
     such as how to recognize (efficiently) whether
     a given (first-order representation of a) nested term graph represents 
     a higher-order term.
     Next, nested term graph rewriting should facilitate a sensible meta-theory.
     Here we may think of suitable notions of orthogonality, or termination techniques
     such as recursive path orders, similar to what has been done for first-order 
     term graph rewriting (see work by Plump~\cite{plum:1999}).    
    \end{description}

\paragraph{Acknowledgement}
  We thank the referees for many valuable comments on earlier versions of this paper, 
  and for insisting on referring to related work. 
  We also thank the editors for their patience.

\bibliographystyle{eptcs}
\bibliography{nested-pp}

\newpage
%---------------------------------------------------------------------------------------
\section*{{\Large Appendix A: Interpretation of nested term graphs by first-order term graphs}}
%---------------------------------------------------------------------------------------

\label{app:start}%
\enlargethispage{0.5ex}
%\begin{example}\label{ex:implementation}
We showcase the transformation process according to Definition~\ref{def:ntg:to:fotg}
of a nested term graph $\antg$ into its first-order term graph interpretation~$\ntgstortgs{\antg}$
for the example of the nested term graph defined in Example~\ref{ex:ntg} and Figure~\ref{fig:ex:rgs:ntg}. 
We start from the \entg\nb-re\-pre\-sen\-ta\-tion $\aedntg$ of $\antg\,$  as illustrated in Figure~\ref{fig:edntg}:
  %\vspace*{-0.5ex}
\begin{center}   
  \scalebox{0.9}{\begin{tikzpicture}[scale=0.75,
                    node/.style={%
                      draw,
                      circle,
                      inner sep=0,
                      outer sep=0,
                      minimum size=0,
                      node distance=0}]
\matrix[row sep=0.4cm,column sep=0.4cm,every node/.style={draw,thick,circle,scale=0.75,minimum size=0.6cm,inner sep=0pt}%,outer sep=.5mm}
                                                                                                                      ]{ 
  & & \node(root_F1){$\ssout$}; & &[5ex] & &[1ex] \node(root_R){$\ssout$}; & & \node(root){$\sntgrootsym$}; & &[5ex] \node(root_G){$\ssout$}; &[5ex] & &[2.5ex] \node(root_F2){$\ssout$};
  \\
  & & \node(0_F1){$\sslabs$}; & & & & \node(0_R){$\sslabs$}; & & & &     \node(0_G){$\sslabs$};  & & & \node(0_F2){$\sslabs$};
  \\
  & & \node(00_F1){$\sslapp$}; & & & & \node(00_R){$\sslapp$};  & & & &  \node(00_G){$\snlvar$}; & & & \node(00_F2){$\sslapp$}; 
  \\
  & \node(000_F1){$\sslapp$}; & & & & \node(000_R){$\afunsymi{1}$}; & & \node(001_R){$\afunsymi{2}$}; & & & & & \node(000_F2){$\sslapp$}; & & & & \node(001_F2){$\sslapp$}; 
  \\
  & & \node(0001_F1){$\snlvar$}; & & & \node(0000_R){$\snlvar$}; & \node(0010_R){$\snlvar$}; & & \node(0011_R){$\bfunsym$}; & & & & & \node(0001_F2){$\snlvar$}; & & \node(0010_F2){$\sslapp$};    
  \\  
  \node(0000_F1){$\ssini{1}$}; & & & & & & & & & & & & & & & & \node(00101_F2){$\snlvar$};
  \\[-2ex]
  & & & & & & & & & & & \node(0000_F2){$\ssini{1}$}; & & & \node(00100_F2){$\ssini{2}$};
  \\
  };
\path (root) ++ (+0.4cm,-0.35cm) node {$\scriptstyle\averti{0}$};
\path (root) ++ (-0.5cm,-0.35cm) node {$\scriptstyle\prefix{\emptyword}$};

\path (root_R) ++ (-0.6cm,-0.2cm) node {$\scriptstyle\prefix{\averti{0}}$};
\path (0_R) ++ (-0.6cm,-0.2cm) node {$\scriptstyle\prefix{\averti{0}}$};
\path (00_R) ++ (-0.6cm,-0.2cm) node {$\scriptstyle\prefix{\averti{0}}$};
\path (000_R) ++ (-0.6cm,-0.2cm) node {$\scriptstyle\prefix{\averti{0}}$};
\path (001_R) ++ (-0.6cm,-0.2cm) node {$\scriptstyle\prefix{\averti{0}}$};
\path (0000_R) ++ (-0.6cm,-0.2cm) node {$\scriptstyle\prefix{\averti{0}}$};
\path (0010_R) ++ (-0.6cm,-0.2cm) node {$\scriptstyle\prefix{\averti{0}}$};
\path (0011_R) ++ (-0.6cm,-0.2cm) node {$\scriptstyle\prefix{\averti{0}}$};

\path (root_F1) ++ (-0.75cm,-0.2cm) node {$\scriptstyle\prefix{\averti{0}\averti{1}}$};
\path (0_F1) ++ (-0.75cm,-0.2cm) node {$\scriptstyle\prefix{\averti{0}\averti{1}}$};
\path (00_F1) ++ (-0.75cm,-0.2cm) node {$\scriptstyle\prefix{\averti{0}\averti{1}}$};
\path (000_F1) ++ (-0.75cm,-0.2cm) node {$\scriptstyle\prefix{\averti{0}\averti{1}}$};
\path (0000_F1) ++ (-0.75cm,-0.2cm) node {$\scriptstyle\prefix{\averti{0}\averti{1}}$};
\path (0001_F1) ++ (-0.75cm,-0.2cm) node {$\scriptstyle\prefix{\averti{0}\averti{1}}$};

\path (root_G) ++ (-0.75cm,-0.2cm) node {$\scriptstyle\prefix{\averti{0}\averti{3}}$};
\path (0_G) ++ (-0.75cm,-0.2cm) node {$\scriptstyle\prefix{\averti{0}\averti{3}}$};
\path (00_G) ++ (-0.75cm,-0.2cm) node {$\scriptstyle\prefix{\averti{0}\averti{3}}$};

\path (root_F2) ++ (-0.75cm,-0.2cm) node {$\scriptstyle\prefix{\averti{0}\averti{2}}$};
\path (0_F2) ++ (-0.75cm,-0.2cm) node {$\scriptstyle\prefix{\averti{0}\averti{2}}$};
\path (00_F2) ++ (-0.75cm,-0.2cm) node {$\scriptstyle\prefix{\averti{0}\averti{2}}$};
\path (000_F2) ++ (-0.75cm,-0.2cm) node {$\scriptstyle\prefix{\averti{0}\averti{2}}$};
\path (0000_F2) ++ (-0.75cm,-0.2cm) node {$\scriptstyle\prefix{\averti{0}\averti{2}}$};
\path (0001_F2) ++ (-0.75cm,-0.2cm) node {$\scriptstyle\prefix{\averti{0}\averti{2}}$};
\path (001_F2) ++ (-0.75cm,-0.2cm) node {$\scriptstyle\prefix{\averti{0}\averti{2}}$};
\path (0010_F2) ++ (-0.75cm,-0.2cm) node {$\scriptstyle\prefix{\averti{0}\averti{2}}$};
\path (00100_F2) ++ (-0.75cm,-0.2cm) node {$\scriptstyle\prefix{\averti{0}\averti{2}}$};
\path (00101_F2) ++ (-0.75cm,-0.2cm) node {$\scriptstyle\prefix{\averti{0}\averti{2}}$};

\path (000_R) ++ (+0.4cm,-0.35cm) node {$\scriptstyle\averti{1}$};
\path (001_R) ++ (+0.5cm,-0.2cm) node {$\scriptstyle\averti{2}$};
\path (0011_R) ++ (+0.5cm,-0.2cm) node {$\scriptstyle\averti{3}$};

\draw[<-,thick,>=latex](root) -- ++ (90:0.7cm);
\draw[->](root_R) to (0_R);
\draw[->](0_R) to (00_R);
\draw[->](00_R) to  (000_R);  \draw[->](00_R) to  (001_R); 
\draw[->](000_R) to (0000_R); \draw[->](001_R) to (0010_R); \draw[->](001_R) to (0011_R); 
\draw[->](root_F1) to (0_F1); 
\draw[->](0_F1) to (00_F1);
\draw[->](00_F1) to (000_F1); \draw[->](00_F1) to[out=-50,in=50,distance=1.75cm] (00_F1);
\draw[->](000_F1) to (0000_F1); \draw[->](000_F1) to (0001_F1);
\draw[->](root_F2) to (0_F2);
\draw[->](0_F2) to (00_F2);
\draw[->](00_F2) to (001_F2);
\draw[->](00_F2) to (000_F2);
\draw[->](000_F2) to (0000_F2);
\draw[->](000_F2) to (0001_F2);
\draw[->](001_F2) to (0010_F2);
\draw[->](001_F2) to[out=-10,in=50,distance=1.5cm] (00_F2);
\draw[->](0010_F2) to (00100_F2);
\draw[->](0010_F2) to (00101_F2);
\draw[->](root_G) to (0_G);
\draw[->](0_G) to (00_G);
\begin{scope}[|->,shorten <=3pt,shorten >=3pt]  
\draw[|->,thick,densely dashed]
  (root) to node[above]{$\svin$} (root_R);
\draw[|->,thick,densely dashed] 
  (000_R) to node[above]{$\;\;\;\;\;\;\;\svin$}  (root_F1); 
\draw[|->,thick,densely dashed,bend right,distance=2cm,shorten <=10pt]
  (001_R) to node[below]{$\svin$}  (root_F2); 
\draw[|->,thick,densely dashed,bend left,distance=1.25cm,shorten >=13pt] 
  (0011_R) to node[above]{$\svin\hspace*{4ex}$} (root_G);  
\draw[|->,thick,densely dashed,shorten >=16pt]
  (0000_F1) to node[below]{$\svout$} (0000_R); 
\draw[|->,thick,densely dashed,shorten <=5pt]
  (0000_F2) to node[below]{$\svout$} (0010_R);
\draw[|->,thick,densely dashed,bend right,distance=1.5cm]
  (00100_F2) to node[above]{$\svout$} (0011_R);
\end{scope}          
\end{tikzpicture} }
\end{center}  
By (i)~removing every vertex $\avert$ with a nested symbol, 
redirecting incoming edges at $\avert$ to the vertex $\vin{\avert}$,
and
(ii)~relabeling every input vertex $\bvert$ with the nullary label $\ssini{k}$ 
    by the binary label $\ssin$, 
    thereby directing the first edge from $\bvert$ to $\vout{\bvert}$
    (the second edge will be dealt with later), 
we obtain: % the first-order term graph:
   %\vspace*{-0.5ex}
\begin{center}   
  \scalebox{0.9}{\begin{tikzpicture}[scale=0.75,
                    node/.style={%
                      draw,
                      circle,
                      inner sep=0,
                      outer sep=0,
                      minimum size=0,
                      node distance=0}]
\matrix[row sep=0.4cm,column sep=0.4cm,every node/.style={draw,thick,circle,scale=0.75,minimum size=0.6cm,inner sep=0pt}%,outer sep=.5mm}
                                                                                                                      ]{ 
  & & \node(root_F1){$\ssout$}; & &[5ex] & &[1ex] \node(root_R){$\ssout$}; & & \node[densely dotted](root){$\sntgrootsym$}; & &[5ex] \node(root_G){$\ssout$}; &[5ex] & &[2.5ex] \node(root_F2){$\ssout$};
  \\
  & & \node(0_F1){$\sslabs$}; & & & & \node(0_R){$\sslabs$}; & & & &     \node(0_G){$\sslabs$};  & & & \node(0_F2){$\sslabs$};
  \\
  & & \node(00_F1){$\sslapp$}; & & & & \node(00_R){$\sslapp$};  & & & &  \node(00_G){$\snlvar$}; & & & \node(00_F2){$\sslapp$}; 
  \\
  & \node(000_F1){$\sslapp$}; & & & & \node[densely dotted](000_R){$\afunsymi{1}$}; & & \node[densely dotted](001_R){$\afunsymi{2}$}; & & & & & \node(000_F2){$\sslapp$}; & & & & \node(001_F2){$\sslapp$}; 
  \\
  & & \node(0001_F1){$\snlvar$}; & & & \node(0000_R){$\snlvar$}; & \node(0010_R){$\snlvar$}; & & \node[densely dotted](0011_R){$\bfunsym$}; & & & & & \node(0001_F2){$\snlvar$}; & & \node(0010_F2){$\sslapp$};    
  \\  
  \node(0000_F1){$\sinvert$}; & & & & & & & & & & & & & & & & \node(00101_F2){$\snlvar$};
  \\[-2ex]
  & & & & & & & & & & & \node(0000_F2){$\sinvert$}; & & & \node(00100_F2){$\sinvert$};
  \\
  };
 
\draw[<-,thick,>=latex](root) -- ++ (90:0.7cm);
\draw[->](root_R) to (0_R);
\draw[->](0_R) to (00_R);
\draw[->,dotted](00_R) to  (000_R);  \draw[->,dotted](00_R) to  (001_R); 
\draw[->,dotted](000_R) to (0000_R); \draw[->,dotted](001_R) to (0010_R); \draw[->,dotted](001_R) to (0011_R); 
\draw[->](root_F1) to (0_F1); 
\draw[->](0_F1) to (00_F1);
\draw[->](00_F1) to (000_F1); \draw[->](00_F1) to[out=-50,in=50,distance=1.75cm] (00_F1);
\draw[->](000_F1) to (0000_F1); \draw[->](000_F1) to (0001_F1);
\draw[->](root_F2) to (0_F2);
\draw[->](0_F2) to (00_F2);
\draw[->](00_F2) to (001_F2);
\draw[->](00_F2) to (000_F2);
\draw[->](000_F2) to (0000_F2);
\draw[->](000_F2) to (0001_F2);
\draw[->](001_F2) to (0010_F2);
\draw[->](001_F2) to[out=-10,in=50,distance=1.5cm] (00_F2);
\draw[->](0010_F2) to (00100_F2);
\draw[->](0010_F2) to (00101_F2);
\draw[->](root_G) to (0_G);
\draw[->](0_G) to (00_G);
\begin{scope}[|->,shorten <=3pt,shorten >=3pt]  
\draw[|->,densely dashed]
  (root) to node[below]{$\scriptstyle\svin$} (root_R);
\draw[|->,densely dashed] 
  (000_R) to node[above]{$\;\;\;\;\;\;\;\scriptstyle\svin$}  (root_F1); 
\draw[|->,densely dashed,bend right,distance=2.5cm]
  (001_R) to node[below]{$\scriptstyle\svin$}  (root_F2); 
\draw[|->,densely dashed,bend left,distance=1.25cm] 
  (0011_R) to node[above]{$\scriptstyle\svin\hspace*{4ex}$} (root_G);  
\draw[|->,densely dashed]
  (0000_F1) to node[below]{$\scriptstyle\svout$} (0000_R); 
\draw[|->,densely dashed]
  (0000_F2) to node[below]{$\scriptstyle\svout$} (0010_R);
\draw[|->,densely dashed,bend right,distance=1.5cm]
  (00100_F2) to node[above]{$\scriptstyle\svout$} (0011_R);
\end{scope}

%\draw[<-,thick,>=latex](root) -- ++ (90:0.7cm);

\path (root) ++ (90:0.7cm) node (rootedgestart){};

\draw[->,thick,bend right,distance=0.5cm]
  (rootedgestart) to (root_R);
\draw[->,thick,bend left,distance=2.5cm]
  (00_R.220) to (root_F1);
\draw[->,thick,bend right,distance=4cm]
  (00_R.320) to (root_F2);  
\draw[->,thick,bend left,distance=1cm]
  (00100_F2.180) to (root_G.305);
\draw[->,thick,bend right,distance=1cm]
  (0000_F1.330) to (0000_R);
\draw[->,thick,bend left,distance=1cm]
  (0000_F2.220) to (0010_R);
       
\end{tikzpicture} }
\end{center}
Then by $\text{(ii)}'$~directing the second edge from 
an input vertex $\bvert$ to become a backlink from $\bvert$ to 
the corresponding output vertex, 
that is, the vertex $\averti{n}$ where $\vanc{\bvert} = \averti{1}\ldots\averti{n}$
    (then $\averti{n}$ is guaranteed to be an output vertex, i.e.\ it is labeled by $\ssout$),
we obtain the term graph: 
  %\vspace*{-0.5ex}
\begin{center}   
  \scalebox{0.9}{\begin{tikzpicture}[scale=0.75,
                    node/.style={%
                      draw,
                      circle,
                      inner sep=0,
                      outer sep=0,
                      minimum size=0,
                      node distance=0}]
\matrix[row sep=0.4cm,column sep=0.4cm,every node/.style={draw,thick,circle,scale=0.75,minimum size=0.6cm,inner sep=0pt}%,outer sep=.5mm}
                                                                                                                      ]{ 
  & & \node(root_F1){$\ssout$}; & &[5ex] & &[1ex] \node(root_R){$\ssout$}; & & \node[draw=none](root){$\phantom{\sntgrootsym}$}; & &[5ex] \node(root_G){$\ssout$}; &[5ex] & &[2.5ex] \node(root_F2){$\ssout$};
  \\
  & & \node(0_F1){$\sslabs$}; & & & & \node(0_R){$\sslabs$}; & & & &     \node(0_G){$\sslabs$};  & & & \node(0_F2){$\sslabs$};
  \\
  & & \node(00_F1){$\sslapp$}; & & & & \node(00_R){$\sslapp$};  & & & &  \node(00_G){$\snlvar$}; & & & \node(00_F2){$\sslapp$}; 
  \\
  & \node(000_F1){$\sslapp$}; & & & & \node[draw=none](000_R){$\phantom{\afunsymi{1}}$}; & & \node[draw=none](001_R){$\phantom{\afunsymi{2}}$}; & & & & & \node(000_F2){$\sslapp$}; & & & & \node(001_F2){$\sslapp$}; 
  \\
  & & \node(0001_F1){$\snlvar$}; & & & \node(0000_R){$\snlvar$}; & \node(0010_R){$\snlvar$}; & & \node[draw=none](0011_R){$\phantom{\bfunsym}$}; & & & & & \node(0001_F2){$\snlvar$}; & & \node(0010_F2){$\sslapp$};    
  \\  
  \node(0000_F1){$\sinvert$}; & & \node[draw=none](helper_F1){};& & & & & & & & & & & & & & \node(00101_F2){$\snlvar$};
  \\[-2ex]
  & & & & & & & & & & & \node(0000_F2){$\sinvert$}; & & \node[draw=none](helper_F2){}; & \node(00100_F2){$\sinvert$};
  \\
  };

\draw[->](root_R) to (0_R);
\draw[->](0_R) to (00_R);
\draw[->,draw=none](00_R) to  (000_R);  \draw[->,draw=none](00_R) to  (001_R); 
\draw[->,draw=none](000_R) to (0000_R); \draw[->,draw=none](001_R) to (0010_R); \draw[->,draw=none](001_R) to (0011_R); 
\draw[->](root_F1) to (0_F1); 
\draw[->](0_F1) to (00_F1);
\draw[->](00_F1) to (000_F1); \draw[->](00_F1) to[out=-50,in=50,distance=1.75cm] (00_F1);
\draw[->](000_F1) to (0000_F1); \draw[->](000_F1) to (0001_F1);
\draw[->](root_F2) to (0_F2);
\draw[->](0_F2) to (00_F2);
\draw[->](00_F2) to (001_F2);
\draw[->](00_F2) to (000_F2);
\draw[->](000_F2) to (0000_F2);
\draw[->](000_F2) to (0001_F2);
\draw[->](001_F2) to (0010_F2);
\draw[->](001_F2) to[out=-10,in=50,distance=1.5cm] (00_F2);
\draw[->](0010_F2) to (00100_F2);
\draw[->](0010_F2) to (00101_F2);
\draw[->](root_G) to (0_G);
\draw[->](0_G) to (00_G); 

% backlinks
%
% for F2
\draw[->,thick]
  (00100_F2) -- ($ (helper_F2) + (5.5cm,0) $);
\draw[->,thick]  
  ($ (helper_F2) + (5.5cm,0) $)
             -- ($ (root_F2) + (5.5cm,0) $)
             -- (root_F2); 
% \draw[->,thick]
%   (0000_F2)  -- ($ (0000_F2.west) + (2cm,0)$)  
%              -- ($ (0000_F2.west) + (2cm,-0.75cm)$)
%              -- ($ (helper_F2.west) + (5cm,-0.75cm) $)
%              -- ($ (helper_F2.west) + (5cm,0) $);
\draw[->,thick]
  (0000_F2)  -- ($ (0000_F2) + (0cm,-0.75cm)$)
             -- ($ (helper_F2) + (5.5cm,-0.75cm) $)
             -- ($ (helper_F2) + (5.5cm,0) $);
                          
% for F1             
\draw[->,thick]
  (0000_F1)  -- ($ (helper_F1.west) + (1.8cm,0) $)
             -- ($ (root_F1.west) + (1.8cm,0) $)
             -- (root_F1);             

% call and return links
\draw[<-,>=latex,thick](root_R) -- ++ (90:0.7cm);
\draw[->,bend left,distance=2.5cm]
  (00_R.220) to (root_F1);
\draw[->,bend right,distance=4cm]
  (00_R.320) to (root_F2);  
\draw[->,bend left,distance=1cm]
  (00100_F2.180) to (root_G.305);
\draw[->,bend right,distance=2.5cm]
  (0000_F1.330) to (0000_R);
\draw[->,bend left,distance=1cm]
  (0000_F2.220) to (0010_R);
       
\end{tikzpicture} }
\end{center}

\newpage
\noindent\enlargethispage{16ex}
\vspace*{-5ex}
\begin{center}
  \scalebox{0.7}{\begin{tikzpicture}[scale=0.75,
                    node/.style={%
                      draw,
                      circle,
                      inner sep=0,
                      outer sep=0,
                      minimum size=0,
                      node distance=0}]
\matrix[row sep=0.4cm,column sep=0.4cm,every node/.style={draw,thick,circle,scale=0.75,minimum size=0.6cm,inner sep=0pt}%,outer sep=.5mm}
                                                                                                                      ]{ 
  & & \node(root_F1){$\ssout$}; & &[5ex] & &[1ex] \node(root_R){$\ssout$}; & & \node[draw=none](root){$\phantom{\sntgrootsym}$}; & &[5ex] \node(root_G){$\ssout$}; &[5ex] & &[2.5ex] \node(root_F2){$\ssout$};
  \\
  & & \node(0_F1){$\sslabs$}; & & & & \node(0_R){$\sslabs$}; & & & &     \node(0_G){$\sslabs$};  & & & \node(0_F2){$\sslabs$};
  \\
  & & \node(00_F1){$\sslapp$}; & & & & \node(00_R){$\sslapp$};  & & & &  \node(00_G){$\snlvar$}; & & & \node(00_F2){$\sslapp$}; 
  \\
  & \node(000_F1){$\sslapp$}; & & & & \node[draw=none](000_R){$\phantom{\afunsymi{1}}$}; & & \node[draw=none](001_R){$\phantom{\afunsymi{2}}$}; & & & & & \node(000_F2){$\sslapp$}; & & & & \node(001_F2){$\sslapp$}; 
  \\
  & & \node(0001_F1){$\snlvar$}; & & & \node(0000_R){$\snlvar$}; & \node(0010_R){$\snlvar$}; & & \node[draw=none](0011_R){$\phantom{\bfunsym}$}; & & & & & \node(0001_F2){$\snlvar$}; & & \node(0010_F2){$\sslapp$};    
  \\  
  \node(0000_F1){$\sinvert$}; & & \node[draw=none](helper_F1){};& & & & & & & & & & & & & & \node(00101_F2){$\snlvar$};
  \\[-2ex]
  & & & & & & & & & & & \node(0000_F2){$\sinvert$}; & & \node[draw=none](helper_F2){}; & \node(00100_F2){$\sinvert$};
  \\
  };

\draw[->](root_R) to (0_R);
\draw[->](0_R) to (00_R);
\draw[->,draw=none](00_R) to  (000_R);  \draw[->,draw=none](00_R) to  (001_R); 
\draw[->,draw=none](000_R) to (0000_R); \draw[->,draw=none](001_R) to (0010_R); \draw[->,draw=none](001_R) to (0011_R); 
\draw[->](root_F1) to (0_F1); 
\draw[->](0_F1) to (00_F1);
\draw[->](00_F1) to (000_F1); \draw[->](00_F1) to[out=-50,in=50,distance=1.75cm] (00_F1);
\draw[->](000_F1) to (0000_F1); \draw[->](000_F1) to (0001_F1);
\draw[->](root_F2) to (0_F2);
\draw[->](0_F2) to (00_F2);
\draw[->](00_F2) to (001_F2);
\draw[->](00_F2) to (000_F2);
\draw[->](000_F2) to (0000_F2);
\draw[->](000_F2) to (0001_F2);
\draw[->](001_F2) to (0010_F2);
\draw[->](001_F2) to[out=-10,in=50,distance=1.5cm] (00_F2);
\draw[->](0010_F2) to (00100_F2);
\draw[->](0010_F2) to (00101_F2);
\draw[->](root_G) to (0_G);
\draw[->](0_G) to (00_G); 

% backlinks
%
% for F2
\draw[->]
  (00100_F2) -- ($ (helper_F2) + (5.5cm,0) $);
\draw[->]  
  ($ (helper_F2) + (5.5cm,0) $)
             -- ($ (root_F2) + (5.5cm,0) $)
             -- (root_F2); 
\draw[->]
  (0000_F2)  -- ($ (0000_F2) + (0cm,-0.75cm)$)
             -- ($ (helper_F2) + (5.5cm,-0.75cm) $)
             -- ($ (helper_F2) + (5.5cm,0) $);             
% for F1             
\draw[->]
  (0000_F1)  -- ($ (helper_F1.west) + (1.8cm,0) $)
             -- ($ (root_F1.west) + (1.8cm,0) $)
             -- (root_F1);             

% call and return links
\draw[<-,>=latex,thick](root_R) -- ++ (90:0.7cm);
\draw[->,bend left,distance=2.5cm]
  (00_R.220) to (root_F1);
\draw[->,bend right,distance=4cm]
  (00_R.320) to (root_F2);  
\draw[->,bend left,distance=1cm]
  (00100_F2.180) to (root_G.305);
\draw[->,bend right,distance=2.5cm]
  (0000_F1.330) to (0000_R);
\draw[->,bend left,distance=1cm]
  (0000_F2.220) to (0010_R);
       
\end{tikzpicture} }
\end{center}
(The downscaled term graph above repeats the last one from the previous page.)
Finally, 
(iii)~by relabeling the output vertex at the root with the special symbol $\ssoutroot$,
(iv)~by changing the arity of variable vertices (labeled by $\snlvar$)
     from zero to one (labeled by $\snlvar^{\prime}$), 
     and letting the outgoing edges target a chain of new binary input vertices 
     through the nesting structure 
     (again the ancestor function can be used for this purpose)
     towards outermost input vertices that get label $\sinroot$ and backlinks to $\soutroot$,
we obtain: 
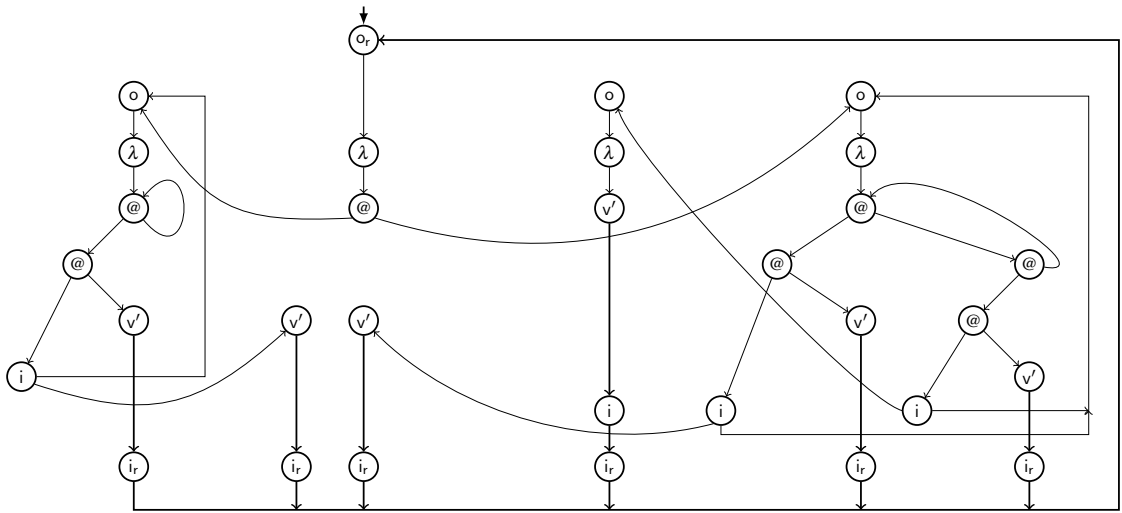
\begin{center} 
  \scalebox{0.85}{\begin{tikzpicture}[scale=0.75,
                    node/.style={%
                      draw,
                      circle,
                      inner sep=0,
                      outer sep=0,
                      minimum size=0,
                      node distance=0}]
\matrix[row sep=0.4cm,column sep=0.4cm,every node/.style={draw,thick,circle,scale=0.75,minimum size=0.6cm,inner sep=0pt}%,outer sep=.5mm}
                                                                                                                      ]{ 
  & &                           & &[5ex] & &[1ex] \node(root_R){$\ssoutroot$};
  \\                                                                                                                      
  & & \node(root_F1){$\ssout$}; & &[5ex] & &[1ex]                          & & \node[draw=none](root){$\phantom{\sntgrootsym}$}; & &[5ex] \node(root_G){$\ssout$}; &[5ex] & &[2.5ex] \node(root_F2){$\ssout$};
  \\
  & & \node(0_F1){$\sslabs$}; & & & & \node(0_R){$\sslabs$}; & & & &     \node(0_G){$\sslabs$};  & & & \node(0_F2){$\sslabs$};
  \\
  & & \node(00_F1){$\sslapp$}; & & & & \node(00_R){$\sslapp$};  & & & &  \node(00_G){$\snlvaracc$}; & & & \node(00_F2){$\sslapp$}; 
  \\
  & \node(000_F1){$\sslapp$}; & & & & \node[draw=none](000_R){$\phantom{\afunsymi{1}}$}; & & \node[draw=none](001_R){$\phantom{\afunsymi{2}}$}; & & & & & \node(000_F2){$\sslapp$}; & & & & \node(001_F2){$\sslapp$}; 
  \\
  & & \node(0001_F1){$\snlvaracc$}; & & & \node(0000_R){$\snlvaracc$}; & \node(0010_R){$\snlvaracc$}; & & \node[draw=none](0011_R){$\phantom{\bfunsym}$}; & & & & & \node(0001_F2){$\snlvaracc$}; & & \node(0010_F2){$\sslapp$};    
  \\  
  \node(0000_F1){$\sinvert$}; & & \node[draw=none](helper_F1){};& & & & & & & & & & & & & & \node(00101_F2){$\snlvaracc$};
  \\[-2ex]
  & & & & & & & & & & \node(000_G){$\ssin$}; & \node(0000_F2){$\sinvert$}; & & \node[draw=none](helper_F2){}; & \node(00100_F2){$\sinvert$};
  \\
   & & \node(00010_F1){$\ssinroot$}; & & & \node(00000_R){$\ssinroot$}; & \node(00100_R){$\ssinroot$}; & & & & \node(0000_G){$\ssinroot$};
    & & & \node(000100_F2){$\ssinroot$}; & & & \node(001010_F2){$\ssinroot$};
  \\
  };

\draw[->](root_R) to (0_R);
\draw[->](0_R) to (00_R);
\draw[->,draw=none](00_R) to  (000_R);  \draw[->,draw=none](00_R) to  (001_R); 
\draw[->,draw=none](000_R) to (0000_R); \draw[->,draw=none](001_R) to (0010_R); \draw[->,draw=none](001_R) to (0011_R); 
\draw[->](root_F1) to (0_F1); 
\draw[->](0_F1) to (00_F1);
\draw[->](00_F1) to (000_F1); \draw[->](00_F1) to[out=-50,in=50,distance=1.75cm] (00_F1);
\draw[->](000_F1) to (0000_F1); \draw[->](000_F1) to (0001_F1);
\draw[->](root_F2) to (0_F2);
\draw[->](0_F2) to (00_F2);
\draw[->](00_F2) to (001_F2);
\draw[->](00_F2) to (000_F2);
\draw[->](000_F2) to (0000_F2);
\draw[->](000_F2) to (0001_F2);
\draw[->](001_F2) to (0010_F2);
\draw[->](001_F2) to[out=-10,in=50,distance=1.5cm] (00_F2);
\draw[->](0010_F2) to (00100_F2);
\draw[->](0010_F2) to (00101_F2);
\draw[->](root_G) to (0_G);
\draw[->](0_G) to (00_G); 

% backlinks
%
% for F2
\draw[->]
  (00100_F2) -- ($ (helper_F2) + (4.75cm,0) $);
\draw[->]  
  ($ (helper_F2) + (4.75cm,0) $)
             -- ($ (root_F2) + (4.75cm,0) $)
             -- (root_F2); 
\draw[->]
  (0000_F2)  -- ($ (0000_F2) + (0cm,-0.5cm)$)
             -- ($ (helper_F2) + (4.75cm,-0.5cm) $)
             -- ($ (helper_F2) + (4.75cm,0) $);    
% for F1             
\draw[->]
  (0000_F1)   -- ($ (helper_F1.west) + (1.8cm,0) $)
              -- ($ (root_F1.west) + (1.8cm,0) $)
              -- (root_F1);  
             
% for R
\draw[->,thick]
  (00010_F1)  -- ($ (00010_F1) + (0,-0.9cm)$)
              -- ($ (00100_R) + (15.75cm,-0.9cm)$) 
              -- ($ (root_R) + (15.75cm,0)$)
              -- (root_R);
\draw[->,thick]
  (00000_R)   -- ($ (00000_R) + (0,-0.9cm)$);  
\draw[->,thick]
  (00100_R)   -- ($ (00100_R) + (0,-0.9cm)$); 
\draw[->,thick]
  (0000_G)    -- ($ (0000_G) + (0,-0.9cm)$);   
\draw[->,thick]
  (000100_F2) -- ($ (000100_F2) + (0,-0.9cm)$); 
\draw[->,thick]
  (001010_F2) -- ($ (001010_F2) + (0,-0.9cm)$);          
\draw[->,thick]
  (0001_F1)   -- (00010_F1); 
\draw[->,thick]
  (0000_R)    -- (00000_R);  
\draw[->,thick]
  (0010_R)    -- (00100_R); 
\draw[->,thick]
  (00_G)      -- (000_G);
\draw[->,thick]  
  (000_G)     -- (0000_G);    
\draw[->,thick]
  (0001_F2)  -- (000100_F2); 
\draw[->,thick]
  (00101_F2) -- (001010_F2);

% call and return links
\draw[<-,>=latex,thick](root_R) -- ++ (90:0.7cm);
\draw[->,bend left,distance=2.5cm]
  (00_R.220) to (root_F1);
\draw[->,bend right,distance=4cm]
  (00_R.320) to (root_F2);  
\draw[->,bend left,distance=1cm]
  (00100_F2.180) to (root_G.305);
\draw[->,bend right,distance=2.5cm]
  (0000_F1.330) to (0000_R);
\draw[->,bend left,distance=2.5cm]
  (0000_F2.240) to (0010_R);
       
\end{tikzpicture} }
\end{center}
This term graph is the result of the transformation process
of $\antg$ into its first-order term graph interpretation~$\ntgstortgs{\antg}$.
It is isomorphic to the term graph below right, and also right in Figure~\ref{fig:nested-and-implementation}. 
In order to facilitate a quick structural comparison with the nested term graph $\antg$ on which the input~$\aedntg$ of this procedure is based,
we also show again, below left, the `pretty print' of
the nested term graph $\antg$ (from Figure~\ref{fig:nested-and-implementation} left).
  % This term graph $\ntgstortgsprime{\antg}$ (see also below right) is %isomorphic
  %                    {a close variant} of the term graph~$\ntgstortgs{\antg}$ right on page~\pageref{fig:nested-and-implementation}
  % {(the difference between $\ntgstortgs{\antg}$ and $\ntgstortgsprime{\antg}$ is indicated below right by two dotted edges 
  %   that are present in $\ntgstortgs{\antg}$,
  %   but absent in $\ntgstortgsprime{\antg}$)}.
  % Note that in three cases backlinks (once from an $\ssout$\nb-vertex, and twice from an $\ssoutroot$\nb-vertex)
  % are only indicated in $\ntgstortgsprime{\antg}$ as illustrated below right.  
  % In order to facilitate a quick structural comparison with the input of this procedure,
  % we also show again, below left, the `pretty print' of
  % the nested term graph $\antg$ (as left in the figure on page~\pageref{fig:nested-and-implementation}).
\begin{flushleft}
  \hspace*{2ex}
  \scalebox{0.78}{\begin{minipage}{\textwidth}
                   \input{figs/nested-and-implementation-ext}
                 \end{minipage}}
    %{\input{figs/nested-named-and-implementation-1}}
\end{flushleft}%
\label{app:end}
% \begin{center}  
%    
%   \input{figs/ex-implementation-ext}
% \end{center}
%\end{example}

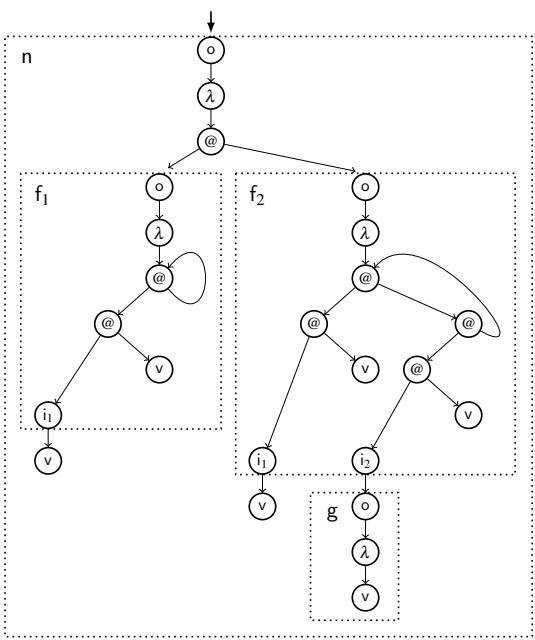
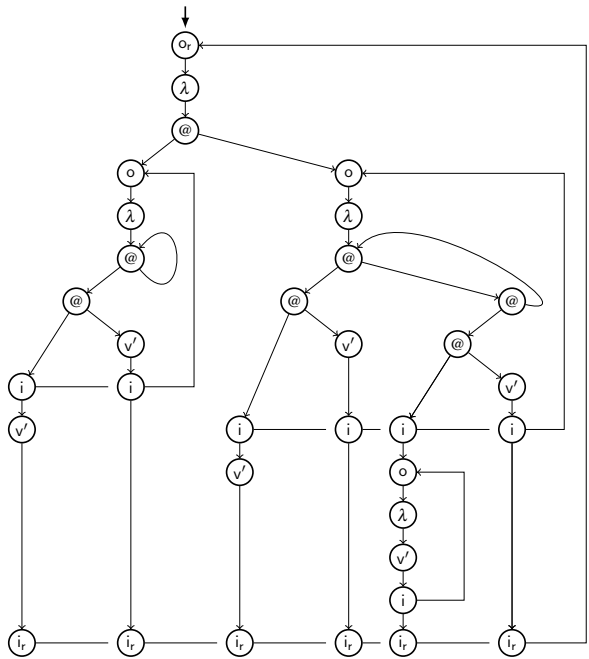
\end{document}